\documentclass[article,10pt]{IEEEtran}
\usepackage{amsmath}
\usepackage{amssymb}
\usepackage{graphicx}
\usepackage{color} 
\usepackage{xcolor}
\usepackage{cite}
\usepackage{bm}
\usepackage{epsfig,psfrag}
\usepackage{tabularx}
\usepackage{acronym}
\usepackage[yyyymmdd,hhmmss]{datetime}
\usepackage{bbm}
\usepackage{notation}
\usepackage{mathrsfs} 
\usepackage{bardiag}
\usepackage{tabularx}
\usepackage{multirow}
\usepackage{colortbl,pgfplotstable}
\usepackage{xr}
\usepackage{pdfpages}
\usepackage{multirow}
\usepackage{makecell}
\usepackage{booktabs}
\usepackage{enumitem} 
\usepackage[caption=false,font=footnotesize]{subfig}
\allowdisplaybreaks
\newcolumntype{L}[1]{>{\raggedright\arraybackslash}p{#1}}
\newcolumntype{C}[1]{>{\centering\arraybackslash}p{#1}}
\newcolumntype{R}[1]{>{\raggedleft\arraybackslash}p{#1}}

\providecommand{\ist}{\hspace*{.3mm}}

\providecommand{\rmv}{\hspace*{-.3mm}}
\providecommand{\iist}{\hspace*{1mm}}

\providecommand{\nn}{\nonumber}

\newcommand{\T}{\text{T}}

\newcommand{\vu}[2]{\mbox{$#1\,\text{#2}$}} %
\newcommand{\atantwo}{\text{atan2}}

\newtheorem{problem*}{Problem}

\acrodef{pnt}[PNT]{positioning, navigation and timing}
\acrodef{pa}[PA]{physical anchor}
\acrodef{va}[VA]{``virtual anchor''}
\acrodef{aoa}[AOA]{angle-of-arrival}
\acrodef{mva}[MVA]{master virtual anchor}
\acrodef{los}[LOS]{line-of-sight}
\acrodef{mpc}[MPC]{multipath component}
\acrodef{snr}[SNR]{signal-to-noise-ratio}
\acrodef{pmva}[PMVA]{potential \ac{mva}}
\acrodef{slam}[SLAM]{simultaneous localization and mapping}
\acrodef{pmf}[PMF]{probability mass function}
\acrodef{pdf}[PDF]{probability density function}
\acrodef{spa}[SPA]{sum-product algorithm}
\acrodef{mmse}[MMSE]{minimum mean-square error}
\acrodef{rmse}[RMSE]{root mean-square error}
\acrodef{ospa}[OSPA]{optimal subpattern assignment}
\acrodef{mospa}[MOSPA]{mean optimal subpattern assignment}
\acrodef{uwb}[UWB]{ultra-wideband}
\acrodef{rt}[RT]{ray-tracing}
\acrodef{rl}[RL]{ray-launching}
\acrodef{rf}[RF]{radio frequency}

\newdateformat{monthyeardate}{\monthname[\THEMONTH] \THEDAY, \THEYEAR} %

\definecolor{colA}{rgb}{0,0,1}%
\definecolor{colB}{rgb}{0.8,0.0,0.5}%
\definecolor{colC}{rgb}{0,0.5,0}%
\definecolor{col_cyan}{rgb}{0,1,1}%
\definecolor{colE}{rgb}{1.00000,0.55,0.00000}%
\definecolor{colF}{rgb}{1.00000,0.0,0.00000}%
\definecolor{col_mag}{rgb}{1.00000,0.00000,1.00000}%
\definecolor{col_grey}{rgb}{0.4,0.4,0.4}%
\definecolor{col_gray2}{rgb}{0.4,0.4,0.4}
\definecolor{lightgray}{rgb}{0.9,0.9,0.9}  

\definecolor{alex}{RGB}{235,134,52}
\definecolor{erik}{RGB}{235,134,52}

\definecolor{matlabBlue}{rgb}{0.00000,0.44700,0.74100}%
\definecolor{matlabOrange}{rgb}{0.85000,0.32500,0.09800}%
\definecolor{matlabYellow}{rgb}{0.92900,0.69400,0.12500}%
\definecolor{matlabLila}{rgb}{0.49400,0.18400,0.55600}%
\definecolor{matlabGreen}{rgb}{0.46600,0.67400,0.18800}%

\definecolor{FGgreen}{RGB}{34,139,34}
\definecolor{FGblue}{RGB}{80,120,255}
\definecolor{FGred}{RGB}{255,110,110}

\colorlet{fgColor_node}{colA}
\colorlet{fgColorUp1}{colB}
\colorlet{fgColor_coop}{colB}
\colorlet{fgColorBG}{colC}
\colorlet{fgColor_facAlpha}{colE}
\colorlet{fgColorPost3}{colF}
\colorlet{fgColorComb}{colE}
\colorlet{fgColorBox}{colE}

\newcommand*\mref[2]{\cite[#1~\ref{M-#2}]{suppl}}

\begin{document}
	\title{\huge Data Fusion for Multipath-Based SLAM:\\Combining Information from Multiple Propagation Paths}
	\author{\normalsize Erik~Leitinger~\IEEEmembership{\normalsize Member,~IEEE}, 
	Alexander~Venus~\IEEEmembership{\normalsize Student Member,~IEEE},\\ 		
	Bryan Teague~\IEEEmembership{\normalsize Member,~IEEE}, 
	Florian~Meyer~\IEEEmembership{\normalsize Member,\hspace{-.3mm} IEEE}
	\vspace*{-8mm}	
	\thanks{ 
	E.\ Leitinger and A.\ Venus are with the Signal Processing and Speech Communication Laboratory, Graz University of Technology, Graz, Austria, and the Christian Doppler Laboratory for Location-aware Electronic Systems (e-mail: (erik.leitinger,a.venus)@tugraz.at).
	B.\ Teague is with the MIT Lincoln Laboratory, Lexington, MA, USA (bryan.teague@ll.mit.edu).
	F.\ Meyer is with the Department of Electrical and Computer Engineering and Scripps Institution of Oceanography, University of California San Diego, San Diego, CA, USA (e-mail: fmeyer@ucsd.edu). 
	This material is based upon work supported by the Under Secretary of Defense for Research and Engineering under Air Force Contract No. FA8702-15-D-0001 and the TU Graz.
	}}	
	\maketitle	
	
	\begin{abstract}	
	Multipath-based \ac{slam} is an emerging paradigm for accurate indoor localization constrained by limited navigation resources. The goal of multipath-based \ac{slam} is to support the estimation of time-varying positions of mobile agents by detecting and localizing radio-reflective surfaces in the environment. In existing Bayesian methods, a propagation surface is represented by the mirror image of each \ac{pa} across that surface -- known as the corresponding \ac{va}. Due to this \acp{va} representation, each propagation path is mapped individually. Existing methods thus neglect inherent geometrical constraints across different paths that interact with the same surface, which limits accuracy and speed. In this paper, we introduce an improved statistical model and estimation method that enables data fusion in multipath-based \ac{slam}. By directly representing each surface with a \ac{mva}, geometrical constraints across propagation paths are also modeled statistically. A key aspect of the proposed method based on \acp{mva} is to check the availability of single-bounce and double-bounce propagation paths at potential agent positions by means of {\ac{rt}}. This availability check is directly integrated into the statistical model as detection probabilities for propagation paths. Estimation is performed by a \ac{spa} derived based on the factor graph that represents the new statistical model. Numerical results based on simulated and real data demonstrate significant improvements in estimation accuracy compared to state-of-the-art multipath-based \ac{slam} methods.
	\vspace*{-5mm}
	\end{abstract}
	
	\acresetall
	\section{Introduction}
	\label{sec:introduction}

	\begin{figure}[!t]
	\centering
	\scalebox{0.67}{\includegraphics{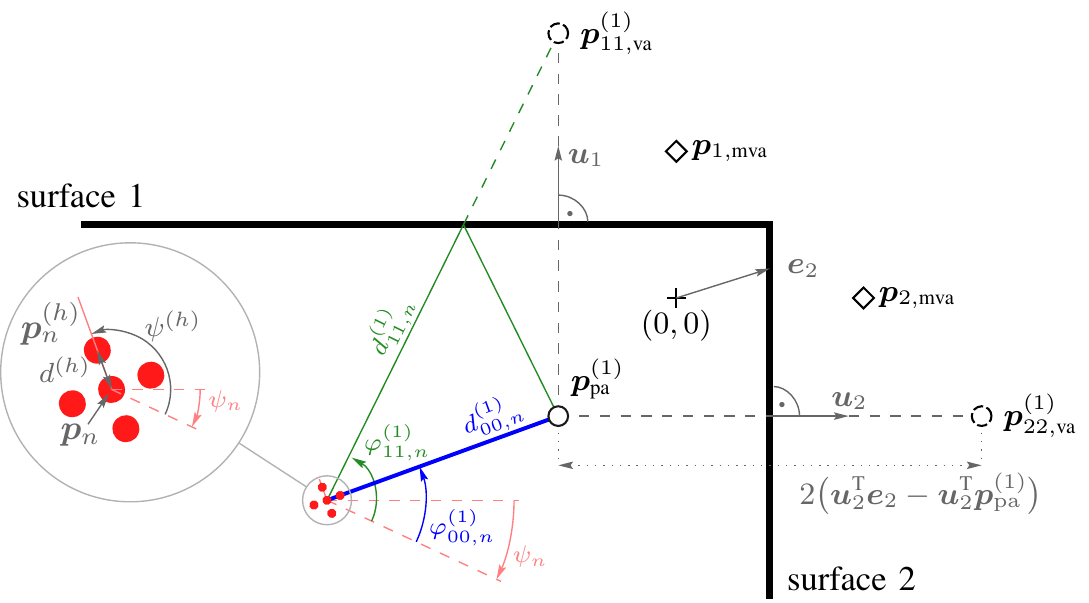}}\\[-3mm]
	\caption{{A multipath-based \acf{slam} scenario at time step $n$, involving two reflective surfaces, a single \acf{pa} with position $\V{p}^{(1)}_{\text{pa}}\rmv\rmv$, and a single mobile agent with position and orientation, $\V{p}_n$ and $\psi_n$, respectively. The mobile agent is depicted with five individual antenna elements at positions $\V{p}^{(h)}_{n}\rmv\rmv$, defined by offset $d^{(h)}$ and angle $\psi^{(h)}$. The reflecting surfaces can be represented by \acfp{va} or \acfp{mva}, e.g., $\V{p}^{(1)}_{\text{11,va}}$ or $\V{p}_{\text{1,mva}}$, respectively. A single-bounce \ac{va} represents the location of the mirror image of a \ac{pa} on the corresponding reflective surface. A \ac{mva} represents the location of the mirror image of a common origin on the corresponding reflective surface, thus creating a unique \ac{mva} for each reflecting surface. For a second \ac{pa} or double-bounce propagation path involving surface 1 (not shown),\ac{va} positions are different than $\V{p}^{(1)}_{\text{11,va}}$ while the unique \ac{mva} position $\V{p}_{\text{1,mva}}$ remains the same. Note that $\V{u}_1$ and  $\V{u}_2$ are the unit vectors perpendicular to surfaces 1 and 2, while $\V{e}_2$ is an arbitrary vector from the origin to surface 2. Distance and angle for \ac{pa} and \ac{va} at surface 1, i.e., $(d_{00,n}^{(1)}, \varphi_{00,n}^{(1)})$ and $(d_{11,n}^{(1)}, \varphi_{11,n}^{(1)})$, are also shown.}}\label{fig:scenario} 
	\vspace*{-4mm} 
	\end{figure}

	Emerging sensing technologies and innovative signal processing methods exploiting multipath propagation will lead to new capabilities for autonomous navigation, asset tracking, and situational awareness in future communication networks. Multipath-based \ac{slam} is a promising approach in wireless networks for obtaining position information of transmitters and receivers as well as information on their propagation environments. In multipath-based \ac{slam}, specular reflections of \ac{rf} signals at flat surfaces are modeled by \acfp{va} that are mirror images of base stations, also referred to as \acfp{pa} \cite{WitMeiLeiSheGusTufHanDarMolConWin:J16}. The number of \acp{va} and their positions are typically unknown. Multipath-based \ac{slam} methods can detect and localize \acp{va} and jointly estimate the time-varying agent position \cite{WitMeiLeiSheGusTufHanDarMolConWin:J16, GentnerTWC2016, LeiMeyHlaWitTufWin:J19, LeiGreWit:ICC2019,MenMeyBauWin:J19}. The availability of \ac{va} location information makes it possible to leverage multiple propagation paths of \ac{rf} signals for agent localization. It can thus significantly improve localization accuracy and robustness \cite{LeitingerJSAC2015,WilGreLeiMueWit:ACSSC2018, ShaGarDesSecWym:TWC2018, MenWymBauAbu:TWC2019}.
	\vspace*{-1mm}
	\subsection{State of the Art}
	\vspace{-1mm}
	Multipath-based \ac{slam} falls under the umbrella of feature-based SLAM approaches that focus on detecting and mapping distinct features in the environment \cite{DurrantWhyte2006, Dissanayake2001,MonThrKolWeg:AAAI2002, MulVoAdaVo:J11,DeuReuDie:SPL2015,CadCarCarLatScaNeiReiLeo:TR2016}. In existing multipath-based \ac{slam} methods, the distinct features of interest are the \acp{va}\cite{LeiMeyHlaWitTufWin:J19,LeiGreWit:ICC2019,MenMeyBauWin:J19,ChuLuGesWanWenWuMuqLi:TWC2022,YanWenJinLi:TWC2022} while ``measurements'' are obtained by extracting parameters from the \acp{mpc} of \ac{rf} signals in preprocessing stage \cite{ShutWanJos:CSTA2013, BadHanFle:TSP2017, HanFleRao:TSP2018,GreLeiFleWit:Arxiv2023}. Measurements can be noisy distances, angles-of-arrival (AoAs), or angles-of-departure (AoDs) \cite{RichterPhD2005, SalRicKoi:TSP2009,LiLeiVenTuf:TWC2022}. 
		
	As typical for feature-based \ac{slam}, a complicating factor in multipath-based \ac{slam} is measurement origin uncertainty, i.e., the unknown association of measurements with features \cite{LeiMeyHlaWitTufWin:J19, LeiGreWit:ICC2019, MenMeyBauWin:J19, MeyWil:J21, LiLeiVenTuf:TWC2022}. In particular, (i) it is not known which \ac{va} was generated by which measurement, (ii) there are missed detections due to low  \ac{snr} or occlusion of features, and (iii) there are false positive measurements due to clutter. Thus, an important aspect of multipath-based SLAM is \emph{data association} between measurements and \acp{va}. Probabilistic data association can increase the robustness and accuracy of multipath-based \ac{slam} but introduces association variables as additional unknown parameters. To avoid the curse of dimensionality related to the high-dimensional parameters space, state-of-the-art methods for multipath-based \ac{slam} perform the \ac{spa} on the factor graph representing the underlying statistical model \cite{LeiMeyHlaWitTufWin:J19, LeiGreWit:ICC2019,MenMeyBauWin:J19}. In addition, since the models for the aforementioned measurements are nonlinear, most methods typically rely on sampling techniques \cite{WitMeiLeiSheGusTufHanDarMolConWin:J16, LeitingerJSAC2015, GentnerTWC2016, LeiMeyHlaWitTufWin:J19, LeiGreWit:ICC2019, MenMeyBauWin:J19}. Recently, multipath-based \ac{slam} was applied to data collected in indoor scenarios by radios with ultra-wide bandwidth \cite{LeitingerICC2015} or multiple antennas \cite{GentnerTWC2016, ZhuTufvesson2015, LeiGreWit:ICC2019}.
	
	In existing methods for multipath-based \ac{slam}, each \ac{va} represents a single propagation path from an agent to a \ac{pa}. In particular, even if a reflective surface takes part in multiple propagation paths, each path is represented by a \ac{va}, and mapped individually \cite{WitMeiLeiSheGusTufHanDarMolConWin:J16, LeitingerJSAC2015, GentnerTWC2016,LeiMeyHlaWitTufWin:J19,LeiGreWit:ICC2019,MenMeyBauWin:J19, KimGraSveKimWym:TVT2022}. Existing methods thus neglect inherent geometrical constraints across different paths that interact with the same surface, which limits accuracy and speed. Note that the multipath-based \ac{slam} methods proposed in \cite{KimGraGaoBatKimWym:TWC2020,ChuLuGesWanWenWuMuqLi:TWC2022, YanWenJinLi:TWC2022} perform fusion of information provided by multiple cooperating agents. However, these methods are limited to single-bounce paths and do not perform fusion across \acp{pa}. {In \cite{SteGanMeiLeiWitZemPed:TAP2016, KoChaHanLeeSeoHua:WC2021, KoiKarHan:EUCAP2022, LiJiaWymWen:WCL2023}, it has been demonstrated that double-bounce paths of \ac{rf} signals exhibit considerable signal strength and should thus be considered in real-world multipath-based \ac{slam} scenarios. This provides compelling evidence that incorporating information from double-bounce paths can lead to an improved multipath-based \ac{slam} performance.}

	\vspace*{-3mm}
	\subsection{Contributions and Notations} \label{sec:contribution}

	The problem studied in this paper can be summarized as follows.
	\vspace{.8mm}
	
	{\textit{Estimate the time-varying position of a mobile agent by making use of the \ac{los} and \acp{mpc} in \ac{rf}  signals. To leverage the position information of \acp{mpc}, the unknown number and positions of reflective surfaces in the environment are estimated during runtime.}}
	\vspace{1.2mm}
	
	We introduce a new statistical model for multipath-based \ac{slam} that considers single-bounce and double-bounce propagation paths and enables data fusion across multiple paths. In existing multipath-based \ac{slam} models, paths represented by \acp{va} are the SLAM features. In the proposed model, however, every reflective surface is directly represented by a \ac{slam} feature referred to as \ac{mva}. The \ac{mva} representation makes it possible to model inherent geometrical constraints across paths statistically. Following the new model, a factor graph is established, and an extension of the \acp{spa} for multipath-based \ac{slam} in \cite{LeiMeyHlaWitTufWin:J19, LeiGreWit:ICC2019, MenMeyBauWin:J19} is developed. The resulting \ac{spa} can infer reflective surfaces by combining information across multiple propagation paths. Particularly appealing is the ability to fuse paths related to different \ac{pa}. Within the \ac{spa}, \acf{rt} \cite{Bor:JASA1984, MckHam:IEEENetwork1991, LuVitDegFusBarBlaBer:TAP2019} is performed to determine the availability of each single-bounce and double-bounce paths at potential agent positions. The availability check is directly integrated into the statistical model as detection probabilities for propagation paths. The resulting multipath-based \ac{slam} method can provide fast and accurate estimates in scenarios with many reflective surfaces. Note that the proposed method uses distance and \ac{aoa} measurements, which have been extracted from the \ac{rf} signal in a preprocessing stage \cite{ShutWanJos:CSTA2013, BadHanFle:TSP2017, HanFleRao:TSP2018,GreLeiFleWit:Arxiv2023}.
	The key contributions of this paper are as follows.	
	\begin{itemize}
		\item {We introduce a statistical model and factor graph for multipath-based \ac{slam} that facilitates data fusion across propagation paths.}
		\vspace{1.3mm}
		
		\item We integrate \ac{rt} into our statistical model to determine the availability of individual single-bounce and double-bounce propagation path. 
		\vspace{1.3mm}
		
		\item We extend the \ac{spa} for multipath-based \ac{slam} \cite{LeiMeyHlaWitTufWin:J19, LeiGreWit:ICC2019, MenMeyBauWin:J19} based on the introduced factor graph to establish data fusion for multipath-based \ac{slam}\footnote{We also provide a pseudocode for the particle-based implementation in the supplementary material{\mref{Section}{sec:pdeudocode}}.}.
		\vspace{1.3mm}
		
		\item We demonstrate significant improvements in estimation performance compared to existing multipath-based \ac{slam} methods based on both simulated and real data.
	\end{itemize}
	
	This paper advances beyond the preliminary account of our method provided in the conference publications \cite{LeiMey:Asilomar2020_DataFusion,LeiTeaZhaLiaMey:Fusion2022} by (i) extending the \acp{mva} model to double-bounce propagation paths, (ii) introducing {\ac{rt}} to determine the availability of paths and integrating it into the statistical model, %
	(iii) presenting a detailed derivation of the factor graph, and (iv) demonstrating performance advantages compared to reference methods. As reference methods, classical multipath-based \ac{slam} \cite{LeiMeyHlaWitTufWin:J19,LeiGreWit:ICC2019}, channel-\ac{slam} \cite{GentnerTWC2016}, and multipath-based positioning assuming known \acp{va} positions \cite{LeiMeyMeiWitHla:GNSS2016} are considered.

	\textit{Notation}: Random variables are displayed in sans serif, upright fonts; their realizations in serif, italic fonts. Vectors and matrices are denoted by bold lowercase and uppercase letters, respectively. For example, a random variable and its realization are denoted by $\rv x$ and $x$, respectively, and a random vector and its realization by $\RV x$ and $\V x$, respectively. Furthermore, $\|\V{x}\|$ and ${\V{x}}^{\text T}$ denote the Euclidean norm and the transpose of vector $\V x$, respectively, and $\langle \V{x}, \V{y} \rangle$ denotes the inner-product between the vectors $\V{x}$ and $\V{y}$; $\propto$ indicates equality up to a normalization factor; $f(\V x)$ denotes the \ac{pdf} of random vector $\RV x$ (this is a short notation for  $f_{\RV x}(\V x)$); $f(\V x | \V y)$ denotes the conditional \ac{pdf} of random vector $\RV x$ conditioned on random vector  $\RV y$  (this is a short notation for  $f_{\RV x | \RV z}(\V x | \V z)$). The four-quadrant inverse tangent of position $\V{p} = [\ist p_1 \ist\ist p_2 \ist]^{\mathrm{T}}$ is denoted as $ \atantwo(p_2, p_1)$. The cardinality of a set $\Set{X}$ is denoted as $ \vert\Set{X}\vert $. $\delta(\cdot)$ denotes the Dirac delta function. Finally, $ \delta_e$ denotes the indicator function of the event $ {e} \rmv=\rmv {0} $ (i.e., $ \delta_e \rmv=\rmv 1 $ if $ e \rmv=\rmv {0} $ and ${0}$ otherwise).

	\section{Geometrical Relations}\label{sec:geometricRel}

	\begin{figure*}[!t]
		\centering
		\hspace*{-2mm}
		\captionsetup[subfigure]{captionskip=-1pt}
		\subfloat[]{
		\scalebox{0.63}{
		\includegraphics{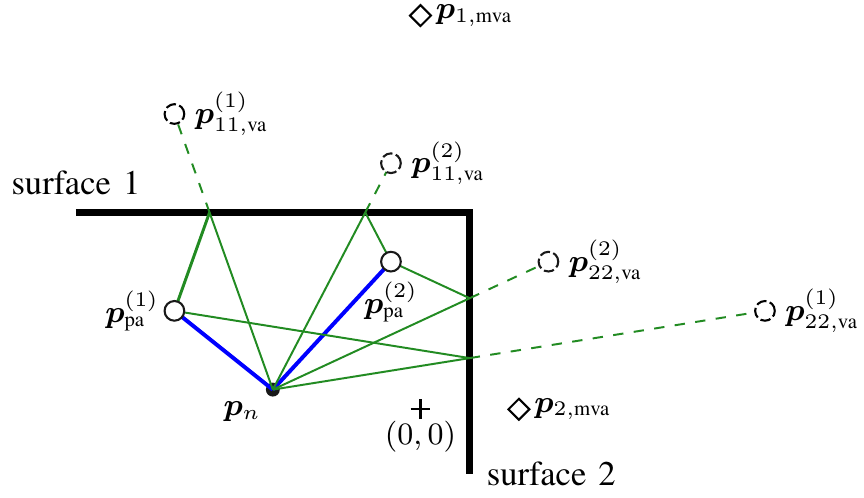}
		}\label{fig:bounceA}}\hspace*{-2mm}
		\subfloat[]{
		\scalebox{0.63}{	
		\includegraphics{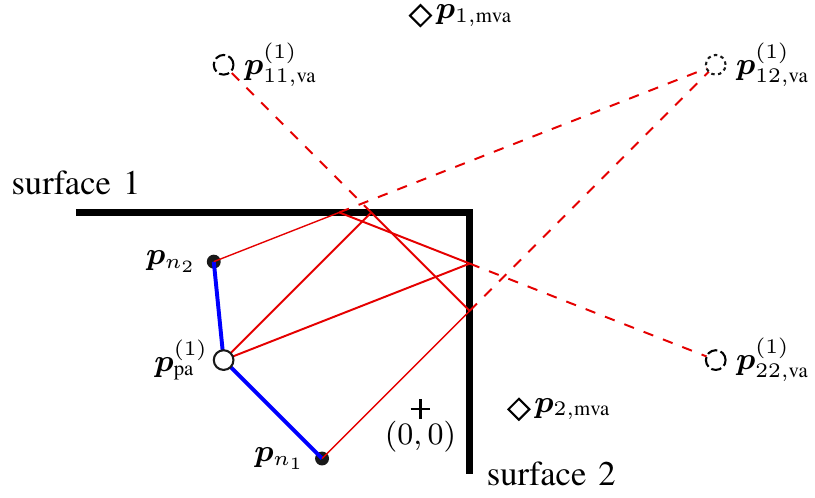}
		}\label{fig:bounceB}}\hspace*{-2mm}		
		\subfloat[]{
		\scalebox{0.63}{
		\includegraphics{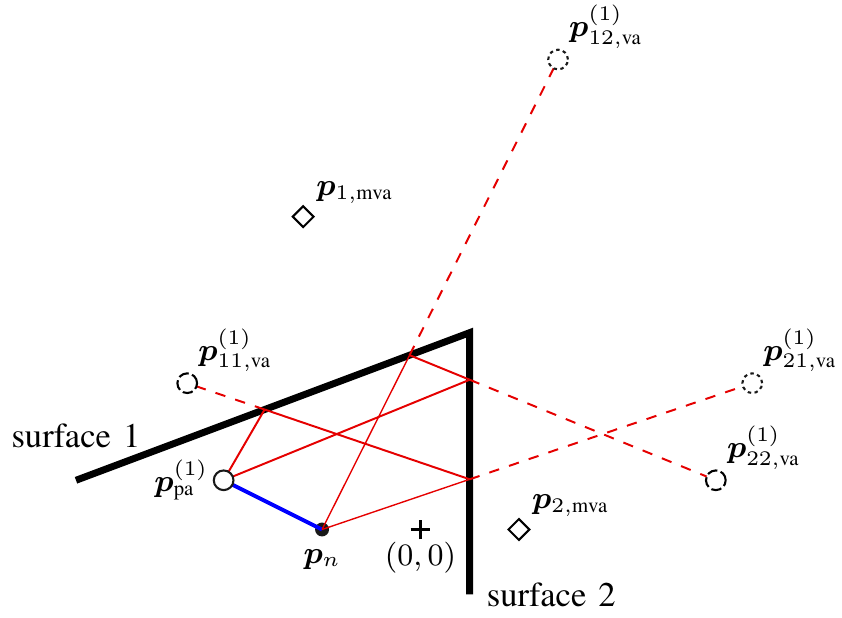}
		}\label{fig:bounceC}}\\
		\centering
		\includegraphics{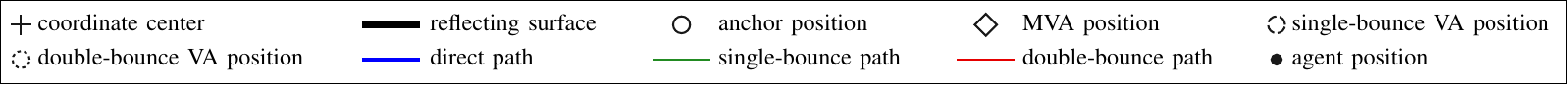}
		\vspace{-2.5mm}
		\caption{{Three multipath-based \acp{slam} scenarios with two reflective surfaces illustrate single- and double-bounce paths in three different geometries. The positions of the mobile agents, \acp{pa}, \acp{va}, and \acp{mva} are shown. (a) depicts a single time step $n$ of a scenario with perpendicular reflective surfaces, two \acp{pa} at positions $\V{p}_\text{pa}^{(1)}$ and $\V{p}_\text{pa}^{(2)}$, and the agent at position $\V{p}_n$; only single-bounce paths are illustrated. (b) shows perpendicular reflective surfaces, one \ac{pa} at position $\V{p}_\text{pa}^{(1)}$, and agent positions $\V{p}_{n_1}$ and $\V{p}_{n_2}$ at two different time steps $n_1$ and $n_2$. Only double-bounce paths are shown. (c) depicts a single time step $n$ of a scenario with reflective surfaces at an acute angle, one \ac{pa} at position $\V{p}_\text{pa}^{(1)}$ and the agent at position $\V{p}_n$. Again, only double-bounce paths are illustrated. Note that in (b), due to perpendicular reflective surfaces, both double-bounce paths have equal lengths, and there is a single double-bounce VA. In (b), however, due to reflective surfaces at an acute angle, the two double-bounce paths have different lengths, and there are two different double-bounce VAs.}}\label{fig:bounce} 
		\vspace{-4mm}
	\end{figure*}

	We consider a mobile agent equipped with an $H$-element antenna array and $J$ \acp{pa} {equipped with a single antenna} at known positions $\V{p}_{\mathrm{pa}}^{(j)} = \big[ {p}_{1,\mathrm{pa}}^{(j)} \ist\ist\ist {p}_{2,\mathrm{pa}}^{(j)} \big]^{\mathrm{T}} \rmv\!\in\rmv \mathbb{R}^2\rmv$, $j \rmv\in\rmv \{1,\ldots,J\}$, where $J$ is assumed to be known. The agent is in an environment with $S$ reflective surfaces indexed by $s \in \Set{S}\triangleq\rmv \{1,\ldots,S\}$. At every discrete time step $n$, the array element locations are denoted by $\V{p}_n^{(h)}$, $h \in \{1,\dots,H\}$. The agent position $\V{p}_n$ refers to the center of gravity of the array. We also define $d^{(h)} = \|\V{p}_n^{(h)} - \V{p}_n\|$ and $\psi^{(h)} = \atantwo\big({p}_{2,n}^{(h)}-{p}_{2,n}, {p}_{1,n}^{(h)}-{p}_{1,n}\big)-\psi_n$, the distance from the reference location $\V{p}_n$ and the orientation, respectively, of the $h$-th element as shown in Fig.~\ref{fig:scenario}. At each discrete time slot $n$, the position $\V{p}_n \rmv\!\in\rmv \mathbb{R}^2$ and the array orientation $\psi_n$  of the agent are unknown. {Each \ac{pa} transmits a \ac{rf} signal, and the agent acts as a receiver.}\footnote{We assume the \acp{pa} to use orthogonal codes, i.e., there is no mutual interference between individual \acp{pa}. Note that the proposed algorithm can be easily reformulated for the case where the agent transmits a \ac{rf} signal and the \acp{pa} act as receivers.} 
	The \ac{rf} signal arrives at the receiver via the \ac{los} path as well as via \acp{mpc} originating from the reflection of surrounding objects. 
	%The \ac{rf} signal arrives at the receiver via the \ac{los} path and \ac{mpc} paths originating from the reflection of surrounding objects.
	We assume time synchronization between all \acp{pa} and the agent. However, our algorithm can be extended to an unsynchronized system according to \cite{GentnerTWC2016, LeitingerJSAC2015, EtzMeyHlaSprWym:TSP2017}. 

	We restrict the representation to \acp{mpc} related to single-bounce and double-bounce paths. In particular, associated with \ac{pa} $j$ there are $|\Set{D}_\text{S}| = S$ single-bounce \acp{va} \cite{Bor:JASA1984,LeiMeyHlaWitTufWin:J19} at unknown positions $\V{p}_{ss,\mathrm{va}}^{(j)} \!\rmv\in\rmv \mathbb{R}^2\rmv$ with index-pair $(s,s) \in  \Set{D}_\text{S} \rmv\triangleq\rmv \{ (s,s) \rmv\in\rmv \Set{S} \rmv\times\rmv \Set{S}\}$ and $|\Set{D}_\text{D}| = S(S-1)$ double-bounce \acp{va} at unknown positions $\V{p}_{ss',\mathrm{va}}^{(j)} \!\rmv\in\rmv \mathbb{R}^2\rmv$ with index-pair $(s,s') \in  \Set{D}_\text{D} \rmv\triangleq\rmv \{ (s,s') \rmv\in\rmv \Set{S} \rmv\times\rmv \Set{S}| \ist s \neq s' \}$. Therefore, the maximum number of \acp{va} for \ac{pa} $j$ is given by $|\mathcal{D}| = S+S(S-1)$ with $\Set{D} \rmv\triangleq\rmv \{ (s,s') \rmv\in\rmv \Set{S} \rmv\times\rmv \Set{S}\} = \Set{D}_\text{D} \cup \Set{D}_\text{S}$. Note that subscript indices denoted as $(\cdot)_{ss'}$ indicate variables related to a pairs $(s,s')$ of reflecting objects. By applying the image-source model \cite{Bor:JASA1984,KulLeiGreWit:TWC2018}, a \ac{va} associated with a single-bounce path is the mirror image of $\V{p}_{\mathrm{pa}}^{(j)}$ at reflective surface $s \in \mathcal{S}$ given by
	\vspace*{-1.5mm}  
	\begin{align}
		\V{p}^{(j)}_{ss,\mathrm{va}} &= \V{p}^{(j)}_{\mathrm{pa}} + 2\big( \V{u}_s^{\T}\V{e}_s - \V{u}_s^{\T}\V{p}^{(j)}_{\mathrm{pa}}\big)\V{u}_s \label{eq:VA1BPAequation}
		 \\[-6mm]
		\nn
	\end{align}
	where $(s,s) \in \Set{D}_\text{S}$, $\V{u}_s$ is the normal vector of reflective surface $s$, and $\V{e}_s$ is an arbitrary point on the considered surface. The second term in \eqref{eq:VA1BPAequation} represents the normal vector w.r.t. to reflective surface $s$ in direction $\V{u}_s$ with the length of two times the distance between \ac{pa} $j$ at position $\V{p}^{(j)}_{\mathrm{pa}}$ and the normal-point at the reflective surface $s$, i.e., $2\big( \V{u}_s^{\T}\V{e}_s - \V{u}_s^{\T}\V{p}^{(j)}_{\mathrm{pa}}\big)$. By again applying the image-source model \cite{Bor:JASA1984, KulLeiGreWit:TWC2018} for \ac{va} at position $\V{p}^{(j)}_{s's',\mathrm{va}}$ another \ac{va}\vspace*{-0.1mm} that is obtained as the mirror image of $\V{p}^{(j)}_{s's',\mathrm{va}}$ at surface $s$, i.e.,
	\vspace*{-1.5mm}  
	\begin{align}
		\V{p}^{(j)}_{ss',\mathrm{va}} &= \V{p}^{(j)}_{s's',\mathrm{va}} + 2\big( \V{u}_s^{\T}\V{e}_s - \V{u}_s^{\T}\V{p}^{(j)}_{s's',\mathrm{va}}\big)\V{u}_s\\[-6mm]
		\nn
	\end{align}
	where $(s,s') \in \Set{D}_\text{D}$. This \ac{va} represents a double-bounce propagation path. For conveniently addressing \ac{pa}-related variables and factors, we also define $\V{p}^{(j)}_{00,\mathrm{va}} \triangleq \V{p}^{(j)}_{\mathrm{pa}}$ and ${\tilde{\mathcal{D}}} = (0,0) \cup {\mathcal{D}}$. The distances and \acp{aoa} related to the propagation paths at the agent position $\V{p}_n$ represented by $\V{p}_{ss',\mathrm{va}}^{(j)}$ with $(s,s') \in {\tilde{\mathcal{D}}}$ are modeled by $d_{ss',n}^{(j)} = \big\|  \V{p}_n - \V{p}^{(j)}_{ss',\mathrm{va}} \big\|$ and $\varphi_{ss',n}^{(j)} =  \atantwo \big ( {p}_{2,n} - {p}^{(j)}_{1,ss',\text{va}}, {p}_{1,n} - {p}^{(j)}_{1,ss',\text{va}} \big ) - \psi_n$. Note that the distance $d_{00,n}^{(j)} $ and the \acp{aoa} $\varphi_{00,n}^{(j)}$  are the parameters related to the \ac{los} path between agent at position $\V{p}_n$ and \ac{pa} $j$ at position $\V{p}^{(j)}_{\mathrm{pa}}$. An example is shown in Fig.~\ref{fig:scenario}. 

	The availability of \acp{va} at certain agent position $\V{p}_n$ is limited due to blockage of associated propagation paths or geometric constraints between reflective surfaces \cite{Bor:JASA1984, MckHam:IEEENetwork1991}.
	Especially, double-bounce paths and their corresponding \ac{va} positions have limited availability depending on the agent position $\V{p}_n$ (see Fig.~\ref{fig:bounceB} and Fig.~\ref{fig:bounceC}). Hence, the number of available \acp{va} for \ac{pa} $j$ is smaller than $|\mathcal{D}| = S+S(S-1)$. As discussed in Section~\ref{sec:Pdvisibility}, the proposed statistical model and method performs an availability check for each VA using \ac{rt} \cite{Bor:JASA1984,MckHam:IEEENetwork1991,LuVitDegFusBarBlaBer:TAP2019}. In this way, for each agent position $\V{p}_n$, it can be determine which of the potential paths in $\mathcal{D}$ is available.

	\begin{table*}[t]	
		\centering
		\captionsetup{width=0.9\columnwidth,labelfont=bf}
		\caption{Notations and definitions of important quantities.}
		\label{tab:symbol_notation}
		\vspace*{-2mm}
		\begin{tabular}{  p{1.3cm}   p{6.9cm} | p{1.3cm} p{6.9cm}}
			\hline
			\textbf{Notation}& \textbf{Definition} 
			&  \textbf{Notation} & \textbf{Definition}\\
			\hline
			
			\rowcolor{blue!10!white} 
			\(\V{p}_{\mathrm{pa}}^{(j)}\) & 2D position of physical anchor \(j\) &
			\(\V{p}_n\) & 2D position of mobile agent at time $n$ \\
			
			\(\psi_n\) & Orientation of mobile agent at time $n$ &
			\(\V{v}_n\) & 2D velocity of mobile agent at time $n$ \\
			
			\rowcolor{blue!10!white}
			\(S\) & Total number of reflecting surfaces (\acp{mva}) &
			\(\rv{M}^{(j)}_n\) & Number of measurements with \ac{pa} \(j\) at time $n$ \\
			
			\(\V{p}_{ss,\mathrm{va}}^{(j)}\) & Single-bounce virtual anchor position &
			\(\V{p}_{ss',\mathrm{va}}^{(j)}\) & Double-bounce virtual anchor position \\
			
			\rowcolor{blue!10!white}
			\(\V{p}_{s,\mathrm{mva}}\) & Position of \ac{mva} corresponding to \ac{pa} \(s\)  &
			\(S_n\) & Number of \acp{pmva} at time $n$ \\
			
			\({\rv{z}_\text{d}}_{m,n}^{(j)}\) & Distance measurement \(m\) with \ac{pa} \(j\) at time $n$ &
			\({\rv{z}_\varphi}_{m,n}^{(j)}\) & \aclu{aoa} measurement \(m\) with \ac{pa} \(j\) at time $n$ \\
			
			\rowcolor{blue!10!white}
			\({\rv{\nu}_\mathrm{\tau}}_{m,n}^{(j)}\) & Distance measurement noise &
			\({\nu_{\mathrm{\varphi}}}_{m,n}^{(j)}\) & \aclu{aoa} measurement noise \\
			
			\(\rv{r}_{s,n}\) & Binary existence variable related to \ac{pmva} \(s\) at time $n$ &
			\( p_{\mathrm{d},ss',n}^{(j)} \) & detection probability of available path\\
			
			\rowcolor{blue!10!white}
			\(\mu_{\mathrm{fp}}\) & False alarm Poisson point process mean &
			\(f_{\mathrm{fp}}(\RV{z}_{m,n}^{(j)})\) & False positive point process \ac{pdf} \\
			
			\(p_\mathrm{s}\) & Probability of survival of legacy \ac{pmva} &
			\(p_{\mathrm{cf}}\) & Confirmation threshold for \acp{pmva} \\
			
			\rowcolor{blue!10!white}
			\(p_{\text{pr}}\) & Pruning threshold to remove \acp{pmva} &
			\(\Delta T\) & Simulation sampling period \\
			
			\(\V{\omega}_{n}\) & Process noise of the agents' motion model &
			\(\) &  \\
			
			\hline
		\end{tabular}
		\vspace*{-0.5cm}
	\end{table*}

	\vspace*{-2mm}
	\subsection{MVA-Based Model of the Environment}
	\label{sec:environmentalModel}

	A reflective surface is involved in multiple propagation paths and thus defines multiple \acp{va}. To enable the consistent combination, i.e., ``fusion'' of map information provided by measurements of different \acp{pa}, we represent reflective surfaces by $S$ unique \acp{mva} at positions $\V{p}_{s,\mathrm{mva}} \!\rmv\in\rmv \mathbb{R}^2\rmv$, $s \in \Set{S}$. The unique \ac{mva} position $\V{p}_{s,\mathrm{mva}} \!\rmv\in\rmv \mathbb{R}^2\rmv$ is defined as the mirror image of $[0 \hspace{.6mm} 0]^{\T}$ on the reflective surface $S$.\footnote{Note that any point can be used, not only the coordinate center. And that the \ac{mva}-model can be extended by the length of the corresponding walls.} By using some algebra, the transformation from \ac{mva} at $\V{p}_{s,\mathrm{mva}}$ to a \ac{va} at $\V{p}^{(j)}_{ss,\mathrm{va}}$, i.e., $\V{p}^{(j)}_{ss,\mathrm{va}} = h_\text{va}\big( \V{p}_{s,\mathrm{mva}}, \V{p}^{(j)}_{\mathrm{pa}} \big)$ can be obtained as
	\vspace*{-1mm} 
	\begin{align}
	\V{p}^{(j)}_{ss,\mathrm{va}} = -\bigg(\frac{ 2 \big\langle \V{p}_{s,\mathrm{mva}},\V{p}^{(j)}_{\mathrm{pa}} \big\rangle}{ \big\|\V{p}_{s,\mathrm{mva}}\big\|^2 } - 1 \bigg) \ist \V{p}_{s,\mathrm{mva}} + \ist \V{p}^{(j)}_{\mathrm{pa}}\ist. \label{eq:nonLinearTransformation}\\[-6mm]\nn
	\end{align}
	The transformation from \acp{mva} at $\V{p}_{s,\mathrm{mva}}$ and $\V{p}_{s',\mathrm{mva}}$ to a {\textit{double-bounce}} \ac{va} at $\V{p}^{(j)}_{ss',\mathrm{va}}$ can be obtained by applying \eqref{eq:nonLinearTransformation} twice, i.e, 
	$\V{p}^{(j)}_{ss',\text{va}} = h_\text{va}\big( \V{p}_{s,\mathrm{mva}}, h_\text{va}\big( \V{p}_{s',\mathrm{mva}}, \V{p}^{(j)}_{\mathrm{pa}} \big)\big)$. The inverse transformation from a \ac{va} to a \ac{mva} is given by
	\vspace*{-2mm} 
	\begin{align}
	\V{p}_{s,\mathrm{mva}} &= h_\text{mva}\big( \V{p}^{(j)}_{ss,\mathrm{va}}, \V{p}^{(j)}_{\mathrm{pa}} \big) \nn\\[1mm]
	&= \frac{ \big\|\V{p}^{(j)}_{\mathrm{pa}}\big\|^2 - \big\|\V{p}^{(j)}_{ss,\mathrm{va}}\big\|^2}{ \big\|(\V{p}^{(j)}_{\mathrm{pa}} - \V{p}^{(j)}_{ss,\mathrm{va}})\big\|^2} \big(\V{p}^{(j)}_{\mathrm{pa}} - \V{p}^{(j)}_{ss,\mathrm{va}}\big)\ist. \label{eq:nonLinearTransformationMVA}\\[-6mm]\nn	
	\end{align}
	Note that the inverse transformation in \eqref{eq:nonLinearTransformationMVA} will be {used} to determine a proposal distribution for \ac{mva} states as discussed in Section~\ref{sec:SPAalgorithm}. Details of the derivation of \eqref{eq:nonLinearTransformation} and  \eqref{eq:nonLinearTransformationMVA} are provided in the supplementary material{\mref{Section}{sec:app_mva_model}}. For example, Fig.~\ref{fig:bounce} shows three scenarios with two reflecting surfaces described by two \acp{mva} at positions $\V{p}_{1,\mathrm{mva}}$ and $\V{p}_{2,\mathrm{mva}}$. Fig.~\ref{fig:bounceA} shows a scenario with two \acp{pa} $j \in \{1,2\}$, the corresponding \acp{va}, and an agent at position $\V{p}_n$. Each \ac{pa} generates one \ac{va} associated with a single-bounce propagation path. Fig.~\ref{fig:bounceB} shows a scenario with one \ac{pa}, the corresponding \acp{va} associated with single-bounce and double-bounce propagation paths, and two agents positions at differnet time steps, i.e., $\V{p}_{n_1}$ and $\V{p}_{n_2}$. Note in case surfaces are perpendicular, a different order of bounces from surfaces, i.e., surface ``$s$ -- surface $s'$'' or ``surface $s'$ -- surface $s$'', does not lead to a different \ac{va} position. Depending on the agent position $\V{p}_{n}$, only one double-bounce propagation path (related to one of the two orders) is available (see also Section~\ref{sec:Pdvisibility}). Fig.~\ref{fig:bounceB} shows a scenario with non-perpendicular surfaces. In this case, different ``bounce orders'' lead to different \ac{va}-positions. In particular, if there is an acute angle between a pair of reflecting surfaces, there exist regions of agent positions $\V{p}_{n}$ for which two double-bounce propagation paths are available at the same time (cf. Section~\ref{sec:Pdvisibility}). These regions depend on the PA position as well as the angle between the two surfaces. Note that when the angle between a pair is obtuse, only one of the two double-bounce propagation paths is available for all positions $\V{p}_{n}$ (an example is given in \mref{Section}{sec:app_results}).

	\section{System Model}
	\label{sec:systemModel}	
	
	At each time $n$, the state of the agent is given by {$\RV{x}_{n} = [\RV{p}_{n}^\T \iist \RV{v}_{n}^\T]$}, where  $\RV{v}_{n}$ is the agent velocity vector. We assume that the array is rigidly coupled with the movement direction{, i.e., array orientation is determined by the direction of the agent velocity vector}. As in \cite{MeyBraWilHla:J17,MeyKroWilLauHlaBraWin:J18,LeiMeyHlaWitTufWin:J19}, we account for the unknown number of \acp{mva} by introducing \acf{pmva} $s \in \Set{S}_n \rmv\triangleq\rmv \{ 1,\dots, \rv{S}_n \}$. The number $S_n$ of \acp{pmva} is the maximum possible number of actual \acp{mva}, i.e., all \acp{mva} that produced a measurement so far \cite{MeyKroWilLauHlaBraWin:J18, LeiMeyHlaWitTufWin:J19} (where $S_n$ increases with time). \ac{pmva} states are denoted as $\RV{y}_{s,n} \rmv\rmv=\rmv\rmv \big[\RV{p}^{\T}_{s,\mathrm{mva}} \; \rv{r}_{s,n} \big ]^\T\rmv\rmv\rmv$. The existence/nonexistence of \ac{pmva} $s$ is modeled by the existence variable $\rv{r}_{s,n} \rmv\rmv\in\rmv\rmv \{0,1\}$ in the sense that \ac{pmva} $s$ exists if $r_{s,n}\rmv\rmv=\rmv\rmv 1$. Formally,  its states is considered even if \ac{pmva} $s$ is nonexistent, i.e., if $r_{s,n} \rmv\rmv=\rmv\rmv 0$. The states $\RV{p}^{\T}_{s,\mathrm{mva}}$ of nonexistent \acp{pmva} are obviously irrelevant. Therefore, all \acp{pdf} defined for \ac{pmva} states, $f(\RV{y}_{s,n}) \rmv\rmv=\rmv\rmv f(\RV{p}_{s,\mathrm{mva}}, \rv{r}_{s,n})$, are of the form $f(\RV{p}_{s,\mathrm{mva}}, 0 )\rmv\rmv=\rmv\rmv f_{s,n} f_{\text{d}}(\RV{p}_{s,\mathrm{mva}})$, where $f_{\text{d}}(\RV{p}_{s,\mathrm{mva}})$ is an arbitrary ``dummy \ac{pdf}'' and $f_{s,n} \!\rmv\in [0,1]$ is a constant and can be interpreted as the probability of non-existence \cite{MeyKroWilLauHlaBraWin:J18, LeiMeyHlaWitTufWin:J19}. A summary of all variables related to agent and PMVA states as well as other variables of the proposed statistical model can be found in Table~\ref{tab:symbol_notation}.
	\vspace*{-1mm}

	\subsection{Measurements and New \acp{pmva}  \vspace{-.6mm}}
	\label{sec:MeasNewPTs}
	
	The distance and \ac{aoa} measurements  related to the path ``agent at position $\RV{p}_n$ -- \ac{va} at position $\RV{p}^{(j)}_{ss',\text{va}}$'' with $(s,s') \in \tilde{\Set{D}}_n$ are given by\vspace*{-2mm}
	\begin{align}
		{\rv{z}_\text{d}}_{m,n}^{(j)} &= \big\|  \RV{p}_n - \V{p}^{(j)}_{ss',\mathrm{va}} \big\| + {\rv{\nu}_{\mathrm{d}}}_{m,n}^{(j)} \label{eq:VADistmeas} \\
		{\rv{z}_\varphi}_{m,n}^{(j)}&=  \atantwo \big ( {p}_{2,n} - {p}^{(j)}_{2,ss',\mathrm{va}}, {p}_{1,n} - {p}^{(j)}_{1,ss',\mathrm{va}} \big )\nn \\
		&\hspace{5mm} {- \atantwo \big ( {v}_{2,n}, {v}_{1,n} \big )}  + {\rv{\nu}_{\mathrm{\varphi}}}_{m,n}^{(j)} \label{eq:VAAnglemeas}\\[-7.5mm]\nn
	\end{align}
	where ${\rv{\nu}_\mathrm{d}}_{m,n}^{(j)}$ and ${\rv{\nu}_{\mathrm{\varphi}}}_{m,n}^{(j)}\vspace*{-0.8mm}$ are, respectively, zero-mean Gaussian measurement noise with standard deviations ${\sigma_{\mathrm{d}}}^{(j)}_{m,n}\vspace*{0.2mm}$ and ${\sigma_{\mathrm{\varphi}}}^{(j)}_{m,n}$. The measurements are combined in the vector $\RV{z}_{m,n}^{(j)} = [{\rv{z}_\text{d}}_{m,n}^{(j)} \iist {\rv{z}_\varphi}_{m,n}^{(j)}]^\T$ with $m \rmv\in\rmv \Set{M}_n^{(j)} \rmv\triangleq\rmv \{1,$ $\dots,\rv{M}^{(j)}_n\}$ for \ac{pa} $j$. (These measurements represent the \ac{mpc} parameters. For details see \cite{LeiGreWit:ICC2019,GentnerTWC2016,LiLeiVenTuf:TWC2022}.) Note that before measurements are acquired, the number of measurements $\rv{M}^{(j)}_n \rmv$ is random.

	\textit{Likelihood function for \ac{los} paths:} Using \eqref{eq:VADistmeas}, \eqref{eq:VAAnglemeas}, and $\V{p}^{(j)}_{00,\text{va}} = \V{p}^{(j)}_{\mathrm{pa}}$, we can directly obtain the likelihood function for the \ac{los} path as $f\big(\V{z}_{m,n}^{(j)}\big|\V{p}_{n}\big)\vspace{.8mm}$.
		
	\textit{Likelihood function for single-bounce paths:} Using \eqref{eq:VADistmeas}, \eqref{eq:VAAnglemeas}, and the transformation in \eqref{eq:nonLinearTransformation}, i.e., $\RV{p}^{(j)}_{ss,\text{va}} = {h_\text{va}}\big( \RV{p}_{s,\mathrm{mva}}, \V{p}^{(j)}_{\mathrm{pa}} \big)$, the likelihood function related to the single-bounce propagation path ``agent -- surface $s$'' with $s \in \Set{S}$ -- \ac{pa} $j$ with \ac{mva} index-pair $(s,s) \in \mathcal{D}_{\text{S},n}$ reads $f(\V{z}_{m,n}^{(j)}|\V{p}_{n},\V{p}_{s,\mathrm{mva}})\vspace{.8mm}$.
	
	\textit{Likelihood function for double-bounce paths:} Using \eqref{eq:VADistmeas}, and \eqref{eq:VAAnglemeas}, and the transformation in \eqref{eq:nonLinearTransformation} twice\vspace{-1mm}, i.e., $\RV{p}^{(j)}_{ss',\text{va}} = {h_\text{va}}\big( \RV{p}_{s,\mathrm{mva}}, {h_\text{va}}\big( \RV{p}_{s',\mathrm{mva}}, \V{p}^{(j)}_{\mathrm{pa}} \big)\big)$, the likelihood function related to the double-bounce path ``agent -- surface $s$ -- surface $s'$ -- \ac{pa} $j$'' with \ac{mva} index-pair $(s,s') \in \mathcal{D}_{\text{D},n}$ can be expressed by $f(\V{z}_{m,n}^{(j)}|,\V{p}_{n},\V{p}_{s,\mathrm{mva}},\V{p}_{s',\mathrm{mva}})\vspace{.8mm}$. 
	
	With the \ac{mva}-based measurement model, at time $n$, the measurements collected by all \acp {pa} $j \rmv\in\rmv \{1,\dots,J\}$ can provide information on the same \ac{mva} and agent positions $\RV{p}_{s,\mathrm{mva}}, s \in \Set{S}$ and $\RV{p}_{n}$, respectively. It is assumed that each \ac{va} (related to a specific \ac{pa}, \ac{mva} or \ac{mva}-\ac{mva} pair) generates at most one measurement and that a\vspace{.5mm} measurement originates from at most one \ac{va}.
	\ac{pa} $j$ at position $\V{p}^{(j)}_{00,\text{va}} = \V{p}^{(j)}_{\mathrm{pa}}$ with $(0,0) \in \tilde{\Set{D}}_n$ generates a measurements $\RV{z}_{m,n}^{(j)}$ with detection probability $p^{(j)}_{\mathrm{d}}\big (\RV{p}_n\big)$. If \ac{mva} $s$ exists ($r_{s,n} \rmv=\rmv 1$), the corresponding single-bounce path $(s,s) \in \Set{D}_{\text{S},n}$ generates a \ac{mva}-originated measurements $\RV{z}_{m,n}^{(j)}$ with detection probability $p^{(j)}_{\mathrm{d}}\big (\RV{p}_n,\RV{p}_{s,\text{mva}} \big)$. The same holds for the double-bounce path $(s,s') \in \Set{D}_{\text{D},n}$ with detection probability $p^{(j)}_{\mathrm{d}}\big (\RV{p}_n,\RV{p}_{s,\text{mva}},\RV{p}_{s',\text{mva}}\big)$. Note that the detection probability is determined by the \ac{snr} of the measurement  \cite{LeiGreWit:ICC2019} as well as the availability check performed by \ac{rt}. In particular, if a path is unavailable, its detection probability is set to zero. A measurement $\RV{z}_{m,n}^{(j)}$ may also not originate from any \ac{mva}. This type of measurement is referred to as a false positive and is modeled as a Poisson point process with mean $\mu_{\mathrm{fp}}$ and \ac{pdf} $f_{\mathrm{fp}}(\RV{z}_{m,n}^{(j)})$.
	
	\textit{Newly detected \acp{mva}}, i.e., \acp{mva} that generated a measurement for the first time, are modeled by a Poisson point process with mean $\mu_{\mathrm{n}}$ and \ac{pdf} $f_{\mathrm{n}}(\V{p}_{m,\mathrm{mva}}|\V{p}_n)$. Newly detected \acp{mva} are represented by \textit{new \ac{pmva} states} $\overline{\RV{y}}^{(j)}_{n,m}$, $m \rmv\rmv \in \{1,\dots, \rv{M}^{(j)}_n \}$ in our statistical model \cite{MeyKroWilLauHlaBraWin:J18, LeiMeyHlaWitTufWin:J19}. Each new \ac{pmva} state corresponds to a measurement $\RV{z}_{m,n}^{(j)}$; $\overline{r}_{m,n} \!=\! 1$ implies that measurement $\RV{z}_{m,n}^{(j)}$ was generated by a newly detected \ac{mva}. All new \ac{pmva} states are introduced, assuming the corresponding measurements originate from a single-bounce path. {This assumption leads to a simpler statistical model and reduced computational complexity as further discussed in Section \ref{sec:ImpNewMVAs}.} This assumption is well motivated by the fact that, due to the lower \ac{snr} of double-bounce paths, new surfaces are typically detected first via a single-bounce measurement. However, the assumption also implies that any reflecting surface can only be mapped if it originates at least one single-bounce measurement.
		 
	We denote by $\overline{\RV{y}}^{(j)}_n \triangleq \big [ \ist \overline{\RV{y}}^{(j)\ist\T}_{1,n} \rmv\cdots\ist \overline{\RV{y}}^{(j)\ist \T}_{\rv{M}^{(j)}_n \rmv\rmv\rmv,n} \ist \big]^{\T}\rmv\rmv\rmv$, the joint vector of all new \ac{pmva} states. Introducing new \ac{pmva} for each measurement leads to a number of \ac{pmva} states that grows with time $n$. Thus, to keep the proposed \ac{slam} algorithm feasible, a sub-optimum pruning step is performed, removing \acp{pmva} with a low probability of existence (see Section~\ref{sec:probFormulation}).
	\vspace{-1mm}

	\subsection{Legacy \acp{pmva}  and State Transition\vspace{-.6mm}}
	\label{sec:MeasNewPTs}
	
	At time $n$, measurements are incorporated sequentially across \acp{pa} $j \rmv\in\rmv \{1,\dots,J\}$. Previously detected \acp{mva}, i.e., \acp{mva} that have been detected either at a previous time $n' \!<\rmv n$ or at the current time $n$ but at a previous PA $j' \!<\rmv  j$, are represented by legacy \ac{pmva} states $\underline{\RV{y}}_{s,n}^{(j)}$. New \acp{pmva} become legacy \acp{pmva} when the next measurements---either of the next \ac{pa} or at the next time instance---are taken into account. In particular, the \ac{mva} represented by the new \ac{mva} state $\overline{\RV{y}}^{(j')}_{m',n'}$ introduced due to measurement $m'$ of \ac{pa} $j'$ at time $n' \leq n$ is represented by the legacy \ac{pmva}\vspace*{-0.5mm} state $\underline{\RV{y}}^{(j)}_{s,n}$ at time $n$, with $s = \rv{S}_{n'-1} + \sum^{j' - 1}_{j'' = 1} \rv{M}^{(j'')}_{n} + m' \vspace{.5mm}$. The number of legacy \ac{pmva} at time $n$, when the measurements of the next \ac{pa} $j$ are incorporated, is updated according to $\rv{S}^{(j)}_{n} = \rv{S}^{(j-1)}_{n} + \rv{M}^{(j-1)}_{n}$, where $\rv{S}^{(1)}_{n} \rmv=\rmv \rv{S}_{n-1}$. Here, $\rv{S}^{(j)}_{n}$ is equal to the number of all measurements collected up to time $n$ and PA $j \rmv-\rmv 1$. 
	The vector of all legacy \ac{pmva} states at time $n$ and up to \ac{pa} $j$ can now be written as $\underline{\RV{y}}^{(j)}_{n} = \big[\underline{\RV{y}}^{(j-1) \T}_{n} \ist\ist\ist \overline{\RV{y}}^{(j-1) \T}_{n} \big]^{\T}\rmv\rmv$. 
	
	Let us denote by $\underline{\RV{y}}^{(1)}_n \rmv\triangleq\rmv \big[ \underline{\RV{y}}^{\T}_{1,n} \rmv\cdots\ist \underline{\RV{y}}^{T}_{S_{n\rmv-\rmv1},n} \big]^{\T}\rmv\rmv$, the vector of all legacy \ac{pmva} states before any measurements at time $n$ have been incorporated. After the measurements of all \acp{pa} $j \in \{1,\dots,J\}$ have been incorporated at time $n$, the total number of \ac{pmva} states is \vspace{-2mm}
	\begin{align}
	\rv{S}_n = \rv{S}_{n-1} + \sum_{j=1}^{J} \rv{M}^{(j)}_n = \rv{S}^{(J)}_{n} + \rv{M}^{(J)}_n \label{eq:NrPMVAtotal-update}\\[-8mm]\nn
	\end{align}
	and the vector of all PMVA states at time $n$ is given by $\RV{y}_{n} \!=\rmv \big[\underline{\RV{y}}^{(J) \T}_{n} \ist\ist \overline{\RV{y}}^{(J) \T}_{n} \big]^{\T}\rmv\rmv$.\footnote{Note that this sequential incorporation of new \ac{pmva} states is based on the multisensor multitarget tracking approach introduced in \cite[Section~VIII]{MeyKroWilLauHlaBraWin:J18}.} We also define the number of \acp{va} for each \ac{pa} given as $|\mathcal{D}^{(j)}_n| = \rv{S}^{(j)}_n+\rv{S}^{(j)}_n(\rv{S}^{(j)}_n-1)$ with $\mathcal{D}^{(j)}_n \in \{ (s,s') \rmv\in\rmv \Set{S}_n \rmv\times\rmv \Set{S}_n \} = \Set{D}_{\text{S},n}^{(j)} \cup \Set{D}_{\text{D},n}^{(j)}$ and $\tilde{\mathcal{D}}^{(j)}_n = (0,0) \cup \mathcal{D}^{(j)}_n $.
	
	Legacy \acp{pmva} states $\underline{\RV{y}}_{s,n}$ and the agent state $\RV{x}_n$ are assumed to evolve independently across time according to state-transition \acp{pdf} $f\big(\underline{\V{y}}_{s,n} \big| \V{y}_{s, n-1}\big)$ and  $f(\V{x}_{n}|\V{x}_{n-1})$, respectively. If \ac{pmva} $k$ exists at time $n \rmv-\! 1$, i.e., $r_{s,n-1} \!=\! 1$, it either disappears, i.e., $\overline{r}_{s,n} \!=\rmv 0$, or survives, i.e., $\overline{r}_{s,n} \!=\! 1$; in the latter case, it becomes a legacy \ac{pmva} at time $n$. The probability of survival is denoted by $p_\mathrm{s}$. Suppose the \ac{pmva} survives. In that case, its position remains unchanged, i.e., the state-transition pdf of the \ac{mva} positions $\underline{\RV{p}}_{s,\mathrm{mva}}$ is given by $f\big(\underline{\V{p}}_{s,\mathrm{mva}}  \ist \big| \ist \V{p}_{s,\mathrm{mva}} \big) = \delta \big(\underline{\V{p}}_{s,\mathrm{mva}} \rmv - \ist \V{p}_{s,\mathrm{mva}} \big)$.
	Therefore, $f\big(\underline{\V{p}}_{s,\mathrm{mva}}\rmv,\underline{r}_{s,n} \ist \big| \ist \V{p}_{s,\mathrm{mva}} , r_{s,n-1} \big) $ for $r_{s,n-1} \rmv=\rmv 1$ is obtained as\vspace*{-2mm}  
	\begin{align}
	&f\big(\underline{\V{p}}_{s,\mathrm{mva}},\underline{r}_{s,n} \ist \big| \ist \V{p}_{s,\mathrm{mva}} , r_{s,n-1} = 1\big) \nn\\[0mm]
	&\hspace{17mm}=\rmv\begin{cases} 
	(1 \!-\rmv p_\mathrm{s}) \ist f_\text{d}\big(\underline{\V{p}}_{s,\mathrm{mva}} \big) , &\!\!\! \underline{r}_{s,n} \!=\rmv 0 \\[0mm]
	p_\mathrm{s} \ist\ist \delta \big(\underline{\V{p}}_{s,\mathrm{mva}} \rmv - \ist \V{p}_{s,\mathrm{mva}} \big) , &\!\!\! \underline{r}_{s,n} \!=\! 1.
	\end{cases}\label{eq:stmpmvarone}\\[-7mm]\nn
	\end{align}
	If \ac{mva} $s$ does not exist at time $n \rmv-\! 1$, i.e., $r_{s,n-1} \!=\! 0$, it cannot exist as a legacy \ac{pmva} at time $n$ \vspace{1mm} either, thus we get
	\vspace*{-2mm}
	\begin{align}
	&f\big(\underline{\V{p}}_{s,\mathrm{mva}},\underline{r}_{s,n} \big| \V{p}_{s,\mathrm{mva}} , r_{s,n-1} \rmv\rmv= \rmv\rmv 0\big) \nn\\ 	
	&\hspace{27mm}=\rmv\begin{cases} 
	f_\text{d}\big( \underline{\V{p}}_{s,\mathrm{mva}} \big) , &\!\!\! \underline{r}_{s,n} \!=\rmv 0 \\[0mm]
	0 , &\!\!\! \underline{r}_{s,n} \!=\! 1.
	\end{cases}\label{eq:stmpmvarzero}\\[-5mm]\nn
	\end{align}
	For $j \geq 2$, we also define $f^{(j)}\big(\underline{\V{y}}^{(j)}_{s,n} \big| \underline{\V{y}}^{(j-1)}_{s, n}\big)$ as
	\begin{align}
		&f^{(j)}\big(\underline{\V{p}}^{(j)}_{s,\mathrm{mva}},\underline{r}^{(j)}_{s,n} \ist \big| {\ist \underline{\V{p}}^{(j-1)}_{s,\mathrm{mva}} , \underline{r}^{(j-1)}_{s,n} = 1}\big) \nn\\[0mm]
		&\hspace{17mm}=\rmv\begin{cases} 
			f_\text{d}\big(\underline{\V{p}}^{(j-1)}_{s,\mathrm{mva}} \big) , &\!\!\! \underline{r}^{(j)}_{s,n} \!=\rmv 0 \\[0mm]
			\delta \big(\underline{\V{p}}^{(j)}_{s,\mathrm{mva}} \rmv - \ist \underline{\V{p}}^{(j-1)}_{s,\mathrm{mva}} \big) , &\!\!\! \underline{r}^{(j)}_{s,n} \!=\! 1
		\end{cases}\label{eq:stmpmvaonepas}\\[-7mm]\nn
	\end{align}
	and
	\begin{align}
		&f^{(j)}\big(\underline{\V{p}}^{(j)}_{s,\mathrm{mva}},\underline{r}^{(j)}_{s,n} \big| {\underline{\V{p}}^{(j-1)}_{s,\mathrm{mva}} , \underline{r}^{(j-1)}_{s,n} \rmv\rmv= \rmv\rmv 0}\big) \nn\\ 	
		&\hspace{27mm}=\rmv\begin{cases} 
			f_\text{d}\big( \underline{\V{p}}^{(j)}_{s,\mathrm{mva}} \big) , &\!\!\! \underline{r}^{(j)}_{s,n} \!=\rmv 0 \\[0mm]
			0 , &\!\!\! \underline{r}^{(j)}_{s,n} \!=\! 1.
		\end{cases}\label{eq:stmpmvazeropas}
	\end{align}
where we introduced $ \underline{\V{y}}^{(j-1)}_{s, n} \triangleq  \underline{\V{y}}_{s, n}$, $\V{p}^{(j-1)}_{s,\mathrm{mva}} \triangleq \V{p}_{s,\mathrm{mva}}$, and $ \underline{r}^{(j-1)}_{s, n} \triangleq  \underline{r}_{s, n}$. It it assumed that at time $n \rmv=\rmv 0$ the initial prior \ac{pdf} $f\big(\V{y}_{s,0} \big)$, $s = \big\{1,\dots,S_0\big\}$ and  $f(\V{x}_{0})$ are known. All (legacy \vspace{0mm} and new) \ac{pmva} states and all agent states up to time $n$ are denoted as $\RV{y}_{0:n} \triangleq \big[\RV{y}^{\T}_{0} \cdots\ist \RV{y}^{\T}_{n} \big]^{\T}\!$ and $\RV{x}_{0:n} \triangleq \big[\RV{x}^{\T}_{0} \cdots\ist \RV{x}^{\T}_{n} \big]^{\T}\!$, respectively.
	
	\subsection{Data Association Uncertainty \vspace{-.5mm}}
	\label{sec:DataAssocUncer}
	Mapping of reflective surfaces modeled by \ac{mva} is complicated by the data association uncertainty: at time $n$ it is unknown which measurement $\RV{z}_{m,n}^{(j)}$ extracted at \ac{pa} $j$ originated from \ac{pa} $j$ itself $(0,0)$, from which \ac{mva} $(s,s) \in \mathcal{D}^{(j)}_{\text{S},n}$, or from which \ac{mva}-\ac{mva} pair $(s,s') \in \mathcal{D}^{(j)}_{\text{D},n}$ associated with single-bounce and double-bounce path. Any \ac{pmva}-to-measurement association (which considers associations to single \acp{pmva} and \ac{pmva}-\ac{pmva} pairs as well as to \ac{pa} $j$ itself) is described by \emph{\ac{pmva}-oriented association \vspace{0mm} variables}
	\begin{equation}
		\hspace*{-2mm}\rv{\underline{a}}^{(j)}_{ss',n} \ist\triangleq \begin{cases} 
		m \rmv\in\rmv \Set{M}_n^{(j)}\ist, & \begin{minipage}[t]{45mm}if legacy \ac{pmva} $ss'$ generates measurement $m$\\[-2.5mm]\end{minipage}\\[4mm]
		0 \ist, & \begin{minipage}[t]{45mm}if legacy \ac{pmva} $ss'$ does not generate any measurement\\[-4.5mm]\end{minipage} 
		\end{cases}\label{eq:pmvaorientedDAV}
	\end{equation}
	with $(s,s') \in \tilde{\mathcal{D}}^{(j)}_n$ and stacked into the \ac{pmva}-oriented association vector as $\vspace*{0.5mm}\underline{\RV{a}}^{(j)}_n \rmv= \big[\rv{\underline{a}}^{(j)}_{00,n} \iist \rv{\underline{a}}^{(j)}_{11,n} \cdots \rv{\underline{a}}^{(j)}_{\rv{S}_n^{(j)}\rv{S}_n^{(j)}\rmv\rmv\rmv,n} \big]^{\T}\rmv\rmv$. 
	To reduce computation complexity, following \cite{BayShaSha:J08,CheKroKrzVerZde:J10, WilLau:J14, MeyBraWilHla:J17, MeyKroWilLauHlaBraWin:J18,LeiMeyHlaWitTufWin:J19}, we use a redundant description of \ac{pmva}-measurement associations, i.e., we introduce \emph{measurement-oriented association variables}
	\vspace{-2mm}
	\begin{equation}
	\rv{\overline{a}}^{(j)}_{m,n} \rmv\rmv\triangleq\rmv\rmv \begin{cases} 
		(s,\rmv s') \rmv\rmv\in\rmv\rmv \tilde{\mathcal{D}}^{(j)}_n \rmv\rmv,\rmv\rmv\rmv\rmv\rmv& \begin{minipage}[t]{45mm}if measurement $m$ is originated by legacy \ac{pmva} $ss'$\end{minipage}\\[2mm]
		0,\rmv\rmv\rmv\rmv\rmv& \begin{minipage}[t]{45mm}if measurement $m$ is not generated by any legacy \ac{pmva} $ss'$\end{minipage}
		\end{cases}\rmv\rmv\label{eq:measorientedDAV}
	\end{equation}
	and stacked into the measurement-oriented association vector as $\overline{\RV{a}}^{(j)}_n \rmv= \big[\rv{\overline{a}}^{(j)}_{1,n} \cdots \rv{\overline{a}}^{(j)}_{M^{(j)}_n \rmv\rmv\rmv,n} \big]^{\T}\rmv\vspace*{0.4mm}$. Note that any data association event that can be expressed by both a joint \ac{pmva}-oriented association vector $\underline{\RV{a}}^{(j)}_n$ and measurement-oriented association vector $\overline{\RV{a}}^{(j)}_n$  is a valid event in the sense that an \ac{pmva} generates at most one measurement. A measurement is originated by at most one \ac{pmva}. This hybrid representation of data association makes it possible to develop scalable \acp{spa} for simultaneous agent localization \ac{mva} mapping \cite{WilLau:J14,MeyBraWilHla:J17,MeyKroWilLauHlaBraWin:J18,LeiMeyHlaWitTufWin:J19}. Finally, we also introduce the joint association \vspace{.5mm} vectors $\underline{\RV{a}}_n \rmv= \big[\underline{\RV{a}}^{(1) \T}_{n} \cdots \underline{\RV{a}}^{(J) \T}_{n} \big]^{\T}\rmv\rmv$, $\overline{\RV{a}}_n \rmv= \big[\overline{\RV{a}}^{(1)\T}_{n} \cdots \overline{\RV{a}}^{(J)\T}_{n} \big]^{\T}\rmv\rmv$, $\underline{\RV{a}}_{1:n} \rmv= \big[\underline{\RV{a}}^{\T}_{1} \cdots \underline{\RV{a}}^{\T}_{n} \big]^{\T}\rmv\rmv$, and $\overline{\RV{a}}_{1:n} \rmv= \big[\overline{\RV{a}}^{\T}_{1} \cdots \overline{\RV{a}}^{\T}_{n} \big]^{\T}\rmv\rmv$.

	\subsection{Detection Probabilites and Availability of Paths \vspace{-.5mm}}
	\label{sec:Pdvisibility}

	\ac{rt} relies on visibility-tree techniques  \cite{Bor:JASA1984, MckHam:IEEENetwork1991, LuVitDegFusBarBlaBer:TAP2019} to determine the visibility of a path in a backward manner as discussed in what follows for the double-bounce case $(s,s') \in \Set{D}_{\text{D},n}^{(j)}$. First, starting from the agent's position, $\V{p}_n$, a straight line is drawn towards the \ac{va} position, $\V{p}^{(j)}_{ss',\mathrm{va}} $, until it intersects with the reflective surface with index $s'$. From the resulting intersection point, another line is drawn towards the  \ac{va} position, $\V{p}^{(j)}_{ss,\mathrm{va}} $, until it intersects with the reflective surface with index $s$. From this second intersection point, a line is finally drawn towards physical anchor position $\V{p}^{(j)}_{\mathrm{pa}} $. If this procedure fails, because (i) there is no intersect first with surface $s'$ and then with surface $s$, or (ii) along the path, there is an intersection with another surface $s'' \in \Set{S} \backslash \{j,j'\}$, the path is considered not available or ``blocked''. Intersections are calculated efficiently based on the fast line intersection algorithm \cite{Antonio:GG1992}. 
	
	For single-bounce and LOS paths, the procedure is simpler. In particular, for the single-bounce case  $(s,s) \in \Set{D}_{\text{D},n}^{(j)}$, starting from the agent's position, the first of two straight lines is already drawn from the agent's position $\V{p}_n$, towards the \ac{va} position, $\V{p}^{(j)}_{ss',\mathrm{va}} $. For the LOS case, the first and only straight line is directly drawn from the agent's position $\V{p}_n$ to $\V{p}^{(j)}_{\mathrm{pa}} $. 
	For each \ac{pa} $j$, this availability check is directly integrated into  detection probabilities. In particular, the \ac{los} path detection probability, $p^{(j)}_{\mathrm{d}}\big (\RV{p}_n\big)$, is given by
	\vspace*{-1mm}
	\begin{align}
		p^{(j)}_{\mathrm{d}}\big (\RV{p}_n\big) \ist\triangleq \begin{cases} 
			p_{\mathrm{d},00,n}^{(j)}\ist, & \begin{minipage}[t]{35mm}path from agent at $\RV{p}_n$ to \ac{va} at $\RV{p}^{(j)}_{\text{pa}}$ is available\\[-2.5mm]\end{minipage}\\[4mm]
			0 \ist, & \begin{minipage}[t]{35mm}path is not available\\[-2.5mm]\end{minipage} 
		\end{cases}\label{eq:visibilityPA}\\[-6mm]\nn
	\end{align}
	where {$p_{\mathrm{d},00,n}^{(j)}$ is the detection probability \cite{MeyKroWilLauHlaBraWin:J18,LeiGreWit:ICC2019,LiLeiVenTuf:TWC2022} of the available \ac{los} path}. Similarly, the single-bounce detection probability, $p^{(j)}_{\mathrm{d}}\big (\RV{p}_n,\RV{p}_{s,\text{mva}}\big)$, $(s,s) \in \Set{D}_{\text{S},n}^{(j)}$, reads
	\vspace*{-1mm}
	\begin{align}
	\hspace*{-1mm} p ^{(j)}_{\mathrm{d}}\big (\RV{p}_n,\RV{p}_{s,\text{mva}}\big) \ist\triangleq \begin{cases} 
		p_{\mathrm{d},ss,n}^{(j)}\ist, & \hspace*{-0.5mm}\begin{minipage}[t]{35mm}path from agent at $\RV{p}_n$ to \ac{va} at $\RV{p}^{(j)}_{ss,\text{va}}$ is  available\\[-2.5mm]\end{minipage}\\[4mm]
		0 \ist, & \hspace*{-0.5mm}\begin{minipage}[t]{35mm}path is not available\\[-2.5mm]\end{minipage} 
	\end{cases}\label{eq:visibilitySB}\\[-7mm]\nn
	\end{align}
	where {$p_{\mathrm{d},ss,n}^{(j)}$ is the detection probability of available single-bounce path. Finally, the double-bounce detection probability $p^{(j)}_{\mathrm{d}}\big (\RV{p}_n,\RV{p}_{s,\text{mva}},\RV{p}_{s',\text{mva}}\big)$, $(s,s') \in \Set{D}_{\text{D},n}^{(j)}$, can be obtained as
	\vspace*{-1mm}
	\begin{align}
	&p^{(j)}_{\mathrm{d}}\big (\RV{p}_n,\RV{p}_{s,\text{mva}},\RV{p}_{s',\text{mva}}\big) \nn\\[1.5mm]
	&\hspace{10mm}\triangleq \begin{cases} 
	p_{\mathrm{d},ss',n}^{(j)}\ist, & \begin{minipage}[t]{43mm}path from agent at $\RV{p}_n$ to \ac{pmva} at $\RV{p}^{(j)}_{ss',\text{va}}$ is  available\\[-2.5mm]\end{minipage}\\[4mm]
		0 \ist, & \begin{minipage}[t]{43mm}path is not available\\[-2.5mm]\end{minipage} 
	\end{cases} \label{eq:visibilityDB}\\[-7mm]\nn
	\end{align}
	where {$p_{\mathrm{d},ss',n}^{(j)}$ is again the detection probability of the available double-bounce path.
	
	\begin{figure*}[!t]
		\centering
		\scalebox{0.80}{
		\includegraphics{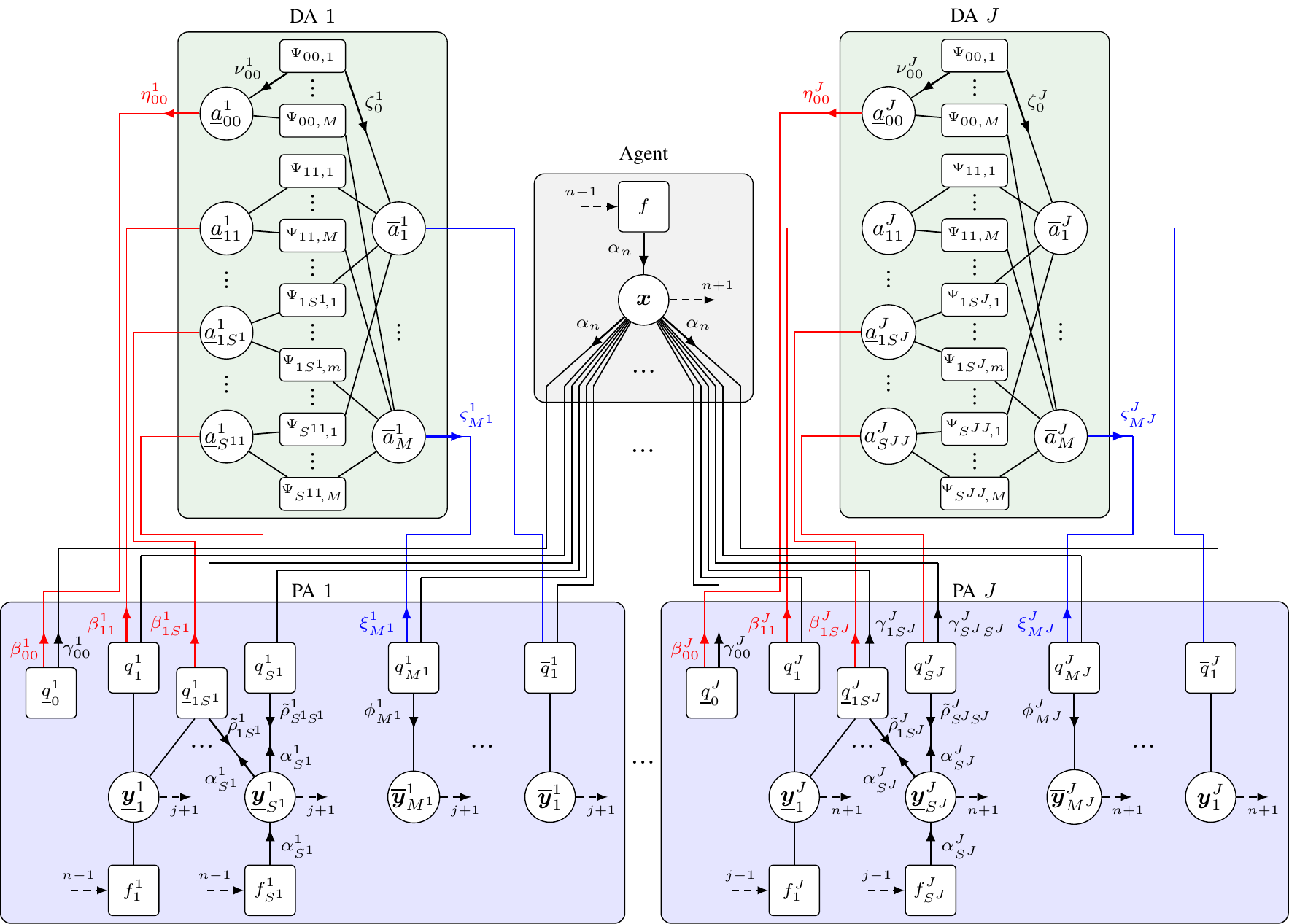}}
		\vspace*{-2mm}
		\caption{
			Factor graph representation of the joint posterior \ac{pdf} (\ref{eq:factorGraph}). Short notations are used. \vspace*{1mm}{In particular, the time index $n$ and the functional dependencies of the factors are neglected:}
			{$\V{x} \rmv\triangleq\rmv \V{x}_n $},		
			$M^j \rmv\triangleq\rmv M_n^{(j)}$,
			$\underline{\V{y}}^j_s \rmv\triangleq\rmv \underline{\V{y}}_{s,n}^{(j)}$,
			$\overline{\V{y}}^j_m \rmv\triangleq\rmv \overline{\V{y}}_{m,n}^{(j)}$,
			$\underline{q}^j_0 \rmv\triangleq\rmv \underline{q}_{\mathrm{P}}\big(  \V{x}_{n} , \underline{a}^{(j)}_{00,n}; \V{z}^{(j)}_{n} \big)$,
			$\underline{q}^j_s \rmv\triangleq\rmv \underline{q}_{\mathrm{S}}\big( \underline{\V{y}}_{s,n}^{(j)},  \underline{a}^{(j)}_{ss,n},  \V{x}_{n} ; \V{z}^{(j)}_{n} \big)$,
			$\underline{q}^j_{ss'} \rmv\triangleq\rmv \underline{q}_\text{D}\big( \underline{\V{y}}_{s,n}^{(j)}, \underline{\V{y}}_{s',n}^{(j)},  \underline{a}^{(j)}_{ss',n},  \V{x}_{n} ; \V{z}^{(j)}_{n} \big)
			{\, \underline{q}_\text{D}\big( \underline{\V{y}}_{s',n}^{(j)}, \underline{\V{y}}_{s,n}^{(j)},  \underline{a}^{(j)}_{s's,n},  \V{x}_{n} ; \V{z}^{(j)}_{n} \big)}$,	
			$\overline{q}^j_m \rmv\triangleq\rmv \overline{q}_{\mathrm{S}}\big( \overline{\V{y}}^{(j)}_{m,n}, \overline{a}^{(j)}_{m,n},\V{x}_{n}; z^{(j)}_{m,n} \big)$, 
			$f \rmv\triangleq\rmv f(\V{x}_n | \V{x}_{n-1})$,
			$f^1_s \rmv\triangleq\rmv f\big(\underline{\V{y}}_{s,n} | \V{y}_{s,n-1}\big)$,
			$f^j_s \rmv\triangleq\rmv f^{(j)}\big(\underline{\V{y}}^{(j)}_{s,n} \big| \underline{\V{y}}^{(j-1)}_{s,n}\big)$,
			$\alpha_n = \alpha( \V{x}_n)$,
			$\alpha_s^j = \alpha_s\big( \underline{\V{y}}^{(j)}_{s,n}\big)$,
			$\beta_{ss'}^j = \beta_{ss'}\big(\underline{a}_{ss',n}^{(j)} \big)  {\,  \beta_{s's}\big(\underline{a}_{s's,n}^{(j)} \big) }$,
			$\gamma_{ss'}^j = \gamma^{(j)}_{ss'}(\V{x}_n)  {\,  \gamma^{(j)}_{s's}(\V{x}_n) }$,
			$\xi_m^j  = \xi \big(\overline{a}^{(j)}_{m,n}\big)$,
			$\eta_{ss'}^j = \eta(\underline{a}_{ss',n}^{(j)} )   {\,   \eta(\underline{a}_{s's,n}^{(j)} )  }$,
			$\nu_{ss'}^j = \nu_{m\rightarrow ss'}^{(p)}(\underline{a}_{ss',n} ^{(j)}) {\,  \nu_{m\rightarrow s's}^{(p)}(\underline{a}_{s's,n} ^{(j)}) }$,
			$\varsigma_{m}^j = \varsigma\big( \overline{a}_{m,n}^{(j)} \big)$,
			$\zeta_{m}^j = \zeta_{ss' \rightarrow m}^{(p)}\big(\overline{a}_{m,n}^{(j)}\big) {\, \zeta_{s's \rightarrow m}^{(p)}\big(\overline{a}_{m,n}^{(j)}\big) }$,
			$\tilde{\rho}_{ss'}^j = \rho_{ss'}\big(\underline{\V{y}}^{(j)}_{s}\big) {\,  \rho_{s's}\big(\underline{\V{y}}^{(j)}_{s}\big)  }$,
			and
			$\phi_{m}^j = \phi(\overline{\V{y}}_{s}^{(j)})${.  For the numbers of \acp{mva}, the short notations reads $S^{1} \rmv\triangleq\rmv S_{n-1}$,
				$S^j \rmv\triangleq\rmv S_n^{(j)}$, and
				$(\cdot)_{S^{jj}} \rmv\triangleq\rmv (\cdot)_{S_n^{(j)}S_n^{(j)}}$. }
			The dashed lines with arrows indicate messages representing the agent and \acp{pmva} beliefs of time $n\rmv\rmv-\rmv\rmv 1$, $n+1$ or of anchors $j\rmv\rmv-\rmv\rmv 1$, $j\rmv\rmv+\rmv\rmv 1${. These messages are either only sent to the next time step (e.g., from $n-1$ to $n$) or only to the next anchor (e.g., from $j-1$ to $j$).} 
		}\label{fig:factorgraph}
		\vspace*{-4mm}
	\end{figure*}
	
	\section{Problem Formulation and Proposed Method}
	In this section, we formulate the estimation problem of interest and present the joint posterior \ac{pdf} and factor graph underlying the proposed \ac{slam} method.
	
	\subsection{Pre-Estimation Stage}\label{sec:channelEstimation}
		
	By applying at each time $ n $ a super-resolution channel estimation algorithm \cite{ShutWanJos:CSTA2013, BadHanFle:TSP2017, HanFleRao:TSP2018, LiLeiVenTuf:TWC2022, GreLeiFleWit:Arxiv2023} to the {observed} \ac{rf} signal vector one obtains, for each anchor $j$, a number of $M_n^{(j)}$ measurements denoted by ${\V{z}^{(j)}_{m,n}}$ with $m \in  \Set{M}_n^{(j)} \triangleq \{1,\,\dots\,,M_n^{(j)}\} $.
	Each  $\V{z}^{(j)}_{m,n} = [{z^{(j)}_\text{d}}_{\rmv\rmv\rmv\rmv\rmv\rmv\rmv m,n} ~ {z^{(j)}_\varphi}_{\rmv\rmv\rmv\rmv\rmv m,n}]^\text{T}$ representing a potential \ac{mpc} parameter estimate contains a distance measurement ${z_\text{d}}_{m,n}^{(j)} \rmv\rmv\in [0, d_\text{max}]$ and an \ac{aoa} measurement ${z_\varphi}_{m,n}^{(j)} \rmv\rmv\in\rmv\rmv [-\pi, \pi )$. The channel estimator decomposes the \ac{rf} signal vector into individual, decorrelated components, reducing the number of dimensions (as ${M}_n^{(j)}$ is usually much smaller than the number of signal samples). It thus compresses the information contained in the \ac{rf} signal vector into $\vspace*{-0.2mm}\V{z}^{(j)}_{n} = [{\bm{z}^{(j)\text{T}}_{1,n}}  \rmv \cdots  {\V{z}^{(j)\text{T}}_{M_n^{(j)},n}}]^\text{T}$. We also define 
	$\V{z}_n \triangleq [ \V{z}_{n}^{(1)\ist \T}  \dots \V{z}_n^{(J)\ist \T} ]^{\T}\rmv\rmv\vspace*{-0.6mm}$ and  $\V{z}_{1:n} \triangleq \big[\V{z}^{\T}_{0} \cdots\ist \V{z}^{\T}_{n} \big]^{\T}$. %
	
	\subsection{Confirmation of \acp{mva} and State Estimation}\label{sec:probFormulation}
	
	We aim to estimate the agent state $\RV{x}_n$ using all available measurements $\V{z}_{1:n}\rmv$ from all \acp{pa} up to time $n$. In particular, we calculate an estimate $\hat{\V{x}}_n$ by using the \ac{mmse} estimator \vspace{.5mm} \cite[Ch.~4]{Poo:B94}\vspace*{-1.5mm}
	\begin{equation}
	\hat{\V{x}}_n \ist\triangleq\, \int \rmv \V{x}_n \ist f( \V{x}_n|\V{z}_{1:n}) \ist \mathrm{d} \V{x}_n \rmv. \label{eq:MMSEagent}
	\vspace*{-1.5mm}
	\end{equation}
	The map of the environment is represented by reflective surfaces described by \acp{pmva}. Therefore, the positions $\RV{p}_{s,\mathrm{mva}}$ of the detected \acp{pmva} $s \!\in\! \{ 1,\dots,S_n \}$ must be estimated. This relies on the marginal posterior existence probabilities $p(r_{s,n} \!=\! 1|\V{z}_{1:n}) = \int f(\V{p}_{s,\text{mva}} , r_{s,n} \!=\! 1| \V{z}_{1:n} ) \mathrm{d}\V{p}_{s,\text{mva}}$ and the marginal posterior \acp{pdf} $f(\V{p}_{s,\text{mva}} | r_{s,n} \!=\! 1, \V{z}_{1:n} ) \rmv\rmv=\rmv\rmv f(\V{p}_{s,\text{mva}} , r_{s,n} \!=\! 1| \V{z}_{1:n} )/p(r_{s,n} \!=\! 1|\V{z}_{1:n})$. A \ac{pmva} $s$ is declared to exist $p(r_{s,n} \!=\! 1|\V{z}_{1:n}) > p_{\mathrm{cf}}$, where $p_{\mathrm{cf}}$ is a {confirmation} threshold. The number $\hat{S}_n$ of \ac{pmva} states that are considered to exist is the estimate of the total number $S$ of \acp{mva}. {To avoid that the number of \acp{pmva} states grows indefinitely, \acp{pmva} states with $p(r_{s,n} \!=\! 1|\V{z}_{1:n}) < p_{\text{pr}}$ are removed from the state space (``pruned'').} For existing \acp{pmva}, an estimate of it's position $\RV{p}_{s,\mathrm{mva}}$ can again be calculated by the \ac{mmse} \cite[Ch.~4]{Poo:B94}
	\vspace*{-1mm}
	\begin{align}
	\hat{\V{p}}_{k,\mathrm{mva}} \,\triangleq \int \rmv \V{p}_{k,\mathrm{mva}}  \ist\ist f(\V{p}_{k,\mathrm{mva}} \ist | \ist r_{s,n} \!=\! 1, \V{z}_{1:n}) \ist\ist \mathrm{d} \V{p}_{k,\mathrm{mva}} \ist. \label{eq:MMSEpmva}
	\vspace*{-4mm}
	\end{align}
	The calculation of $f( \V{x}_n|\V{z}_{1:n})$, $p(r_{s,n} \!=\! 1 |\V{z})$, and $f(\V{p}_{k,\mathrm{mva}} |$ $r_{s,n} \rmv=\rmv 1, \V{z}_{1:n})$ from the joint posterior $f( \V{y}_{0:n}, \V{x}_{0:n},\underline{\V{a}}_{1:n},\overline{\V{a}}_{1:n} $ $| \ist \V{z}_{1:n}) $ by direct marginalization is not feasible. By performing sequential message passing using the SPA rules \cite{KscFreLoe:TIT2001, MeyHliHla:TSPIN2016, MeyBraWilHla:J17, LeiMeyHlaWitTufWin:J19} on the factor graph in Fig.~\ref{fig:factorgraph}, approximations (``beliefs'') $\tilde{f}\big(\V{x}_{n} \big)$ and $\tilde{f}_s\big(\V{y}_{s,n} \big)$ of the marginal posterior \acp{pdf} $f( \V{x}_n|\V{z}_{1:n})$, $p(r_{s,n} \!=\! 1 |\V{z}_{1:n})$, and $f(\V{p}_{s,\mathrm{mva}} |$ $r_{s,n} \rmv=\rmv 1, \V{z}_{1:n})$ can be obtained efficiently for the agent state as well as all legacy and new \acp{pmva} states $s \in \Set{S}_n$. 
	
	\begin{figure*}[!t]		
		\begin{align}
			&f( \V{y}_{0:n}, \V{x}_{0:n}, \underline{\V{a}}_{1:n},\overline{\V{a}}_{1:n} \ist | \ist \V{z}_{1:n})\nn\\[-.5mm]
			&\hspace{0mm} \propto \hspace{-3mm} \underbrace{\bigg( \ist f(\V{x}_{0}) \ist \prod^{S_0}_{s = 1} \ist f(\V{y}_{s,0})  \bigg)}_{\footnotesize \text{Initial prior \acp{pdf} of \ac{pmva} and agent states}} \hspace{-5mm}\prod^{n}_{n' = 1}  \underbrace{f\big(\V{x}_{n'} | \V{x}_{n'-1}\big)}_{\footnotesize \text{Agent state transition}} \hspace{.8mm} \underbrace{\bigg(\prod^{J}_{j = 1} \ist \underline{q}_{\mathrm{P}}\big( \V{x}_{n'}, \underline{a}^{(j)}_{00,n'}; \V{z}^{(j)}_{n'} \big) \prod^{M^{(j)}_{n'}}_{m' = 1} \rmv\rmv \Psi\big(\underline{a}^{(j)}_{00,n'},\overline{a}^{(j)}_{m',n'} \big) \bigg)}_{\footnotesize \text{Factors related to \acp{pa}}} \underbrace{\bigg(  \prod^{S_{n'\rmv-\rmv1}}_{s' = 1} f \big(\underline{\V{y}}_{s'\rmv\rmv,n'} | \V{y}_{s'\rmv\rmv,n'-1}\big) \rmv \bigg)}_{\footnotesize \text{Legacy \ac{pmva} state transition}}\nn\\[-1mm]
			&\hspace{1mm}\times \underbrace{\bigg( \prod^{J}_{j' = 2} \ist\ist \bigg(\prod^{S^{(j')}_{n'}}_{s' = 1}\ist f^{(j)}\big(\underline{\V{y}}^{(j')}_{s',n'} \big| \underline{\V{y}}^{(j'-1)}_{s', n'}\big)\bigg)\bigg) \prod^{J}_{j = 1} \bigg( \ist \prod^{S^{(j)}_{n'}}_{s = 1} \underline{q}_{\mathrm{S}}\big( \underline{\V{y}}^{(j)}_{s,n'}\rmv, \underline{a}^{(j)}_{ss,n'},  \V{x}_{n'} ; \V{z}^{(j)}_{n'} \big) \bigg(\prod^{M^{(j)}_{n'}}_{m' = 1} \rmv\rmv \Psi\big(\underline{a}^{(j)}_{ss,n'},\overline{a}^{(j)}_{m',n'} \big) \rmv\rmv \bigg)}_{\footnotesize  \text{Factors related to legacy \ac{pmva} states}} \nn\\[-1mm]
			& \hspace{1mm} \times \rmv\rmv\rmv\underbrace{\prod^{S^{(j)}_{n'}}_{s' = 1,s' \neq s}  \underline{q}_{\mathrm{D}}\big(\underline{\V{y}}^{(j)}_{s,n'}, \underline{\V{y}}^{(j)}_{s',n'},\underline{a}^{(j)}_{ss',n'},  \V{x}_{n'} ; \V{z}^{(j)}_{n'} \big) \rmv\rmv\prod^{M^{(j)}_{n'}}_{m' = 1} \rmv\rmv \Psi\big(\underline{a}^{(j)}_{ss',n'},\overline{a}^{(j)}_{m',n'} \big) \bigg)}_{\footnotesize \text{Factors related to legacy \ac{pmva} states}}  \underbrace{\bigg( \prod^{M^{(j)}_{n'}}_{m = 1} \overline{q}_{\mathrm{S}}\big( \overline{\V{y}}^{(j)}_{m,n'}, \overline{a}^{(j)}_{m,n'}, \V{x}_{n'} ; \V{z}^{(j)}_{n'} \big) \rmv \bigg)}_{\footnotesize  \text{Prior \acp{pdf} and related parameters of new \ac{pmva} states}}
			\label{eq:factorGraph}\\[-9mm] \nn
		\end{align}
		\hrulefill
		\vspace*{-3mm}
	\end{figure*}

	\subsection{The Factor Graph}\label{sec:FactorGraph}

	By using common assumptions \cite{MeyKroWilLauHlaBraWin:J18,LeiMeyHlaWitTufWin:J19}, and for fixed {(observed)} measurements $\V{z}_{1:n}$, %
	the joint posterior \ac{pdf} of $\RV{y}_{0:n}$, $\RV{x}_{0:n}$, $\underline{\RV{a}}_{1:n}$, and $\overline{\RV{a}}_{1:n}$, conditioned on $\V{z}_{1:n}$ is given by \eqref{eq:factorGraph} as shown on top of the page, where we introduced the \vspace{-.2mm} functions $\underline{q}_{\mathrm{P}}\big( \V{x}_{n'}, \underline{a}^{(j)}_{00,n'}; \V{z}^{(j)}_{n'} \big)$, $\underline{q}_{\mathrm{S}}\big( \underline{\V{y}}^{(j)}_{s,n}\rmv, \underline{a}^{(j)}_{ss,n},  \V{x}_{n} ; \V{z}^{(j)}_{n} \big)$, $\underline{q}_{\mathrm{D}}\big( \underline{\V{y}}^{(j)}_{s,n}\rmv,\underline{\V{y}}^{(j)}_{s',n}\rmv, \underline{a}^{(j)}_{ss',n},  \V{x}_{n} ; \V{z}^{(j)}_{n} \big)$, $\overline{q}_{\mathrm{S}}\big( \overline{\V{y}}^{(j)}_{m,n}, \overline{a}^{(j)}_{m,n}, \V{x}_{n} ; z^{(j)}_{m,n} \big) $, $\bar{f}\big(\overline{\V{y}}_{s,n}\big)$, and \vspace{.2mm}  $\Psi\big(a^{(j)}_{s,n},\overline{a}^{(j)}_{m,n} \big) $ that will be discussed next. {A detailed derivation of \eqref{eq:factorGraph} is provided in the supplementary material \mref{Section}{sec:app_statistical_model}.}
	
	The \textit{pseudo likelihood functions} of \ac{pa} $j$ $\underline{q}_{\mathrm{P}}\big( \V{x}_{n}, \underline{a}^{(j)}_{00,n}; \V{z}^{(j)}_{n} \big)$, of legacy \acp{pmva} related to single-bounce paths $\underline{q}_{\mathrm{S}}\big( \underline{\V{y}}^{(j)}_{s,n}\rmv,\underline{a}^{(j)}_{ss,n},  \V{x}_{n} ; \V{z}^{(j)}_{n} \big) = \underline{q}_{\mathrm{S}}\big( \underline{\V{p}}^{(j)}_{s,\mathrm{mva}}\rmv,\underline{r}^{(j)}_{s,n}\rmv,$ $\underline{a}^{(j)}_{ss,n},  \V{x}_{n} ; \V{z}^{(j)}_{n} \big)$ and to double-bounce paths $\underline{q}_{\mathrm{D}}\big( \underline{\V{y}}^{(j)}_{s,n}\rmv,\underline{\V{y}}^{(j)}_{s',n}\rmv, \underline{a}^{(j)}_{ss',n},  \V{x}_{n} ; \V{z}^{(j)}_{n} \big) = \underline{q}_{\mathrm{D}}\big( \underline{\V{p}}^{(j)}_{s,\mathrm{mva}}\rmv, \underline{r}^{(j)}_{s,n}\rmv,\underline{\V{p}}^{(j)}_{s',\mathrm{mva}}\rmv,$ $\underline{r}^{(j)}_{s',n}\rmv, \underline{a}^{(j)}_{ss',n}, \V{x}_{n} ; \V{z}^{(j)}_{n} \big)\vspace*{0.4mm}$ are, respectively, given for $(0,0)$ by
	\begin{align}
		&\underline{q}_{\mathrm{P}}\big( \V{x}_{n}, \underline{a}^{(j)}_{00,n}; \V{z}^{(j)}_{n} \big)  \nn\\
		&\hspace{4mm} \triangleq \begin{cases}
			\displaystyle \ist \frac{ p^{(j)}_{\mathrm{d}} (\V{p}_n)  f\big( \V{z}_{m,n}^{(j)} \big|\ist \V{p}_n \big)}{ \mu_{\mathrm{fp}} \ist f_{\mathrm{fp}}\big( \V{z}_{m,n}^{(j)} \big)} \ist, 
			& \!\!\rmv \underline{a}^{(j)}_{00,n} \!=\rmv m \rmv\in\rmv \Set{M}_n^{(j)}\\[3.5mm]
			1 \!-\rmv p^{(j)}_{\mathrm{d}} (\V{p}_n) \ist, & \!\!\rmv \underline{a}^{(j)}_{00,n} \!=\rmv 0   
		\end{cases} \label{eq:factorqP}\\[-5mm]\nn
	\end{align}
	for $(s,s) \in \mathcal{D}^{(j)}_{\text{S},n}$ by
	\begin{align}
		&\underline{q}_{\mathrm{S}}\big( \underline{\V{p}}^{(j)}_{s,\mathrm{mva}}\rmv,  \underline{r}^{(j)}_{s,n} \rmv=\rmv1, \underline{a}^{(j)}_{ss,n},  \V{x}_{n} ; \V{z}^{(j)}_{n} \big)  \nn\\
		&\hspace{0mm} \triangleq \begin{cases}
			\displaystyle \ist \frac{ 	p^{(j)}_{\mathrm{d}}(\V{p}_n,\V{p}^{(j)}_{s,\text{mva}})  f\big( \V{z}_{m,n}^{(j)} \big|\ist \V{p}_n, \V{p}_{s,\text{mva}}^{(j)} \big)}{ \mu_{\mathrm{fp}} \ist f_{\mathrm{fp}}\big( \V{z}_{m,n}^{(j)} \big)} \ist, 
			& \!\!\rmv \underline{a}^{(j)}_{ss,n} \!=\rmv m \rmv\in\rmv \Set{M}_n^{(j)}\\[3.5mm]
			1 \!-\rmv 	p^{(j)}_{\mathrm{d}}(\V{p}_n,\V{p}^{(j)}_{s,\text{mva}}) \ist, & \!\!\rmv \underline{a}^{(j)}_{ss,n} \!=\rmv 0   
		\end{cases} \label{eq:factorqS}
	\end{align}
	and $\underline{q}_{\mathrm{S}}\big( \underline{\V{p}}^{(j)}_{s,\mathrm{mva}}\rmv, \underline{r}^{(j)}_{s,n} \rmv=\rmv 0, \underline{a}^{(j)}_{ss,n},  \V{x}_{n} ; \V{z}^{(j)}_{n} \big)   \rmv\triangleq\rmv {\delta_{\underline{a}^{(j)}_{ss,n}}}$ as well as for $(s,s') \in \mathcal{D}^{(j)}_{\text{D},n}$ by
	\begin{align}
		&\underline{q}_{\mathrm{D}}\big( \underline{\V{p}}^{(j)}_{s,\mathrm{mva}}\rmv, \underline{r}^{(j)}_{s,n}\rmv=\rmv1\rmv,\underline{\V{p}}^{(j)}_{s',\mathrm{mva}}\rmv, \underline{r}^{(j)}_{s',n}\rmv=\rmv1\rmv,  \underline{a}^{(j)}_{ss',n},  \V{x}_{n} ; \V{z}^{(j)}_{n} \big)\nn\\
		&\hspace{0mm}\triangleq \begin{cases}
			\displaystyle \ist \frac{ p^{(j)}_{\mathrm{d}} (\V{p}_n,\V{p}^{(j)}_{s,\text{mva}},\V{p}^{(j)}_{s',\text{mva}})}{\mu_{\mathrm{fp}}} \\[3mm] 
			\hspace{4mm}\times \frac{ f( \V{z}_{m,n}^{(j)} |\ist \V{p}_n, \underline{\V{p}}^{(j)}_{s,\mathrm{mva}},  \underline{\V{p}}^{(j)}_{s',\mathrm{mva}})}{ f_{\mathrm{fp}}( \V{z}_{m,n}^{(j)} )} 
			\ist, \iist
			& \!\!\rmv a^{(j)}_{ss'} \!=\rmv m \rmv\in\rmv \Set{M}_n^{(j)} \\[3.5mm]
			1 \!-\rmv p^{(j)}_{\mathrm{d}} (\V{p}_n,\V{p}^{(j)}_{s,\text{mva}},\V{p}^{(j)}_{s',\text{mva}})  \ist, \iist & \!\!\rmv a^{(j)}_{ss'} \!=\rmv 0    
		\end{cases} \label{eq:factorqD}\\[-5mm]\nn
	\end{align}
	and $\underline{q}_{\mathrm{D}}\big( \underline{\V{p}}^{(j)}_{s,\mathrm{mva}}\rmv, \underline{r}^{(j)}_{s,n},\underline{\V{p}}^{(j)}_{s',\mathrm{mva}}\rmv, \underline{r}^{(j)}_{s',n},  \underline{a}^{(j)}_{ss',n},  \V{x}_{n} ; \V{z}^{(j)}_{n} \big)   \rmv\triangleq\rmv \delta_{\underline{a}^{(j)}_{ss',n}}$ if any $\underline{r}^{(j)}_{s'',n} \rmv=\rmv 0$ for $s'' \in \{s,s'\}$.
	
	The \textit{pseudo likelihood functions} related to new \acp{pmva} $\overline{q}_{\mathrm{S}}\big(  \overline{\V{y}}^{(j)}_{m,\mathrm{mva}},\rmv \overline{a}^{(j)}_{m,n},\rmv \V{x}_{n} ;\rmv \V{z}^{(j)}_{n} \big) \rmv\rmv=\rmv\rmv \overline{q}_{\mathrm{S}}\big(  \overline{\V{p}}^{(j)}_{m,\mathrm{mva}},\rmv \overline{r}^{(j)}_{s,n},\rmv \overline{a}^{(j)}_{m,n},\rmv \V{x}_{n};\rmv \V{z}^{(j)}_{n} \big)$ is given by
	\begin{align}
		&\overline{q}_{\mathrm{S}}\big(  \overline{\V{p}}^{(j)}_{m,\mathrm{mva}}, \overline{r}^{(j)}_{s,n} \rmv=\rmv1, \overline{a}^{(j)}_{m,n}, \V{x}_{n} ; \V{z}^{(j)}_{n} \big) \nn\\[1.5mm]
		&\hspace{0mm} \triangleq \begin{cases}
			0    \ist, 
			& \hspace{-1mm} \overline{a}^{(j)}_{m,n}  \rmv\rmv\in\rmv \tilde{\Set{D}}_{n}^{(j)} \\[1mm] %
			\frac{ \mu_{\mathrm{n}}f_{\mathrm{n}}(\overline{\V{p}}^{(j)}_{m,\mathrm{mva}}\ist | \ist\V{p}_{n}) f(\V{z}_{m,n}^{(j)}\ist | \V{p}_{n}, \overline{\V{p}}^{(j)}_{m,\mathrm{mva}})}{\mu_{\mathrm{fp}}  f_{\mathrm{fp}}(\V{z}_{m,n}^{(j)})} \ist,  & \hspace{-1mm} \overline{a}^{(j)}_{m,n} \rmv=\rmv 0  
		\end{cases}  \label{eq:factorvNewPMVAs}\\[-6mm]\nn
	\end{align}
	and $\overline{q}_{\mathrm{S}}\big(  \overline{\V{p}}^{(j)}_{m,\mathrm{mva}}, \overline{r}^{(j)}_{s,n} \rmv=\rmv 0,  \overline{a}^{(j)}_{m,n}, \V{x}_{n} ; z^{(j)}_{m,n} \big) \rmv\triangleq\rmv 	f_\text{d}\big(\overline{\V{p}}^{(j)}_{m,\mathrm{mva}}\big)$, respectively. Note that the first line in \eqref{eq:factorvNewPMVAs} is zero because per definition, the new \ac{pmva} with index $m$ exists ($\overline{r}_{m,n} \rmv=\rmv 1$) if and only if the measurement is not associated to a legacy PMVA. For $\overline{a}^{(j)}_{m,n} = 0$, for each measurement $z^{(j)}_{m,n}$ a new \ac{pmva} is introduced, implying that each measurement $\V{z}_{m,n}^{(j)}$ is assumed to originate from a single-bounce path corresponding to exactly one \ac{pmva} $s \in \Set{M}_{n}^{(j)}$ (not a pair of \acp{pmva}).
	
	Finally, the binary \textit{check functions} that validates consistency for any pair $(\rv{\underline{a}}^{(j)}_{ss',n},\overline{\rv{a}}^{(j)}_{m,n})$ of \ac{pmva}-oriented and measurement-oriented association variable at time $n$, \vspace{-.4mm} read
	\begin{align}
		&\Psi(\underline{a}^{(j)}_{ss',n},\overline{a}^{(j)}_{m,n}) %
		\hspace{2mm}\triangleq \begin{cases} 
			0, & \begin{minipage}{5cm}
					$\underline{a}^{(j)}_{ss',n} =  m, \overline{a}^{(j)}_{m,n} \neq (s,s') \text{ or }\\ \underline{a}^{(j)}_{ss',n} \neq  m, \overline{a}^{(j)}_{m,n} = (s,s')$
			\end{minipage} \\[.1mm]
			1 \ist, & \text{otherwise}. 
		\end{cases} \nn \\[-6mm]
		\nn
	\end{align}	
	In case the joint \ac{pmva}-oriented association vector $\underline{\RV{a}}_n$ and the measurement-oriented association vector $\overline{\RV{a}}_n$ do not describe the same association event, at least one check function in \eqref{eq:factorGraph} is zero. Thus $f( \V{y}_{0:n}, \V{x}_{0:n}, \underline{\V{a}}_{1:n},\overline{\V{a}}_{1:n} \ist | \ist \V{z}_{1:n})$ is zero as well. The factor graph representing factorization \eqref{eq:factorGraph} is shown in Fig.~\ref{fig:factorgraph}. {A detailed derivation of related factorization structures developed in the context of multitarget tracking and their relationship to the TOMB/P filter is presented in \cite{MeyKroWilLauHlaBraWin:J18}. The problem and resulting factorization structure considered in this paper are more complicated than multisensor multitarget tracking because (i) there is also an unknown agent state and (ii) measurements can originate from two different types of propagation paths.}
	\vspace*{-2mm}
	
	\section{Proposed Sum-Product Algorithm}\label{sec:SPAalgorithm}
	
	Since our factor graph in Fig.~\ref{fig:factorgraph} has cycles, we have to decide on a specific order of message computation \cite{KscFreLoe:TIT2001,Loe:SMP2004_FG}. We choose the order according to the following rules \cite{LeiMeyHlaWitTufWin:J19,MenMeyBauWin:J19,MeyWil:J21,LiLeiVenTuf:TWC2022,MeyBraWilHla:J17,MeyKroWilLauHlaBraWin:J18,MeyHliHla:TSPIN2016}: (i) messages are only sent forward in time; (ii) messages are only sent from \ac{pa} $j\rmv-\rmv 1$ to \ac{pa} $j$, i.e., the measurements of \ac{pa} are processed serial, thus, \ac{pa} $j\rmv-\rmv 1$ establishes new \acp{pmva} that are acting as legacy \acp{pmva} for \ac{pa} $j$; 
	(iii) \emph{iterative} message passing is only performed for data association \cite{LeiMeyHlaWitTufWin:J19, LeiGreWit:ICC2019}, i.e., in particular, for the loops connecting different \acp{pmva}, we only perform a single message passing iteration; and (iv) along an edge connecting an agent state variable node and a new \ac{pmva} state variable node, messages are only sent from the former to the latter. 
	With these rules, the message passing equations of the SPA \cite{KscFreLoe:TIT2001} yield the following operations at each time step. The corresponding messages are shown in Fig.~\ref{fig:factorgraph}. Note that this message passing order has been developed for real-time processing. Sending messages also backward in time, referred to as  ``smoothing,'' will improve post-processing performance but lead to increased computational complexity. 
	
	We note that similarly to the ``dummy \acp{pdf}'' introduced in Section~\ref{sec:systemModel}, we consider messages $\varphi\big(\V{y}_{s,n}\big) \rmv\rmv=\rmv\rmv \varphi\big(\V{p}_{s,\mathrm{mva}}\ist, r_{s,n}\big)$ for non-existing \ac{pmva} states, i.e., for $r_{s,n} \!=\rmv 0$. We define these messages by $\varphi\big( \V{p}_{s,\mathrm{mva}} \ist, 0\big) \triangleq \varphi^{(j)}_{k,n}$ (note that these messages are not \acp{pdf} and thus are not required to integrate to $1$). To keep the notation concise, we also define the sets $\Set{M}_{0,n}^{(j)} \triangleq \Set{M}_n^{(j)} \cup \{0\}$ and $ \tilde{\Set{D}}^{(j)}_{0,n} \in \tilde{\Set{D}}^{(j)}_{n} \cup \{0\}$.
		
	\subsection{Prediction Step}
	First, a \textit{prediction step} is performed for the agent and all legacy \acp{pmva} $s \in S_{n-1}$. Based on the \ac{spa} rule, the prediction message for the agent state is given by
	\vspace*{-2mm}
	\begin{align}
		\alpha( \V{x}_n) =\rmv \int \rmv f(\V{x}_n|\V{x}_{n-1}) \ist \tilde{f}(\V{x}_{n-1}) \ist \mathrm{d}\V{x}_{n-1} 
		\label{eq:stateTransitionMessageAgent}\\[-7mm]\nn
	\end{align}
	and the prediction message for the legacy \acp{pmva} is given by\vspace*{-2mm}
	\begin{align}
		\hspace*{-0.5mm}\alpha_s\big( \underline{\V{p}}_{s,\text{mva}} \ist, \underline{r}_{s,n}\big) &=\rmv\rmv\rmv\rmv   \sum_{r_{k,n-1} \in \{0,1\}}\rmv\int \rmv\rmv f\big(\underline{\V{p}}_{s,\text{mva}}\ist, \underline{r}_{s,n} \big| \V{p}_{s,\text{mva}}\ist, r_{s,n-1} \big) \nn \\
		&\hspace{5mm}\times\tilde{f}\big(\V{p}_{s,\text{mva}}\ist, r_{s,n-1} \big) \ist \mathrm{d}\V{p}_{s,\text{mva}}
		\label{eq:stateTransitionMessageFeature} \\[-7mm]\nn
	\end{align}
	$s \rmv\rmv\in\rmv\rmv \{1,\dots,S_{n-1}\}$, where the beliefs of the agent state, $\tilde{f}(\V{x}_{n-1})$, and of the \ac{pmva} states, $\tilde{f}\big(\V{p}_{s,\text{mva}}\ist, r_{s,n-1} \big)$, were calculated at the preceding time $n \rmv-\rmv 1$. Inserting \eqref{eq:stmpmvarone} and \eqref{eq:stmpmvarzero} for $f\big(\underline{\V{p}}_{s,\text{mva}}\ist, \underline{r}_{s,n} \big| \V{p}_{s,\text{mva}} \ist, r_{s,n-1}\rmv\rmv=\rmv\rmv 1\big)$ and $f\big(\underline{\V{p}}_{s,\text{mva}}\ist, \underline{r}_{s,n} \big|\V{p}_{s,\text{mva}}\ist,$\linebreak$ r_{s,n-1}\rmv=\rmv 0\big)$, respectively, we obtain\vspace*{-2mm}
	\begin{align}
		\alpha_s\big(\underline{\V{p}}_{s,\text{mva}}\ist, 1\big) 
		=\ist p_\text{s} \rmv \int \rmv \delta\big(\underline{\V{p}}_{s,\text{mva}} - \V{p}_{s,\text{mva}}\big) \ist \tilde{f}_s\big(\V{p}_{s,\text{mva}}\ist, 1 \big) \ist \mathrm{d}\V{p}_{s,\text{mva}} 
		\label{eq:stateTransitionMessageFeature_r1}\\[-7mm]\nn
	\end{align}
	and $\alpha_s\big(\underline{\V{p}}_{s,\text{mva}}\ist, 0\big)  = \alpha_{s,n} f_\text{d}\big( \V{p}_{s,\mathrm{mva}} \big)$ with\vspace*{-2mm}
	\begin{align}
		\alpha_{s,n} = (1 \!-\rmv p_\text{s}\big) \rmv\int\rmv \tilde{f}_s\big(\V{p}_{s,\text{mva}}\ist, 1\big) \ist \mathrm{d} \V{p}_{s,\text{mva}} +\ist 	\tilde{f}_{s,n-1}\ist.
		\label{eq:stateTransitionMessageFeature_r0}\\[-7mm]\nn
	\end{align}
	We note that $\tilde{f}_{s,n-1} \triangleq \int \rmv \tilde{f}_s\big( \V{p}_{s,\text{mva}}\ist, 0\big) \ist \mathrm{d}\V{p}_{s,\text{mva}}$ approximates the probability of non-existence of legacy \ac{pmva} $s$ at the previous time step $n\rmv-\rmv1$. 
	
	\vspace*{-2mm}
	\subsection{Sequential \ac{pa} Update}
	
	At iteration $j \in \{1,\dots,J\}$, the following operations are calculated for all legacy and new \acp{pmva}. 
	
	\subsubsection{Transition of New and Legacy \ac{pmva} States Between \acp{pa}} For $j = 1$, the number of legacy \acp{pmva} is $S^{(1)}_{n} = S_{n-1}$ with the corresponding state $\underline{\V{y}}^{(1)}_n \rmv\triangleq\rmv \big[ \underline{\V{y}}^{\T}_{1,n} \rmv\cdots\ist \underline{\V{y}}^{T}_{S_{n\rmv-\rmv1},n} \big]^{\T}\rmv\rmv$. Furthermore, the state $\overline{\V{y}}^{(j-1)}$ is empty and the prediction message of legacy \acp{pmva} is $\alpha_s\big( \underline{\V{p}}^{(j)}_{s,\text{mva}} \ist, \underline{r}_{s,n}\big) \triangleq \alpha_s\big( \underline{\V{p}}_{s,\text{mva}} \ist, \underline{r}_{s,n}\big)$ as well as $\alpha^{(j)}_{s,n} \triangleq \alpha_{s,n}$. For $j > 1$, we have $S^{(j)}_{n} = S^{(j-1)}_{n} + M^{(j-1)}_{n}$ legacy \acp{pmva} with states $\underline{\V{y}}^{(j)}_{n} = \big[\underline{\V{y}}^{(j-1) \T}_{n} \ist\ist\ist \overline{\V{y}}^{(j-1) \T}_{n} \big]^{\T}$ and $M^{(j)}_{n}$ new \acp{pmva} with states $\overline{\V{y}}^{(j)}$. For $1 \leq s \leq S^{(j-1)}_{n}$, using \eqref{eq:stmpmvaonepas}, \eqref{eq:stmpmvazeropas}, and \eqref{eq:stateTransitionMessageFeature} the prediction message of former legacy \acp{pmva} is given by
	\vspace*{-1mm}
	\begin{equation}
		\alpha_s\big(\underline{\V{p}}^{(j)}_{s,\text{mva}}\ist, 1\big) 
		=\rmv\rmv \int \rmv\rmv \delta\big(\underline{\V{p}}^{(j)}_{s,\text{mva}} - \V{p}^{(j-1)}_{s,\text{mva}}\big) \ist \gamma\big(\underline{\V{p}}^{(j-1)}_{s,\mathrm{mva}},1\big) \ist \mathrm{d}\V{p}^{(j-1)}_{s,\text{mva}}
		\vspace{1.2mm}
		\label{eq:statePATransitionMessageFeature_r1}
	\end{equation}
	and $\alpha_s\big(\underline{\V{p}}^{(j)}_{s,\text{mva}}\ist, 0\big)  = \alpha^{(j)}_{s,n} f_\text{d}\big( \underline{\V{p}}^{(j)}_{s,\mathrm{mva}} \big)$ with $\alpha^{(j)}_{s,n} =  \gamma_{s,n}^{(j-1)}$.
	The prediction message of former new \acp{pmva} is $\alpha_s\big( \underline{\V{p}}^{(j)}_{s,\text{mva}} \ist, \underline{r}_{s,n}\big) \triangleq \phi\big(\overline{\V{y}}^{(j)}_{s,n}\big)$ and $\alpha^{(j)}_{s,n} \triangleq \phi^{(j)}_{s,n}$.\\[-1mm]	
	\subsubsection{Checking the Availability of Propagation Paths}\label{sec:visibilityAlgo} 
	
	The proposed SPA algorithm performs an availability check for each propagation path using \textit{{\ac{rt}}} \cite{Bor:JASA1984, MckHam:IEEENetwork1991,LuVitDegFusBarBlaBer:TAP2019} to determine whether a \ac{va} can provide map information. First, the \ac{va} positions $\V{p}^{(j)}_{ss',\text{va}}$ are determined by applying \eqref{eq:nonLinearTransformation} directly to \ac{pa} $j$ at position $\V{p}^{(j)}_{\text{pa}}$ to get single-bounce-related \acp{va} at positions $\V{p}^{(j)}_{ss,\text{va}}$. Next, \eqref{eq:nonLinearTransformation} is also applied to this single-bounce-related \acp{va} to get double-bounce-related \acp{va} at positions $\V{p}^{(j)}_{ss',\text{va}}$. {\Ac{rt} is performed as described in Section~\ref{sec:Pdvisibility}.}
	In case the path between the agent position and a \ac{va} position or between two \ac{va} positions (double-bounce path) intersects with a reflective surface (e.g., it is blocked), the corresponding path is not available at  $\V{p}_n$. Hence, the corresponding \ac{va} cannot be associated with measurements at time $n$.

	\subsubsection{Measurement Evaluation for the \ac{los} Path}\label{sec:betaPAMess} The messages $\beta_{00}\big(\rmv \underline{a}^{(j)}_{00,n} \big)$ passed from the factor node $\rmv\underline{q}_{\mathrm{P}}\rmv\big(\rmv \V{x}_{n}, \underline{a}^{(j)}_{00,n}; \rmv \V{z}^{(j)}_{n} \rmv\big)\rmv\rmv$ to the feature-oriented association variables $ \underline{a}^{(j)}_{00,n}\rmv\rmv$ are calculated as 
	\vspace*{-2mm}
	\begin{align}
		\beta_{00}\big( \underline{a}^{(j)}_{00,n} \big) &\rmv\rmv\rmv=\rmv\rmv\rmv \int \rmv\rmv\rmv \alpha(\V{x}_n) \ist \underline{q}_{\mathrm{P}}\big(  \V{x}_{n}, \underline{a}^{(j)}_{00,n} ; \V{z}^{(j)}_{n} \big) \mathrm{d}\V{x}_{n}\ist. 
		\label{eq:bp_measevalutionPA}
		\\[-7mm]\nn
	\end{align}
	\subsubsection{Measurement Evaluation for Legacy \acp{pmva}}\label{sec:betaMess} For $s=s'$, the messages $\beta_{ss}\big(\underline{a}_{ss,n}^{(j)} \big)$ passed from the factor node $\underline{q}_{\mathrm{S}}\big( \underline{\V{p}}^{(j)}_{s,\mathrm{mva}}\rmv, \underline{r}^{(j)}_{s,n}\rmv,  \underline{a}^{(j)}_{ss,n},  \V{x}_{n} ; \V{z}^{(j)}_{n} \big)$ of single \acp{pmva} to the feature-oriented association variables $a_{ss,n}^{(j)}$ given by\vspace*{-1mm}
	\begin{align}	
		&\beta_{ss}\big(\underline{a}_{ss,n}^{(j)} \big) \nn\\
		&=\int\!\!\!\int \rmv\rmv\rmv \alpha_s\big( \underline{\V{p}}_{s,\text{mva}}^{(j)},1\big) \ist \alpha(\V{x}_n) \underline{q}_{\mathrm{S}}\big( \underline{\V{p}}^{(j)}_{s,\mathrm{mva}}\rmv, 	\underline{r}^{(j)}_{s,n}\rmv,  \underline{a}^{(j)}_{ss,n},  \V{x}_{n} ; \V{z}^{(j)}_{n} \big) \nn \\
		&\rmv\rmv\rmv \hspace{5mm} \times \mathrm{d}\V{x}_{n} \ist \mathrm{d}\underline{\V{p}}_{s,\text{mva}}^{(j)}  \ist+\ist 1\big(\underline{a}_{ss,n}^{(j)}\big)\ist\alpha_{s,n}^{(j)} \ist. 
		\label{eq:bp_measevalutionLPMVA1}
		\\[-7mm]\nn
	\end{align}
	For $s\neq s'$, the message $\beta_{ss'}\big(\underline{a}_{ss',n}^{(j)} \big)$ passed from the factor node $\underline{q}_{\mathrm{D}}\big( \underline{\V{p}}^{(j)}_{s,\mathrm{mva}}\rmv, \underline{r}^{(j)}_{s,n},\underline{\V{p}}^{(j)}_{s',\mathrm{mva}}\rmv, \underline{r}^{(j)}_{s',n},  \underline{a}^{(j)}_{ss',n},  \V{x}_{n} ; \V{z}^{(j)}_{n} \big)$ of pairs of \acp{pmva} to the feature-oriented association variables $a_{ss',n}^{(j)}$ are obtained as
	\begin{align}
		\hspace*{-2mm}\beta_{ss'}\big(\underline{a}_{ss',n}^{(j)} \big) &\rmv\rmv=\rmv\rmv\rmv \int\!\!\!\int\!\!\!\int \rmv\rmv\rmv \alpha_s\big( \underline{\V{p}}_{s,\text{mva}}^{(j)},1\big) \ist\alpha_s\big( \underline{\V{p}}_{s',\text{mva}}^{(j)},1\big)  \alpha(\V{x}_n) \nn \\ 
		&\hspace{-1mm}\times \underline{q}_{\mathrm{D}}\big( \underline{\V{p}}^{(j)}_{s,\mathrm{mva}}\rmv, \underline{r}^{(j)}_{s,n},\underline{\V{p}}^{(j)}_{s',\mathrm{mva}}\rmv, \underline{r}^{(j)}_{s',n},  \underline{a}^{(j)}_{ss',n},  \V{x}_{n} ; \V{z}^{(j)}_{n} \big) \nn \\
		&\hspace{-1mm}\times \mathrm{d}\V{x}_{n} \ist \mathrm{d}\underline{\V{p}}_{s,\text{mva}}^{(j)} \mathrm{d}\underline{\V{p}}_{s',\text{mva}}^{(j)}\rmv\rmv+\rmv\rmv 1\big(\underline{a}_{ss',n}^{(j)}\big)\ist\alpha_{s,n}^{(j)}\alpha_{s',n}^{(j)} \ist. 
		\label{eq:bp_measevalutionLPMVA2}
		\\[-7mm]\nn
	\end{align}

	\subsubsection{Measurement Evaluation for New \acp{pmva}}\label{sec:xiMess} For \ac{pa} $j$, the messages $\xi\big(\overline{a}^{(j)}_{m,n}\big)$ sent from the factor node $\overline{q}_{\mathrm{S}}\big(  \overline{\V{p}}^{(j)}_{m,\mathrm{mva}}, \overline{r}^{(j)}_{k,\mathrm{mva}}, \overline{a}^{(j)}_{m,n}, \V{x}_{n} ; z^{(j)}_{m,n} \big)$ to the variable nodes corresponding to the measurement-oriented  association variables $\overline{a}^{(j)}_{m,n}$ are given by\vspace*{-2mm}
	\begin{align}
		\xi\big(\overline{a}^{(j)}_{m,n}\big) &=\rmv\rmv \sum_{\overline{r}^{(j)}_{m,n} \in \{0,1\}} \int\!\!\!\int \rmv  \overline{q}_{\mathrm{S}}\big(  \overline{\V{p}}^{(j)}_{m,\mathrm{mva}}, \overline{r}^{(j)}_{k,\mathrm{mva}}, \overline{a}^{(j)}_{m,n}, \V{x}_{n} ; z^{(j)}_{m,n} \big) \nn \\
		&\hspace{5mm}\times  \alpha(\V{x}_n) \, \mathrm{d}\V{x}_{n} \ist \mathrm{d}\overline{\V{p}}^{(j)}_{m,\mathrm{mva}} \ist.
		\label{eq:bp_measevalutionNF1}\\[-6.5mm]
		\nn
	\end{align}
	Using the expression of $ \overline{q}_{\mathrm{S}}\big(  \overline{\V{p}}^{(j)}_{m,\mathrm{mva}}, \overline{r}^{(j)}_{k,\mathrm{mva}}, \overline{a}^{(j)}_{m,n}, \V{x}_{n} ; z^{(j)}_{m,n} \big)$ in Section \ref{sec:FactorGraph} in \eqref{eq:factorvNewPMVAs}, Eq.\ \eqref{eq:bp_measevalutionNF1} simplifies to $\xi\big(\overline{a}^{(j)}_{m,n}\big) \!=\! 1$
	for $\overline{a}^{(j)}_{m,n} \!\in\! \Set{D}^{(j)}_{n}$, and for $\overline{a}^{(j)}_{m,n} \!=\rmv 0$ it becomes
	\begin{align}
		\xi\big(\overline{a}^{(j)}_{m,n}\big) &= 1 + \frac{ \mu_{\text{n}} }{\mu_\text{fp} \ist f_{\text{fp}}\big( \V{z}^{(j)}_{m,n} \big)} 
		\int\!\!\!\int\rmv \alpha(\V{x}_n) \ist f_{\text{n}}\big(\overline{\V{p}}^{(j)}_{m,\mathrm{mva}}\big| \V{x}_n \big) \nn \\[1.5mm]
		&\hspace{5mm}\times  f\big( \V{z}^{(j)}_{m,n} \big|\V{x}_{n}, \overline{\V{p}}^{(j)}_{m,\mathrm{mva}} \big) \, \mathrm{d}\V{x}_{n} \ist \mathrm{d}\overline{\V{p}}^{(j)}_{m,\mathrm{mva}}\ist.
		\label{eq:bp_measevalutionNF}
	\end{align}

	\subsubsection{Iterative Data Association}\label{sec:DAmain} Next, from $\beta\big(\underline{a}_{ss',n}^{(j)} \big)$ and $\xi\big(\overline{a}^{(j)}_{m,n}\big)$, messages 
	$\eta\big(\underline{a}_{ss',n}^{(j)} \big)$ and $\varsigma\big( \overline{a}_{m,n}^{(j)} \big)$ are obtained using loopy (iterative) BP.
	First, for each measurement, $m \!\in\! \Set{M}_n^{(j)}$, messages $\nu_{m\rightarrow s}^{(p)}\big(\underline{a}_{ss',n}^{(j)}\big)$ 
	and, then, for $ss' \rmv\in\rmv \tilde{\mathcal{D}}_{n}^{(j)}$ messages $\zeta_{s \rightarrow m}^{(p)}\big(\overline{a}_{m,n}^{(j)}\big)$ are calculated iteratively according to \cite{WilLau:J14,MeyKroWilLauHlaBraWin:J18}, for each iteration index $p \in \{ 1,\ldots, P\}$. After the last iteration $p \rmv=\rmv P\rmv$, the messages $\eta\big(\underline{a}_{ss',n}^{(j)} \big)$ and $\varsigma\big( \overline{a}_{m,n}^{(j)} \big)$ are calculated according to \cite{WilLau:J14,MeyKroWilLauHlaBraWin:J18}. Details are provided in the supplementary material{\mref{Section}{sec:DA}}.

	\subsubsection{Measurement Update for the Agent} The message $\gamma^{(j)}_{00}(\V{x}_n)$ sent from the factor node $\underline{q}_{\mathrm{P}}\big( \V{x}_{n},  \underline{a}^{(j)}_{00,n}  ; \V{z}^{(j)}_{n} \big)$ to the agent variable node is computed as \vspace*{-1mm}
	\begin{align}
		\gamma^{(j)}_{00}(\V{x}_n) &=\! \sum_{\underline{a}^{(j)}_{00,n} \in \Set{M}_{0,n}^{(j)}} \!\! \eta\big( \underline{a}^{(j)}_{00,n} \big) 
		\underline{q}_{\mathrm{P}}\big( \V{x}_{n},  \underline{a}^{(j)}_{00,n}  ; \V{z}^{(j)}_{n} \big)\ist.
		\label{eq:measurementUpdateAgentPA}\\[-6.5mm]
		\nn
	\end{align}
	 For $s = s'$, the messages $\gamma^{(j)}_{ss}(\V{x}_n)$ sent from the factor node $\underline{q}_{\mathrm{S}}\big( \underline{\V{p}}^{(j)}_{s,\mathrm{mva}}\rmv, 1,  \underline{a}^{(j)}_{ss,n},  \V{x}_{n} ; \V{z}^{(j)}_{n} \big)$ to the agent variable node are given by
	\begin{align}
		\gamma^{(j)}_{ss}(\V{x}_n) &=\hspace{-4mm} \sum_{a_{ss,n}^{(j)} \in \Set{M}_{0,n}^{(j)}} \!\!  \eta\big( \underline{a}_{ss,n}^{(j)} \big) 
		\rmv\int \!\underline{q}_{\mathrm{S}}\big( \underline{\V{p}}^{(j)}_{s,\mathrm{mva}}\rmv, 1,  \underline{a}^{(j)}_{ss,n},  \V{x}_{n} ; \V{z}^{(j)}_{n} \big) \nn \\
		&\hspace{2mm} \times \alpha_s\big(\underline{\V{p}}_{s,\text{mva}}^{(j)} , 1\big) \mathrm{d} \underline{\V{p}}_{s,\text{mva}}^{(j)} \ist+\ist \eta\big( \underline{a}_{ss,n}^{(j)} \!\rmv=\! 0\big) \ist\alpha_{s,n}^{(j)}.
		\label{eq:measurementUpdateAgent1}
	\end{align}
	Furthermore, for $s \neq s'$, the messages passed from the factor node $\underline{q}_{\mathrm{D}}\big( \underline{\V{p}}^{(j)}_{s,\mathrm{mva}}\rmv, 1,\underline{\V{p}}^{(j)}_{s',\mathrm{mva}}\rmv, 1,$ $\underline{a}^{(j)}_{ss',n},  \V{x}_{n} ; \V{z}^{(j)}_{n} \big)$ to the agent variable node are obtained as\vspace*{-1mm}
		\begin{align}
		\hspace{-2mm}\gamma^{(j)}_{ss'}(\V{x}_n) &=\!\!\!\! \sum_{\underline{a}_{ss',n}^{(j)} \in \Set{M}_{0,n}^{(j)}} \!\!\!\!\! \eta\big( \underline{a}_{ss',n}^{(j)} \big) \int\!\!\!\!\int\!\! \alpha_s\big(\underline{\V{p}}_{s,\text{mva}}^{(j)} , 1\big) \alpha_{s'}\big(\underline{\V{p}}_{s',\text{mva}}^{(j)} , 1\big) \nn \\
		&\hspace{2mm}\times \rmv \!\underline{q}_{\mathrm{D}}\big( \underline{\V{p}}^{(j)}_{s,\mathrm{mva}}\rmv, 1,\underline{\V{p}}^{(j)}_{s',\mathrm{mva}}\rmv, 1,  \underline{a}^{(j)}_{ss',n},  \V{x}_{n} ; \V{z}^{(j)}_{n} \big) \nn \\
		&\hspace{2mm}\times \mathrm{d} \underline{\V{p}}_{s,\text{mva}}^{(j)}\mathrm{d} \underline{\V{p}}_{s',\text{mva}}^{(j)} \ist+\ist \eta\big( \underline{a}_{ss',n}^{(j)} \!\rmv=\! 0\big) \ist\alpha_{s,n}^{(j)} \ist\alpha_{s',n}^{(j)}\ist. 
		\label{eq:measurementUpdateAgent1} \\[-5mm]
		\nn
	\end{align}
	\subsubsection{Measurement Update for Legacy \acp{pmva}} Similarly, for $s = s'$, the messages  $\rho_{ss'}\big(\underline{\V{y}}^{(j)}_{s}\big)$ sent to the legacy \acp{pmva} variable nodes are given by\vspace*{-1mm}
	\begin{align}
		\rho_{ss}\big(\underline{\V{p}}^{(j)}_{s,\mathrm{mva}},1\big)  &=\rmv\rmv\rmv \sum_{\underline{a}_{ss,n}^{(j)} \in \Set{M}_{0,n}^{(j)}} \rmv\rmv\int\!\! \eta\big( \underline{a}_{ss,n}^{(j)} \big) \alpha(\V{x}_n)\nn \\
		&\hspace{2mm}\times \!\underline{q}_{\mathrm{S}}\big( \underline{\V{p}}^{(j)}_{s,\mathrm{mva}}\rmv, 1,  \underline{a}^{(j)}_{ss,n},  \V{x}_{n} ; \V{z}^{(j)}_{n} \big)  \mathrm{d} \V{x}_n\\[1.5mm]
		\rho_{ss}^{(j)} &\triangleq \rho_{ss}\big(\underline{\V{p}}^{(j)}_{s,\mathrm{mva}},0\big) = \eta\big( \underline{a}_{ss,n}^{(j)} \!\rmv=\! 0\big)  \\[-6mm]
		\nn
	\end{align}
	and for $s\neq s'$ by
	\vspace*{-2mm}
	\begin{align}
		\rho_{ss'}\big(\underline{\V{p}}^{(j)}_{s,\mathrm{mva}},1\big)  &=\rmv\rmv\rmv\rmv\rmv\rmv\rmv\rmv\rmv\rmv\rmv\rmv\rmv \sum_{\underline{a}_{ss',n}^{(j)} \in \Set{M}_{0,n}^{(j)}} \rmv\int\rmv\rmv\rmv\rmv\rmv\rmv\rmv\rmv\rmv\int\rmv\rmv \eta\big( \underline{a}_{ss',n}^{(j)} \big)  \alpha(\V{x}_n) \alpha_{s'}\big(\underline{\V{p}}_{s',\text{mva}}^{(j)} , 1\big) \nn \\
		&\hspace{0mm}\times  \!\underline{q}_{\mathrm{D}}\big( \underline{\V{p}}^{(j)}_{s,\mathrm{mva}}\rmv, 1,\underline{\V{p}}^{(j)}_{s',\mathrm{mva}}\rmv, 1,  \underline{a}^{(j)}_{ss',n},  \V{x}_{n} ; \V{z}^{(j)}_{n} \big) \nn \\
		&\hspace{0mm}\times \mathrm{d}\V{x}_n \mathrm{d} \underline{\V{p}}_{s',\text{mva}}^{(j)}\\[1.5mm]
		\rho_{ss'}^{(j)} &\triangleq \rho_{ss'}\big(\underline{\V{p}}^{(j)}_{s,\mathrm{mva}},0\big) = \eta\big( \underline{a}_{ss',n}^{(j)} \!\rmv=\! 0\big) \ist.\\[-7mm]\nn
	\end{align}
	Based on these messages, the message sent to the next \ac{pa} $\gamma\big(\underline{\V{y}}_{s}^{(j)}\big) \triangleq \gamma\big(\underline{\V{p}}_{s,\text{mva}}^{(j)}, \underline{r}_{k,n}^{(j)}\big)$ is computed as
	\vspace*{-3mm}
	\begin{align}
		\label{eq:measurementUpdateLegacy1}
		\hspace{-2mm}\gamma\big(\underline{\V{p}}^{(j)}_{s,\mathrm{mva}},1\big)  &=\! \rho_{ss}\big(\underline{\V{p}}^{(j)}_{s,\mathrm{mva}},1\big) \rmv\rmv+\rmv\rmv \prod^{S_n^{(j)}}_{s' = 1, s\neq s'} \rho_{ss'}\big(\underline{\V{p}}^{(j)}_{s,\mathrm{mva}},1\big)\\[-1mm]
		\gamma_{s,n}^{(j)} &\triangleq\ist \gamma\big(\underline{\V{p}}^{(j)}_{s,\mathrm{mva}},0\big) = \rho_{ss}^{(j)} + \prod^{S_n^{(j)}}_{s' = 1, s\neq s'} \rho_{ss'}^{(j)} \ist.
		\label{eq:measurementUpdateLegacy2}\\[-7mm]\nn
	\end{align}

	\subsubsection{Measurement Update for New \acp{pmva}} Finally, the messages $\phi\big(\overline{\V{y}}_{m}^{(j)}\big) \triangleq \phi\big(\overline{\V{p}}_{m,\text{mva}}^{(j)},\overline{r}_{m,n}^{(j)}\big)$ sent to the new \ac{pmva} variable nodes are obtained as
		\begin{align}
			\label{eq:measurementUpdateNew1}
			\phi\big(\overline{\V{p}}_{m,\text{mva}}^{(j)} , 1\big) &=\ist\int\! \overline{q}_{\mathrm{S}}\big(  \overline{\V{p}}^{(j)}_{m,\mathrm{mva}}, 1, \overline{a}^{(j)}_{m,n}, \V{x}_{n} ; z^{(j)}_{m,n} \big)  \, \nn \\
			&\hspace{5mm}\times
			\ist \alpha(\V{x}_n) \mathrm{d}\V{x}_n \varsigma\big( \overline{a}_{m,n}^{(j)} \!\rmv=\! 0\big) \\[.5mm] 
			\phi_{m,n}^{(j)} &\triangleq\ist \phi\big(\overline{\V{p}}_{m,\text{mva}}^{(j)},0\big) =\rmv \sum_{\overline{a}_{m,n}^{(j)} \in  \tilde{\Set{D}}^{(j)}_{0,n}} \!\! \varsigma\big( \overline{a}_{m,n}^{(j)}\big) \ist.
			\label{eq:measurementUpdateNew2}\\[-10mm]\nn
		\end{align}

	\subsection{Belief Calculation} \label{sec:beliefCalc} After the messages for all \acp{pa}, $j \in \{1,\dots,J\}$ are computed, the belief $\tilde{f}(\V{x}_n)$ of the agent state can be calculated as normalized production of all incoming messages \cite{KscFreLoe:TIT2001}, i.e.,
	\begin{align}
		\tilde{f}(\V{x}_n) \ist=\ist C_n\alpha( \V{x}_n)  \prod_{(s,s') \in \tilde{\Set{D}}_{n}^{(J)}}\!\!\rmv\gamma^{(J)}_{ss'}(\V{x}_n)
		\label{eq:f_x}\\[-7mm]\nn
	\end{align}
	with normalization constant $C_n = \big(\int \alpha( \V{x}_n) \prod_{(s,s') \in \tilde{\Set{D}}_{n}^{(J)}}$ $\gamma^{(J)}_{ss'}( \V{x}_n) \mathrm{d}\V{x}_n\big)^{-1}$ that guarantees that \eqref{eq:f_x} is a valid probability distribution. Similarly, the beliefs $ \tilde{f}_s\big(\underline{\V{y}}^{(J)}_{s,n}\big) = \tilde{f}_s\big(\underline{\V{p}}^{(J)}_{s,\mathrm{mva}}, \underline{r}_{s,n}^{(J)}\big)\vspace*{0.5mm}$ of legacy \ac{pmva} $s \in S_n^{J}$, is given by 
	\vspace*{-.5mm}
	\begin{align}
		\tilde{f}_s(\underline{\V{y}}^{(J)}_{s,n}) \ist=\ist \underline{C}_{s,n}\gamma\big(\underline{\V{y}}^{(J)}_{s}\big)
		\label{eq:fs_legacy} \\[-6mm]\nn
	\end{align}
	with constant $\underline{C}_{s,n} \rmv\rmv=\rmv\rmv \big(\int \gamma\big(\underline{\V{p}}^{(J)}_{s,\mathrm{mva}}, \underline{r}_{s,n}^{(J)} \rmv= \rmv1\big) \mathrm{d}\underline{\V{p}}^{(J)}_{s,\mathrm{mva}}\rmv\rmv+ \gamma_{s,n}^{(J)}\big)^{-1}$. Similarly, the $ \tilde{f}_m\big(\overline{\V{y}}^{(J)}_{m,n}\big) = \tilde{f}_m\big(\overline{\V{p}}^{(J)}_{m,\mathrm{mva}}, \overline{r}_{m,n}^{(J)}\big) $ of new \ac{pmva} $m \in M_n^{J}$, is obtained as
		\begin{align}
		\tilde{f}_m\big(\overline{\V{y}}^{(J)}_{m,n}\big) \ist=\ist \overline{C}_{m,n} \phi\big(\overline{\V{y}}^{(J)}_{m}\big)
		\label{eq:fs_new}\\[-5mm]\nn
	\end{align}
	where $\overline{C}_{m,n} \rmv\rmv\rmv\rmv = \rmv \rmv\rmv \big(\int \phi\big(\overline{\V{y}}^{(J)}_{m}\big) \mathrm{d}\overline{\V{y}}^{(J)}_{m} \rmv +\rmv  \phi^{(J)}_{m,n}\big)^{-1}$ is again a constant.
	
	A computationally feasible sequential particle-based message passing implementation can be obtained following \cite{MeyHliHla:TSPIN2016,MeyBraWilHla:J17,LeiMeyHlaWitTufWin:J19}. In particular, we adopted the approach in \cite{MeyHliHla:TSPIN2016} using the ``stacked state'' comprising the agent state and the \ac{pmva} states. To avoid that the number of \ac{pmva} states grows indefinitely, \acp{pmva} states with $p(r^{(j)}_{s,n} \!=\! 1 |\V{z}_{1:n})$ below a threshold $p_{\text{pr}}$ are removed from the state space (``pruned'') after processing the measurements of each \ac{pa} $j$. {To limit computation complexity, one might limit the maximum number of \ac{pmva} states, i.e., in case $S_n > S_{\mathrm{max}}$, where $S_{\mathrm{max}}$ is another predefined threshold, only the $S_{\mathrm{max}}$ \ac{pmva} states with the highest existence probability are considered when the measurement of the next \ac{pa} is processed.} {Pseudocode for the particle-based implementation is provided in the supplementary material{\mref{Section}{sec:pdeudocode}}.}	
	
	\subsection{Implementation Aspects and Computational Complexity } \label{sec:ImpNewMVAs}
	 
	When beliefs of new PMVAs (cf.~\eqref{eq:fs_new}, \eqref{eq:measurementUpdateNew1}, \eqref{eq:measurementUpdateNew2}, and \eqref{eq:factorvNewPMVAs}) are introduced, contrary to \cite{MeyBraWilHla:J17,LeiMeyHlaWitTufWin:J19},  we do not use the conditional prior \ac{pdf} of newly detected MVAs, $f_{\mathrm{n}}\big(\V{p}^{(j)}_{m,\mathrm{mva}}|\V{p}_n\big)$, as a proposal \ac{pdf}. We develop an alternative proposal \ac{pdf} where first samples of $\V{p}^{(j)}_{ss,\mathrm{va}}$ are obtained by using the inverse transformation of \eqref{eq:VADistmeas} and \eqref{eq:VAAnglemeas}. Based on the resulting samples of $\V{p}^{(j)}_{ss,\mathrm{va}}$, samples of $\V{p}^{(j)}_{m,\mathrm{mva}}$ can then be computed by exploiting the inverse transformation in \eqref{eq:nonLinearTransformationMVA}.	
	
	New \ac{pmva} states are introduced based on the assumption that the measurements come from a single-bounce path. However, if a measurement from a double-bounce path is used, the corresponding \ac{pmva} will be pruned after a few time steps. This is because, due to the assumption that the measurement is from a single bounce path, the spatial distribution of the corresponding new \ac{pmva} has high probability mass at incorrect locations. As a result, the probability of its existence will vanish. Dropping this assumption would require the introduction of new \acp{pmva}-pairs for each measurement, significantly complicating the data association.

	The computational complexity of the $j$th processing block that performs probabilistic data association can be analyzed as follows. As discussed in \cite{LeiMeyHlaWitTufWin:J19,MeyKroWilLauHlaBraWin:J18}, the computational complexity of such a processing block scales as $\mathcal{O}(L^{(j)}M)$, where $L^{(j)}$ is the number of PMVA-oriented association variables, and $M$ is the number of measurement-oriented association variables. It can easily be verified that in the proposed model, $L^{(j)}$ is upper bounded by $S^{(j)\ist2}$. The computational complexity of the $j$th processing block thus scales as $\mathcal{O}(S^{(j) 2} \hspace{.1mm} M)$. If conventional probabilistic data association \cite{BarWilTia:B11} would be used, i.e., if the graph structure related to PMVA-oriented association variables and measurement-oriented association variables were not exploited, the computational complexity would scale exponentially in the number of PMVAs and the number of measurements. It is straightforward to see that after appropriate pruning, as discussed in the previous Section \ref{sec:beliefCalc}, the computation complexity is linear in the number of processing blocks and, thus, in the number of PAs. Note that even a moderate number of \acp{pmva} leads to a large number of corresponding \acp{va}. A key feature of the proposed method that makes this possible is that probabilistic data association can be solved in a scalable way.

	\begin{figure*}[t!]
		\centering
		\subfloat[\label{fig:floorplan2}]{\scalebox{0.8}{\includegraphics{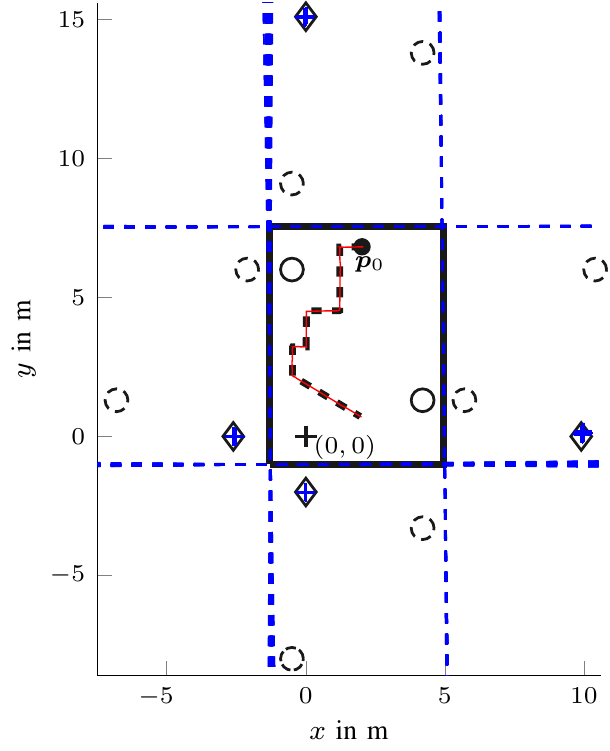}}}\hfil
		\subfloat[\label{fig:floorplan2walls}]{\scalebox{0.8}{\includegraphics{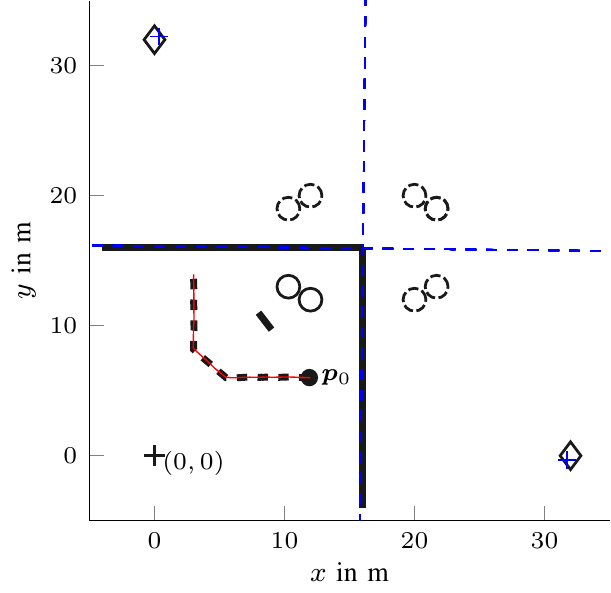}}}\hfil
		\subfloat[\label{fig:floorplanMeas}]{\scalebox{0.8}{\includegraphics{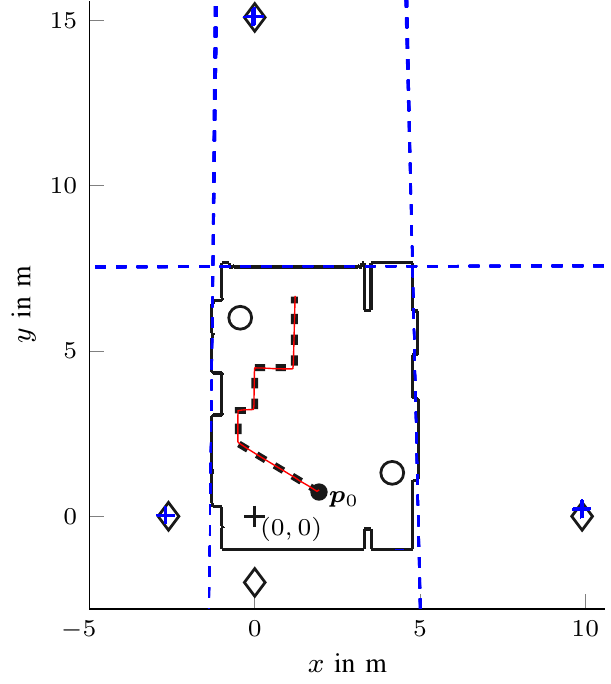}}}\\
		\centering
		\includegraphics{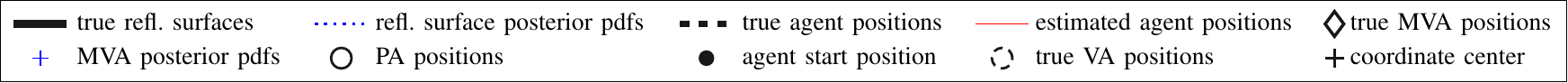}
		\vspace*{-2mm}
		\caption{Scenarios used for numerical evaluation: (a) {shows the scenario for Experiments $1$ and $2$ in Sections~\ref{sec:results_compare} and \ref{sec:results_variances}.} This scenario consists of a rectangular room with two \acp{pa}, i.e., there are four reflective surfaces. MVAs and VAs corresponding to single-bounce paths are shown. (b) shows the scenario for Experiment $3$ in Section~\ref{sec:results_olos} which consists of two \acp{pa}, two reflective surfaces and an obstructing wall segment. MVAs and VAs corresponding to single-bounce paths are shown. (c) {shows the scenario for Experiment 4 in Section~\ref{sec:result_meas}}. This scenario consists of a classroom with two \acp{pa} and four main reflective surfaces. {Particle representations of the beliefs of MVA positions are shown as blue crosses. A line representing a reflecting surface is computed and shown as a dashed blue line for each particle. The geometric relations used to calculate a line from an MVA position have been presented in Section \ref{sec:geometricRel}.}  }
		\vspace*{-5mm}
	\end{figure*}

	\section{Evaluation}
	\label{sec:results}
		
	The performance of the proposed \ac{mva}-based \ac{slam} algorithm is validated and compared with the multipath-based \ac{slam} algorithm from \cite{LeiMeyHlaWitTufWin:J19} that has been extended to use \ac{aoa} measurements described here and in \cite{LeiGreWit:ICC2019}, the channel \ac{slam} algorithm from \cite{GentnerTWC2016}, and the multipath assisted positioning algorithm from \cite{LeiMeyMeiWitHla:GNSS2016} using synthetic measurements as well as real \ac{rf} measurements. Additional simulation results can be found in the supplementary material\mref{Section}{sec:app_results}.
	
	\vspace*{-2mm}			 	
	\subsection{Common Setup and Performance Metrics}
	\label{sec:commonSetup}
		
	The agent's state-transition pdf $f(\V{x}_n|\V{x}_{n-1})$, with $\V{x}_n \rmv=\rmv  [\V{p}_n^{\T} \; \V{v}_n^{\T} ]^{\T}\rmv$, is defined by a linear, near constant-velocity motion model 
	\cite[Sec. 6.3.2]{BarShalomBook:Book2001}, i.e., $\V{x}_n =  \V{A} \V{x}_{n-1} + \V{B} \V{w}_n$.
	Here, $\V{A} \rmv\in\rmv \mathbb{R}^{4 \times 4}$ and $\V{B} \rmv\in\rmv \mathbb{R}^{4 \times 2}$ are as defined in \cite[Sec. 6.3.2]{BarShalomBook:Book2001} 
	(with sampling period $\vu{\Delta T \rmv=\rmv 1\rmv}{s}$), and the driving process $\V{w}_n$ is iid across $n$, zero-mean, and Gaussian with covariance matrix $\sigma_w^2\bold{I}_2$, where $\bold{I}_2$ denotes the $2 \rmv\times\rmv 2$ identity matrix and $\sigma_w$ denotes the acceleration noise standard deviation. For the sake of numerical stability, we introduced a small regularization noise to the PMVA state $\V{p}_{s,\mathrm{mva}}$ at each time $n$, i.e., $\underline{\V{p}}_{s,\mathrm{mva}} \rmv\rmv=\rmv\rmv \V{p}_{s,\mathrm{mva}} \rmv+\rmv \V{\omega}_{s}$, where $\V{\omega}_{s}$ is iid across $s$, zero-mean, and Gaussian with covariance matrix $\sigma_a^2\bold{I}_2$. The particles for the initial agent state are drawn from a 4-D uniform distribution with center $\V{x}_0 = [\V{p}_{0}^{\T}\;0\;\, 0]^{\T}\rmv$, where $\V{p}_{0}$ is the starting position of the actual agent trajectory, and the support of each position component about the respective center is given by $[-0.5\,\text{m}, 0.5\,\text{m}]$ and of each velocity component is given by $[-0.1\,\text{m/s}, 0.1\,\text{m/s}]$. At time $n \rmv=\rmv 0$, the number of \acp{pmva} is $S_0 = 0$, i.e., no prior map information is available. The prior distribution for new \ac{pmva} states $f_\mathrm{n}(\overline{\V{y}}_{m,n})$ is uniform on the square region given by $[-15\,\text{m},\ist 15\,\text{m}]\ist\times\ist[-15\,\text{m},\ist 15\,\text{m}]$ around the center of the floor plan shown in Fig.~\ref{fig:floorplan2} and the mean number of new \ac{pmva} is $\mu_n = 0.05$. The probability of survival is $p_{\mathrm{s}} = 0.999$, the {confirmation} and pruning thresholds are respectively ${p_{\mathrm{cf}}} = 0.5$ and $p_{\mathrm{pr}} = 10^{-3}$. We performed $500$ simulation runs. The performance of the different methods discussed is measured in terms of the \ac{rmse} of the agent position, as well as the \acf{ospa} error \cite{SchVoVo:TSP2008} of \acp{va} and \acp{mva}. {Since the proposed method estimates MVAs, we first map MVA estimates to VA estimates following equation \eqref{eq:nonLinearTransformation}, before we compute \ac{ospa} errors of \acp{va}.} {\ac{ospa} is a multi-object tracking metric that combines a localization error and a cardinality error into a single scalar score. As a result, it penalizes both state estimation inaccuracy and incorrect numbers of estimated features}. We calculate the \ac{ospa} errors based on the Euclidean metric with cutoff parameter $c = 5\,$m and order $p = 1$. The mean \ac{ospa} (MOSPA) errors, \acp{rmse} of each unknown variable are obtained by averaging over all converged simulation runs. We declare a simulation run to be converged if $\{\forall n: ||\hat{\V{x}}_n - \V{x}_n|| < 5\,\text{m} \}$.

	{For synthetic measurements (\textit{Experiment $1$--$3$}), we use the following common parameters. The detection probability of all paths is $\vspace{0.2mm} p_{\mathrm{d},ss,n}^{(j)} = p_{\mathrm{d},ss',n}^{(j)} = p_\mathrm{d} = 0.95$ for $(s,s) \in \mathcal{D}_\text{S}$ and $(s,s') \in \tilde{\mathcal{D}}_\text{D}$, respectively. In addition, a mean number $\mu_{\mathrm{fp}} = 1$ of false positive measurements $\V{z}_{m,n}^{(j)}$ were generated according to the pdf $f_{\mathrm{fp}}(\V{z}_{m,n}^{(j)})$ that is uniformly distributed on $[0\,\text{m},30\,\text{m}]$ for distance measurements and uniformly distributed on $[-\pi, \pi]$ for \ac{aoa} measurements.  In each simulation run, we generated noisy distance and \ac{aoa} measurements according to \eqref{eq:VADistmeas} and \eqref{eq:VAAnglemeas} stacked into the vector $\V{z}_{m,n}^{(j)}$.}

	\vspace*{-2mm}
	\subsection{{Reference Methods}} \label{sec:reference_methods}
	
	In the following sections, we compare the proposed \ac{mva}-based \ac{slam} algorithm (\textit{PROP}) to four different reference methods as described in the following:
	\begin{enumerate}[leftmargin=4mm]
		\item \textit{MP-SLAM}: The multipath-based \ac{slam} algorithm from \cite{LeiMeyHlaWitTufWin:J19}. The method considered here is a combination of \cite{LeiMeyHlaWitTufWin:J19} and \cite{LeiGreWit:ICC2019} since in \cite{LeiGreWit:ICC2019} statistical model of \cite{LeiMeyHlaWitTufWin:J19} is extended to \ac{aoa} measurements of \acp{mpc}. However, in contrast to \cite{LeiGreWit:ICC2019}, we do not use the component \acp{snr} estimates of \acp{mpc}. Contrary to the method proposed in this paper, the reference method does rely on a much simpler statistical model without \acp{mva} and ray-tracing. It thus cannot fuse data across propagation paths and VAs.
		\item \textit{CH-SLAM}: The channel \ac{slam} algorithm from \cite{GentnerTWC2016}. We only implemented the channel \ac{slam} algorithm proposed in \cite{GentnerTWC2016}, not the full two-stage method that includes a channel estimator/tracker. Since the measurement-feature association is unknown, we perform Monte-Carlo data association for each particle separately, as it is commonly done in classical Rao-Blackwellized \ac{slam} \cite[Section 13]{ThrFoxBur:B05}, \cite{DurrantWhyte2006, MonThrKolWeg:AAAI2002}.
		\item \textit{MINT}: The multipath-based positioning algorithm from \cite{LeiMeyMeiWitHla:GNSS2016}, which assumes known map features (i.e., the \ac{va} positions are known). 
		\item \textit{LOS-MINT}: A reduced version of MINT, where we only consider the \acp{pa} (i.e., no \acp{va}).
	\end{enumerate}
    We used $50000$ particles for PROP, MP-SLAM, MINT, and LOS-MINT. For CH-SLAM, we used $2000$ particles for the agent and $1000$ for the \acp{va}. To analyze the performance gain due to exploiting double-bounce reflections, we generate two different datasets: (Setup-I) the full setup considering all \acp{va} corresponding to single-bounce and double-bounce paths and (Setup-II) a reduced setup considering only \acp{va} corresponding to single-bounce paths. If not stated differently, measurements are generated according to Setup-I.

	\subsection{{Experiment 1: Comparison with Reference Methods}}\label{sec:results_compare}
	
	In this experiment, we compare PROP to MP-SLAM and CH-SLAM. Furthermore, we compare PROP with measurements generated without false positive measurements and missed detections termed ground-truth (GT) for Setup-I and Setup-II. We consider the indoor scenario shown in Figure~\ref{fig:floorplan2}. We chose the scenario to be identical to \cite{LeiMeyHlaWitTufWin:J19} for easy comparison. The scenario consists of four reflective surfaces, i.e., $K=4$ \acp{mva} and two \acp{pa}. The noise standard deviations for the \ac{los} path are ${\sigma_{\mathrm{d}}}^{(j)}_{m,n} = 0.05\,$m and ${\sigma_{\mathrm{\varphi}}}^{(j)}_{m,n} = 10^{\circ}$, for single-bounce path are ${\sigma_{\mathrm{d}}}^{(j)}_{m,n} = 0.10\,$m and ${\sigma_{\mathrm{\varphi}}}^{(j)}_{m,n} = 15^{\circ}$, and for double-bounce path are ${\sigma_{\mathrm{d}}}^{(j)}_{m,n} = 0.15\,$m and ${\sigma_{\mathrm{\varphi}}}^{(j)}_{m,n} = 25^{\circ}$. The acceleration noise standard deviation is $\sigma_w = 9\cdot10^{-3}\,\text{m}/\text{s}^2$. As an example, Fig.~\ref{fig:floorplan2} depicts for one simulation run the posterior \acp{pdf} represented by particles of the \ac{mva} positions and corresponding reflective surfaces as well as estimated agent tracks. Fig.~\ref{fig:VAserrorB} shows the \acs{mospa} error for the two \acp{pa} and all associated \acp{va}, Fig.~\ref{fig:MVAserrorB} shows the \acs{mospa} error for all \acp{mva}, and Fig.~\ref{fig:agenterrorposB} shows the \ac{rmse} of the agent's position all versus time $n$. Fig.~\ref{fig:agenterrorposCDF} shows the cumulative frequency of the agent errors (not excluding the diverged runs). Note that for all algorithms, none of the $500$ simulation runs diverged.
	
		\begin{figure}[!t]
		\centering
		\captionsetup[subfigure]{captionskip=0pt}
		\subfloat[\label{fig:VAserrorB}]{\includegraphics{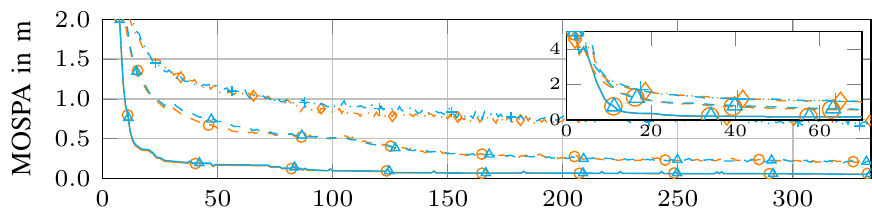}}\\[-1.3mm]
		\subfloat[\label{fig:MVAserrorB}]{\includegraphics{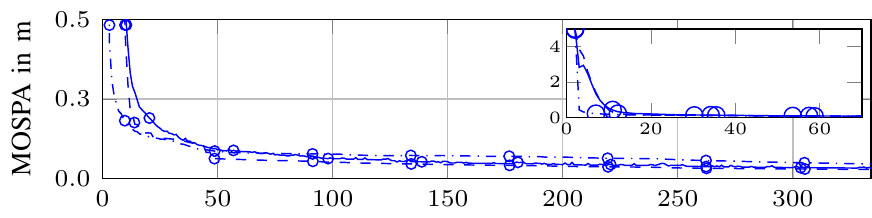}}\\[0.5mm]
		\captionsetup[subfigure]{captionskip=-5pt}	
		\subfloat[\label{fig:agenterrorposB}]{\includegraphics{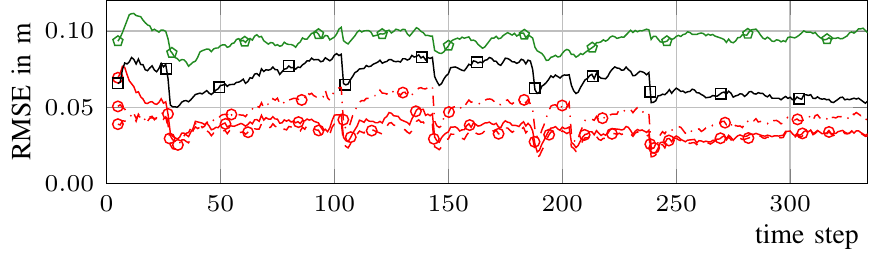}}\\[-1.5mm]
		\captionsetup[subfigure]{captionskip=-5pt}
		\subfloat[\label{fig:agenterrorposCDF}]{\includegraphics{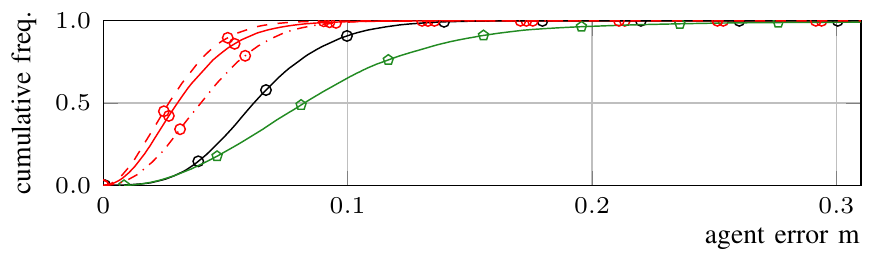}}\\[1mm]
		\centering
		\includegraphics{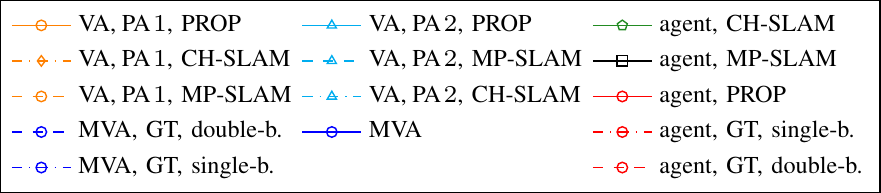}
		\vspace*{-6mm}
		\caption{{Performance results for Experiment 1 in Section~\ref{sec:results_compare}}: (a) MOSPA errors of the \acp{va} of each PA, (b) MOSPA errors versus time of the \acp{mva}, (c) RMSEs of the agent position versus time, (d) RMSEs versus time of the agent orientation{, and (e) cumulative frequency of the RMSEs of the agent position.}} \label{fig:errorsB}
		\vspace*{-5mm}
	\end{figure}
	
	The MOSPA error of PROP related to VAs and PAs is shown in Fig.~\ref{fig:VAserrorB}. It can be seen that the MOSPA drops significantly after only a few time steps. In contrast, MP-SLAM only converges rather slowly to a small mapping error. Furthermore, the MOSPA error of PROP converges along the agent track to a smaller value, i.e., to a smaller mapping error. This shows that PROP efficiently exploits all measurements for the map features provided by the PAs. Fig.~\ref{fig:MVAserrorB} shows the MOSPA error of the \ac{mva} positions, which confirms the results seen in Fig.~\ref{fig:VAserrorB}. The RMSE of the agent position in \ref{fig:agenterrorposB} {of PROP is considerably smaller than that of MP-SLAM along the whole agent track. Moreover, the RMSE of the agent position of PROP consistently decreases over time $n$, while that of MP-SLAM increases slightly during changes in the agent's direction.} 
	%Both methods show similar performance in terms of the agent orientation RMSE. 
	PROP consistently demonstrates a statistically significant improvement in accuracy across all metrics, as illustrated in Figure \ref{fig:agenterrorposCDF}, facilitating the statistical dependencies across multiple paths and \acp{pa}. Furthermore, the results comparing Setup I and II using GT measurements illustrate that leveraging double-bounce paths systematically improves the performance of PROP. The average runtimes per time step for MATLAB implementations on a single core of Intel i7 CPUs (computer cluster with different versions of CPUs) were measured $4\,$s for PROP, $1.2\,$s for MP-SLAM, and $11\,$s for CH-SLAM.\footnote{{Rao-Blackwellized \ac{slam} explicitly considers the dependency of map features on the agent state in its posterior representation. Thus, the particle-based representation in CH-SLAM \cite{GentnerTWC2016} is very computationally demanding.}}

	\subsection{Experiment 2: Varying Measurement Uncertainties}\label{sec:results_variances}

	In this experiment, we analyze the performance of PROP with varying measurement noise standard deviations and compare it to MP-SLAM. 
	In particular, we introduce factors $f_\text{std, all}$ and $f_\text{std,los}$. Both factors equally increase the base values of all measurement standard deviations in distance ${\sigma_{\mathrm{d}}}^{(j)}_{m,n}$ and \ac{aoa} ${\sigma_{\mathrm{\varphi}}}^{(j)}_{m,n}$ for all $m \in M_n^{(j)} \, | \, j\in J,\, n \in \{1,\,...\,, N\}$ in a multiplicative way. Base values are set as the measurement standard deviations of \textit{Experiment 1}. While $f_\text{std,all}$ affects measurements corresponding to \acp{pa} and MVAs, $f_\text{std,los}$ affects only measurement standard deviations corresponding to PAs (i.e., LOS measurements). The scenario is identical to experiment 1 of Section~\ref{sec:results_compare}. Note that for all algorithms none of the $500$ simulation runs diverged.
	
	\begin{figure}[!t]
		\centering
		\captionsetup[subfigure]{captionskip=0pt}
		\subfloat[\label{fig:VAserrorCLow}]{\includegraphics{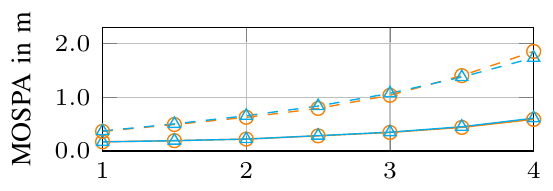}}\hspace*{-1mm}
		\subfloat[\label{fig:VAserrorCHigh}]{\includegraphics{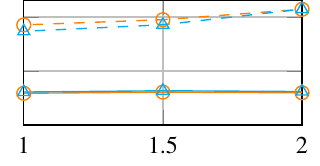}}\\[-2.5mm]
		\subfloat[\label{fig:MVAserrorCLow}]{\includegraphics{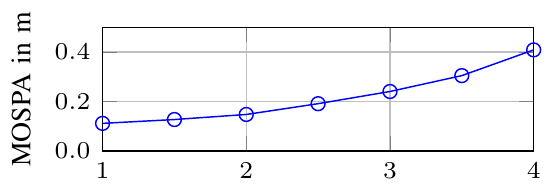}}\hspace*{-1mm}	
		\subfloat[\label{fig:MVAserrorCHigh}]{\includegraphics{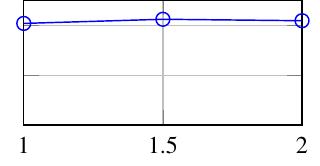}}\\[-1.4mm]
		\captionsetup[subfigure]{captionskip=-5pt}	
		\subfloat[\label{fig:agenterrorCLow}]{\includegraphics{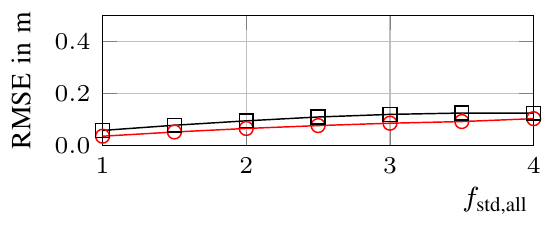}}\hspace*{-1mm}
		\subfloat[\label{fig:agenterrorCHigh}]{\includegraphics{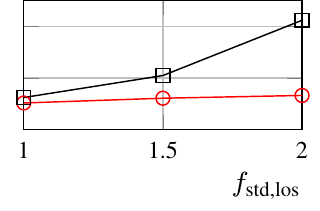}}\\[1mm]
		\centering
		\includegraphics{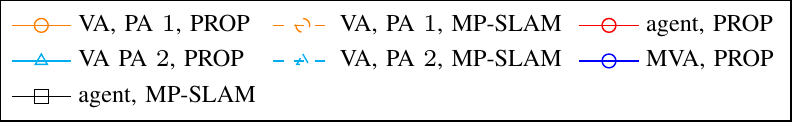}		
		\vspace*{-2mm}
		\caption{{Performance results for Experiment 2 in Section~\ref{sec:results_variances}} for different noise standard deviations. Errors are averaged across time steps. Different noise standard deviations are used for all VAs and PAs:  (a) MOSPA errors of VAs of each PA, (c) MOSPA errors of MVAs, and (e) RMSEs of the agent positions. Different noise standard deviations are used for all PAs: (b) MOSPA errors of VAs of each PA, (d) MOSPA errors of MVAs, and (f) RMSEs of the agent position.  } \label{fig:errorsC}
		\vspace*{-5mm}
	\end{figure}
	
	Fig.~\ref{fig:VAserrorCLow} to \ref{fig:agenterrorCHigh} show the mean MOSPA of VAs and PAs, the mean MOSPA of MVAs or the mean agent RMSE, respectively, over all time steps $n$ as a function of $f_\text{std,all}$ or $f_\text{std,los}$. In particular, in Fig.~\ref{fig:VAserrorCLow}, \ref{fig:MVAserrorCLow} and \ref{fig:agenterrorCLow} we varied $f_\text{std,all}$ and kept $f_\text{std,los} \triangleq 1$ fixed, while in Fig.~\ref{fig:VAserrorCHigh}, \ref{fig:MVAserrorCHigh} and \ref{fig:agenterrorCHigh} we varied $f_\text{std,los}$ and kept $f_\text{std,all} \triangleq 4$ fixed. The \acp{va} MOSPA errors in Figs.~\ref{fig:VAserrorCLow} to \ref{fig:MVAserrorCHigh} emphasize {the observations of Experiment 1 (Section~\ref{sec:results_compare})}. All error values increase with increasing $f_\text{std,all}$. However, the MOSPA errors of the proposed method increases much slower. This is because the proposed method can fuse information provided by different \acp{pa} and different propagation paths. In contrast, the agent RMSE in Fig.~\ref{fig:agenterrorCLow} remains constant for both methods as there is still enough information available for proper localization, mainly provided by the PAs. Thus, when additionally increasing $f_\text{std,los}$ in Fig.~\ref{fig:agenterrorCHigh}, the RMSE of MP-SLAM significantly increases. In contrast, the agent RMSE of PROP remains approximately constant due to the increased map stability.	

	\vspace*{-1mm}
	\subsection{{Experiment 3: Low Information and Obstructed LOS}}\label{sec:results_olos}
	
	\begin{figure}[t!]
		\centering
		\captionsetup[subfigure]{captionskip=0pt}
		\subfloat[\label{fig:MVAserrorC}]{\includegraphics{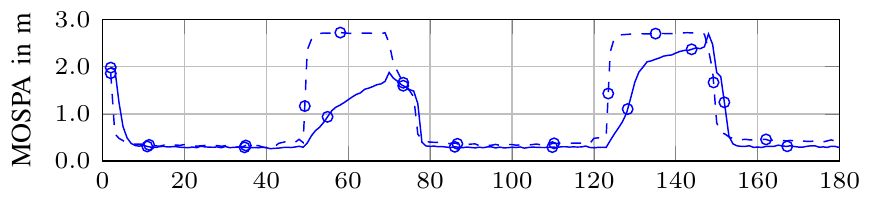}}\\[-1.5mm]
		\captionsetup[subfigure]{captionskip=-5pt}
		\subfloat[\label{fig:agenterrorposC}]{\includegraphics{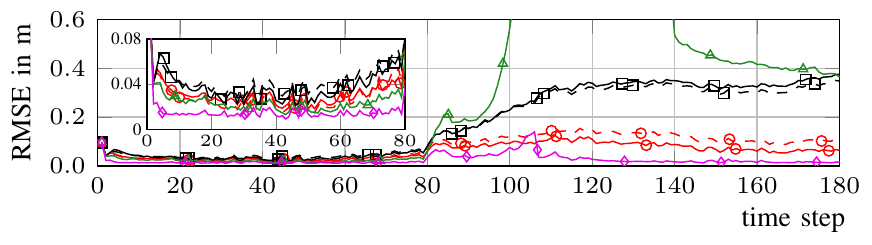}}\\[-2.5mm]
		\captionsetup[subfigure]{captionskip=-5pt}
		\subfloat[\label{fig:agenterrorposCDFC}]{\includegraphics{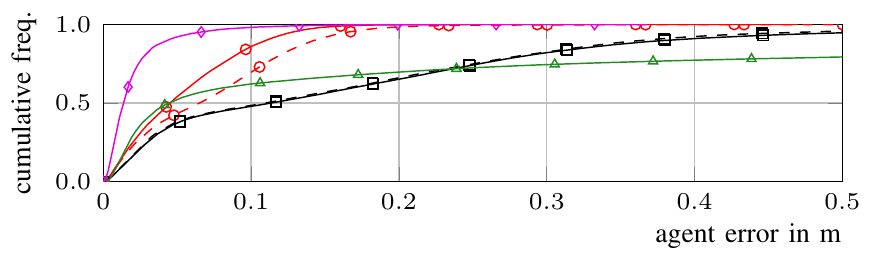}}\\[2mm]
		\centering
		\vspace*{-1mm}
		\includegraphics{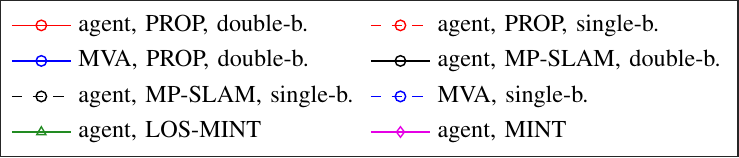}
		\vspace*{-2mm}
		\caption{{Performance results for Experiment 3 in Section~\ref{sec:results_olos}: (a) MOSPA errors versus time of \acp{mva}, (b) RMSEs versus time of the agent position, (c) RMSEs versus time of the agent orientation, (d) cumulative frequency of the RMSEs of the agent position (with outliers).}} \label{fig:errorsD}
		\vspace*{-5mm}
	\end{figure}
	
	In this experiment, we analyze the performance of PROP in the scenario shown in Figure~\ref{fig:floorplan2walls}. It contains only two reflective walls ($K=2$) while one short wall obstructs the \ac{los} path, i.e., the path between \acp{pa} and the agent as well as the paths between \acp{va} and agent\cite{VenusAsilomar2022}. The obstructing wall does not cause any \acp{va} due to the geometric constellation. We compare MP-SLAM, MINT, and LOS-MINT. The scenario contains two \acp{pa}. We generate the distance and the \ac{aoa} of the individual \ac{mpc} parameters according to a \ac{rf} signal model \cite{LiLeiVenTuf:TWC2022}. We assume the agent has a $ 3 \times 3 $ uniform rectangular array ($H=9$) with an inter-element spacing of $ 2 $\,cm. The transmit signal spectrum has a root-raised-cosine shape, with a roll-off factor of $0.6$ and a $3$-dB bandwidth of $B=500\,\mathrm{MHz}$ centered at $6\,\mathrm{GHz}$ resulting in a sampling time of $T_\text{s} = 1/(1.6 \, B)$. The amplitude of each \ac{mpc} is assumed to follow free-space path loss and is attenuated by $ 3 $\,dB after each reflection. The \ac{snr} output at $1$ m distance to the agent is assumed to be $38$ dB. The measurement noise standard deviations are calculated based on the Fisher information \cite{WilGreLeiMueWit:ACSSC2018, LeiGreWit:ICC2019, LiLeiVenTuf:TWC2022}. The acceleration noise standard deviation is $\sigma_w = 0.02\,\text{m}/\text{s}^2$.
	
	Fig.~\ref{fig:MVAserrorC} shows the MOSPA error for all \acp{mva}, Fig.~\ref{fig:agenterrorposC} shows the \ac{rmse} of the agent's position for converged simulation runs, all versus time $n$. Finally, Fig.~\ref{fig:agenterrorposCDFC} shows the cumulative frequency of all agent's position errors (not excluding the diverged runs). We show results for all investigated algorithms, where solid lines correspond to Setup-I and dashed lines correspond to Setup-II as described above. For PROP and MP-SLAM, none of the $500$ simulation runs diverged, but $30\,$\% of the simulation runs diverged for LOS-MINT. LOS-MINT performs poorly, as in the central part of the track ($n=93$ to $n=107$), the \ac{los} to all anchors is obstructed. This leads to LOS-MINT tending to choose an \ac{mpc} as the \ac{los} hypothesis as the agent state gradually becomes more uncertain. Figure~\ref{fig:agenterrorposC} illustrates the benefits of PROP with respect to MP-SLAM, as it systematically leverages both \acp{pa} as well as both single-bounce and double-bounce propagation paths to infer the map features leading to a reduction in the \ac{rmse} of the agent's position. Furthermore, Figure~\ref{fig:agenterrorposCDFC} statistically shows that this fusion results in fewer instances of large agent errors and that PROP consistent outperforms MP-SLAM. A possible explanation is the increased presence of \acp{mva} and their corresponding reflective surfaces, which are more likely to exist due to the additional double-bounce measurement update.Although the single-bounce paths may not be visible, their absence is compensated by the fact that the proposed method performs data fusion across multiple propagation paths, as observed in Figure~\ref{fig:MVAserrorC}. In contrast, MP-SLAM independently estimates each \ac{va}, thus lacking this advantageous feature. MINT has perfect (prior) knowledge of the \ac{va} positions and, thus, provides a lower bound for \ac{va}-based \ac{slam} algorithms. PROP, which exploits double-bounce propagation paths, comes close to approaching this lower bound. Note that, in Figure \ref{fig:agenterrorposC}, between $n=0$ and $n=80$, LOS-MINT shows a slightly higher positioning accuracy compared to the proposed method. This difference is due to the uncertainty in \ac{mva} positions. A theoretical analysis on how uncertainty of map information affects the positioning accuracy of the agent is provided in \cite{ShaGarDesSecWym:TWC2018, MenWymBauAbu:TWC2019}.
	
	\subsection{{Experiment 4:} Validation Using Measured Radio Signals} \label{sec:result_meas}
	
	\begin{figure}[t!]
		\centering
		\includegraphics[width=0.85\columnwidth,trim={0 0.7cm 0 0.4cm},clip]{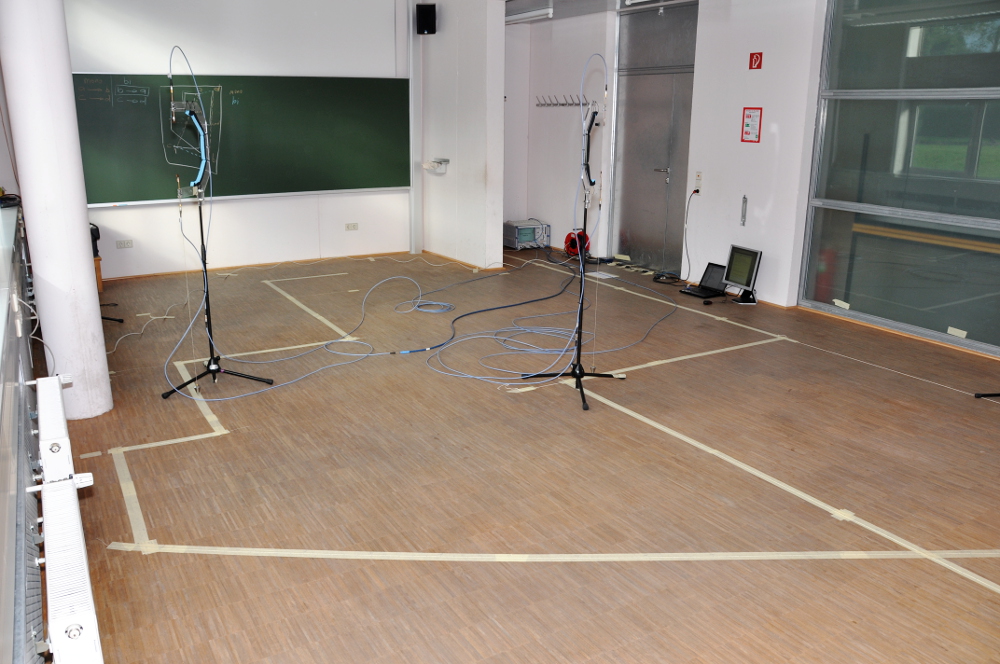}
		\vspace*{-1mm}
		\caption{Picture of the classroom used for data collection.}
		\label{fig:pic}
		\vspace*{-3mm}
	\end{figure}
		
	To validate the applicability of the proposed \ac{mva}-based \ac{slam} algorithm to real \ac{rf} measurements, we use data collected in a classroom shown in Fig.~\ref{fig:pic} at TU Graz, Austria. More details about the measurement environment and \ac{va} calculations can be found in \cite{MeasureMINT2013,LeiMeyHlaWitTufWin:J19,LeitingerJSAC2015}. On the \ac{pa} side, a dipole-like antenna with an approximately uniform radiation pattern in the azimuth plane and zeros in the floor and ceiling directions was used. At each agent position, the same antenna was deployed multiple times on a $3\times3$, 2-D grid to yield a virtual uniform rectangular array with an inter-element spacing of $2$\,cm. The UWB signals are measured at $ 180 $ agent positions along a trajectory with position spacing of {approx.} $5$\,cm as shown in Fig.~\ref{fig:floorplanMeas} using an M-sequence correlative channel sounder with frequency range $3.1–10.6\,$GHz. Within the measured frequency-band, the actual signal spectrum was selected by a filter with root-raised-cosine shape, with a roll-off factor of $0.6$ and a $3$-dB bandwidth of $B=1\,\mathrm{GHz}$ centered at $6\,\mathrm{GHz}$. The received signal is critically sampled with $T_\text{s} = 1/(1.6 \, B)$ and artificial AWGN is added such that the output signal-to-noise-ratio is $\ac{snr}=30$\,dB. We apply a variational sparse Bayesian parametric channel estimation algorithm \cite{ShutWanJos:CSTA2013} to acquire the $M_n^{(j)}$ distance estimates ${z_\text{d}}_{m,n}^{(j)}$ and \ac{aoa} estimates ${z_\varphi}_{m,n}^{(j)}$ of \acp{mpc}. The corresponding noise standard deviations are calculated based on the Fisher information \cite{WilGreLeiMueWit:ACSSC2018, LeiGreWit:ICC2019, LiLeiVenTuf:TWC2022}. Compared to the synthetic setup, we changed the mean number of false alarm measurements to $\mu_\mathrm{fp}=3$, the detection probability to $p_\text{d} = 0.7$, regularization noise standard deviation to $\sigma_a = 2 \cdot 10^{-3}\,\text{m}$, and the acceleration noise standard deviation to $\sigma_w = 0.0114\,\text{m}/\text{s}^2$. Note that for all algorithms, none of the $500$ simulation runs diverged.

	Fig.~\ref{fig:floorplanMeas} depicts for one simulation run the posterior \acp{pdf} represented by particles of the \ac{mva} positions and corresponding reflective surfaces as well as estimated agent tracks. PROP can identify the main reflective surfaces of the room (The lower wall is only visible at the beginning of the agent track since the reflection coefficient is very low). Although the walls have a rich geometric structure (windows, doors, etc.) and generate many \acp{mpc} estimates, i.e., measurements, PROP robustly estimates the main walls.\footnote{Note that \cite[Figure~7]{LeiMeyHlaWitTufWin:J19} shows results using real measured radio signals in the same environment. This figure shows the presence of single-bounce and multiple-bounce propagation paths. For instance, in the case of the \ac{pa} indicated in blue, the double-bounce path related to left and top VA is clearly visible. It is particularly noteworthy that single-bounce and double-bounce paths are consistently observable along a significant portion of the agent track.} Fig.~\ref{fig:errorsMeas} compares PROP and MP-SLAM in terms of the agent \ac{rmse}. Fig.~\ref{fig:errorsMeasA} shows the \ac{rmse} of the agent's position, and Fig.~\ref{fig:errorsMeasB} shows the \ac{rmse} of the agent's orientation for simulation runs, all versus time $n$. The comparison of the position \acp{rmse} shows a similar behavior as for synthetic measurements (see Fig.~\ref{fig:errorsB}), i.e., PROP outperforms MP-SLAM. The mapping capability and low agent \acp{rmse} of PROP, when applied to real \ac{rf} signals, demonstrate the high potential of PROP for accurate and robust \ac{rf}-based localization.
	
	\begin{figure}[t!]
		\centering
		\captionsetup[subfigure]{captionskip=0pt}
		\subfloat[\label{fig:errorsMeasA}]{\includegraphics{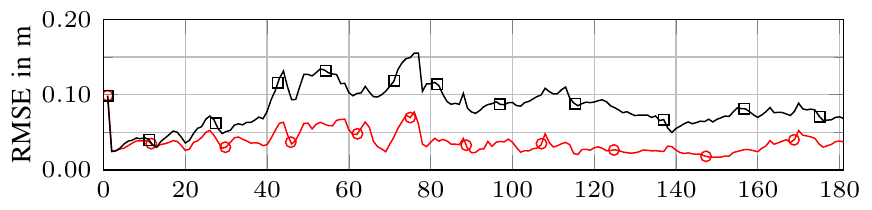}}\\[-1mm]
		\captionsetup[subfigure]{captionskip=-5pt}
		\subfloat[\label{fig:errorsMeasB}]{\includegraphics{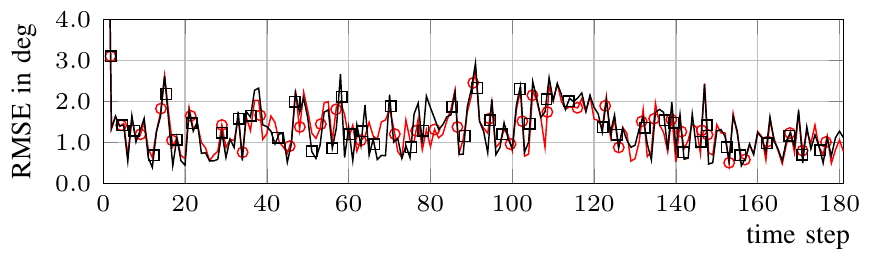}}
		\vspace*{-2mm}
		\caption{{Performance results for Experiment 4 in Sec~\ref{sec:result_meas}}: (a) RMSEs versus time of the agent position and (b) RMSEs versus time of the agent orientation.} \label{fig:errorsMeas}
		\vspace*{-3mm}
	\end{figure}	

	\section{Conclusions} \label{sec:conclusion}
	\acresetall
	In this paper, we introduced data fusion for multipath-based \ac{slam}. A key novelty of our approach is to represent each reflective surface in the propagation environment by a single \ac{mva}. In this way, we address a key limitation of existing multipath-based \ac{slam} methods, which represent every propagation path by a \ac{va} and thus neglect inherent geometrical constraints across different paths that interact with the same reflective surface. As a result, the accuracy and speed of existing multipath-based \ac{slam} methods are limited. A key aspect in leveraging the advantages of the introduced \ac{mva}-based model was to check the availability of single-bounce and double-bounce propagation paths at potential agent positions by means of {\acf{rt}}. Availability checks were directly integrated into the statistical model as detection probabilities of paths. Our numerical simulation results demonstrated significant improvements in estimation accuracy and mapping speed compared to state-of-the-art multipath-based \ac{slam} methods. Looking forward, we expect to extend our approach to large-scale scenarios and more realistic 3-D environments. We expect such an extension to yield significantly increased computational complexity due to an increased dimensionality of the states to be estimated and an increased number of \acp{mva} due to floor and ceiling surfaces. Promising directions for future research also include an extension to multiple-measurement-to-feature data association \cite{MeyWil:J21,WieVenWilLeiArxiv2023} and an advanced \ac{mva} model, where the length and shape of reflective surfaces \cite{ChuLuGesWanWenWuMuqLi:TWC2022} are also taken into account. Another future research venue aims at incorporating amplitude information to make detection probabilities and measurement variances adaptive \cite{LeiGreWit:ICC2019, LiLeiVenTuf:TWC2022,VenusAsilomar2022}. 
	
	\vspace*{-0.5mm}
	\section*{Acknowledgement}
	DISTRIBUTION STATEMENT A: Approved for public release. This material is based upon work supported by the Under Secretary of Defense for Research and Engineering under Air Force Contract No. FA8702-15-D-0001. Any opinions, findings, conclusions, or recommendations expressed in this material are those of the author(s) and do not necessarily reflect the views of the Under Secretary of Defense for Research and Engineering.
	
	\vspace*{-0.5mm}
	\renewcommand{\baselinestretch}{.945}
	\selectfont
	\bibliographystyle{IEEEtran}
	\bibliography{IEEEabrv,StringDefinitions,Books,References,XRefs}

	\begin{IEEEbiography}[{\includegraphics[width=25mm,height=32.15mm,clip,keepaspectratio]{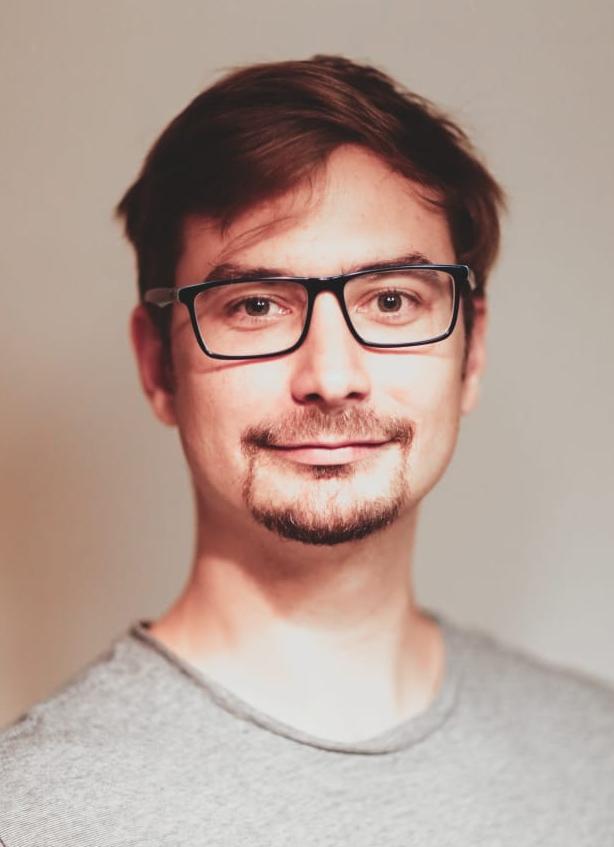}}]{Erik~Leitinger} (Member, IEEE) received his MSc and PhD degrees (with highest honors) in electrical engineering from Graz University of Technology, Austria in 2012 and 2016, respectively. He was postdoctoral researcher at the department of Electrical and Information Technology at Lund University from 2016 to 2018. He is currently a University Assistant at Graz University of Technology. Dr. Leitinger served as co-chair of the special session "Synergistic Radar Signal Processing and Tracking" at the IEEE Radar Conference in 2021. He is co-organizer of the special issue "Graph-Based Localization and Tracking" in the Journal of Advances in Information Fusion (JAIF). Dr.\ Leitinger received an Award of Excellence from the Federal Ministry of Science, Research and Economy (BMWFW) for his PhD Thesis. He is an Erwin Schr\"odinger Fellow. His research interests include inference on graphs, localization and navigation, machine learning, multiagent systems, stochastic modeling and estimation of radio channels, and estimation/detection theory. 
\end{IEEEbiography}

\begin{IEEEbiography}[{\includegraphics[width=25mm,height=32.15mm,clip,keepaspectratio]{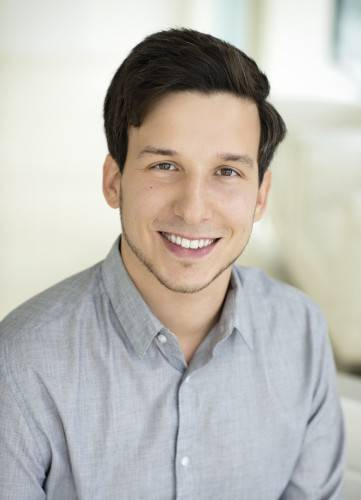}}]{Alexander~Venus} (Student Member, IEEE) received his BSc and  MSc degrees (with highest honors) in biomedical engineering and information and communication engineering from Graz University of Technology, Austria in 2012 and 2015, respectively. He was a research and development engineer at Anton Paar GmbH, Graz from 2014 to 2019. He is currently a project assistant at Graz University of Technology, where he is pursuing his Ph.D. degree. His research interests include radio-based localization and navigation, statistical signal processing, estimation/detection theory, machine learning and error bounds. 
\end{IEEEbiography}

\begin{IEEEbiography}[{\includegraphics[width=25mm,height=32.15mm,clip,keepaspectratio]{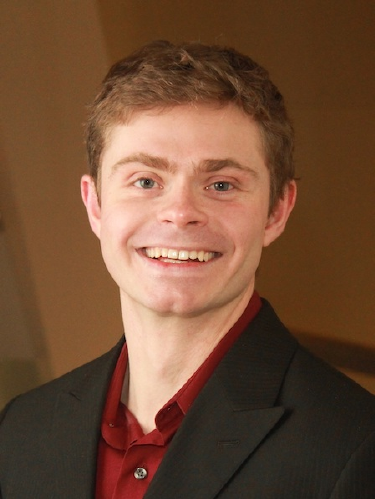}}]{Bryan~Teague} (Member, IEEE) is a technical staff member at MIT Lincoln Laboratory. He received an MS degree in aerospace engineering from the Massachusetts Institute of Technology (MIT), Cambridge, Massachusetts, in 2017, and a BS (with highest honors) degree in engineering from Harvey Mudd College, Claremont, California, in 2010. He was a member of Wireless Information and Network Sciences Laboratory, Massachusetts Institute of Technology (MIT) from 2015 to 2017. He is a winner of a 2018 R\&D100 innovation award. His research interests include probabilistic inference, optimal control, radio frequency technologies, and decentralized intelligence.
\end{IEEEbiography}

\begin{IEEEbiography}[{\includegraphics[width=25mm,height=32.15mm,clip,keepaspectratio]{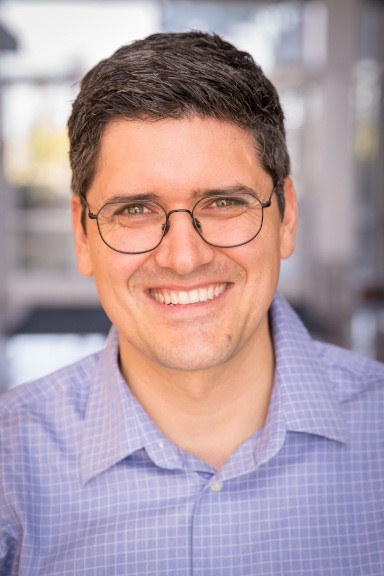}}]{Florian~Meyer} (Member, IEEE) received the MSc and PhD degrees (with highest honors) in electrical engineering from TU Wien, Vienna, Austria in 2011 and 2015, respectively. He is an Assistant Professor with the University of California San Diego, La Jolla, CA, jointly between the Scripps Institution of Oceanography and the Electrical and Computer Engineering Department. From 2017 to 2019 he was a Postdoctoral Fellow and Associate with the Laboratory for Information \& Decision Systems at the Massachusetts Institute of Technology, Cambridge, MA, and from 2016 to 2017 he was a Research Scientist with the NATO Centre for Maritime Research and Experimentation, La Spezia, Italy. Prof. Meyer is the recipient of the 2021 ISIF Young Investigator Award, a 2022 NSF CAREER Award, a 2022 DARPA Young Faculty Award, and a 2023 ONR Young Investigator Award. He is an Associate Editor with the IEEE Transactions on Aerospace and Electronic Systems and the ISIF Journal of Advances in Information Fusion and was a keynote speaker at the IEEE Aerospace Conference in 2020.  His research interests include statistical signal processing, high-dimensional and nonlinear estimation, inference on graphs, machine perception, and graph neural networks.

\end{IEEEbiography}

\end{document}

\title{\huge  Data Fusion for Multipath-Based SLAM:\\[-3mm]
Combining Information from Multiple Propagation Paths:\\[-3mm] 
Supplementary Material}
\author{\large{Erik Leitinger, Alexander Venus, Bryan Teague, Florian Meyer}}
\maketitle
\frenchspacing

This manuscript provides additional analysis for the publication ``Data Fusion for Multipath-Based SLAM: Combining Information from Multiple Propagation Paths'' by the same authors \cite{Main}. 

\section{Geometrical Transformations}

In this section, we derive the non-linear transformations \meqref{eq:nonLinearTransformation} and \meqref{eq:nonLinearTransformationMVA} in Section~\mref{Sec.}{sec:geometricRel}.

\subsection{Derivation of Transformation from \ac{mva} to \ac{va}}\label{sec:app_mva_model}

To calculate \meqref{eq:nonLinearTransformation}, i.e., $\V{p}^{(j)}_{ss,\mathrm{va}} = h_{\mathrm{va}}\big( \V{p}_{s,\mathrm{mva}}, \V{p}^{(j)}_{\mathrm{pa}} \big)$, we define the point $\V{p}^{(j)}_{s,\mathrm{wp}}$ given by the intersection of reflective surface $s$ and the line between $\V{p}^{(j)}_{ss,\mathrm{va}}$ and $\V{p}^{(j)}_{\mathrm{pa}}$. $\V{p}^{(j)}_{s,\mathrm{wp}}$ can expressed as function of $\V{p}_{s,\mathrm{mva}} = [p_{1,s,\mathrm{mva}} \iist p_{2,s,\mathrm{mva}}]^{\T}$, i.e.,
\begin{equation}
	\V{p}^{(j)}_{s,\mathrm{wp}} = \gamma_1 \iist [-p_{2,s,\mathrm{mva}} \iist p_{1,s,\mathrm{mva}}]^{\T} + \frac{\V{p}_{s,\mathrm{mva}}}{2} \label{eq1}
	\vspace{2mm}
\end{equation}
as well as function of $\V{p}^{(j)}_{\mathrm{pa}}$ and $\V{p}_{s,\mathrm{mva}}$
\begin{equation}
	\V{p}^{(j)}_{s,\mathrm{wp}} = \gamma_2 \iist \V{p}_{s,\mathrm{mva}} + \V{p}^{(j)}_{\mathrm{pa}} \label{eq2}
	\vspace{2mm}
\end{equation}
where the constants $\gamma_1$ and $\gamma_2$ will be defined in what follows. Furthermore, we express the position of the \ac{va} $\V{p}^{(j)}_{ss,\mathrm{va}} = [p^{(j)}_{1,ss,\mathrm{va}} \iist p^{(j)}_{2,ss,\mathrm{va}}]^{\T}$ as function of $\V{p}^{(j)}_{\mathrm{pa}} = [p^{(j)}_{1,\mathrm{pa}} \iist p^{(j)}_{2,\mathrm{pa}}]^{\T}$ and $\V{p}_{s,\mathrm{mva}}$, i.e.,
\begin{equation}
	\V{p}^{(j)}_{ss,\mathrm{va}} = 2 \gamma_2 \iist \V{p}_{s,\mathrm{mva}} + \V{p}^{(j)}_{\mathrm{pa}}. \label{eq:VAgamma2}
\end{equation}
By combining \eqref{eq1} and \eqref{eq2}, we obtain the following expression for $\gamma_1$ and $\gamma_2$
\begin{equation}
	\gamma_1 = \frac{- (1/2 + \gamma_2) \ist\ist p_{2,s,\mathrm{mva}} + p^{(j)}_{2,\mathrm{pa}}}{p_{1,s,\mathrm{mva}}} \label{eq:gamma1}
\end{equation}
and 
\begin{align}
	\gamma_2 = -\frac{ p^{(j)}_{1,\mathrm{pa}} \iist p_{1,s,\mathrm{mva}} + p^{(j)}_{2,\mathrm{pa}} \iist p_{2,s,\mathrm{mva}}}{p^2_{1,s,\mathrm{mva}} + p^2_{2,s,\mathrm{mva}}} + \frac{1}{2}\ist. \label{eq:gamma2}
\end{align}
By plugging \eqref{eq:gamma2} into \eqref{eq:VAgamma2}, the nonlinear transformation from \ac{mva} to \ac{va} is given by
\begin{align}
	\V{p}^{(j)}_{ss,\mathrm{va}} &= h_\text{va}\big( \V{p}_{s,\mathrm{mva}}, \V{p}^{(j)}_{\mathrm{pa}} \big) \nn \\
	&= -\Bigg(\frac{ 2 p^{(j)}_{1,\mathrm{pa}} \iist p_{1,s,\mathrm{mva}} + 2 p^{(j)}_{2,\mathrm{pa}} \iist p_{2,s,\mathrm{mva}}}{p^2_{1,s,\mathrm{mva}} + p^2_{2,s,\mathrm{mva}}}  - 1 \Bigg) \ist \V{p}_{s,\mathrm{mva}} + \V{p}^{(j)}_{\mathrm{pa}} \nn \\
	&= -\bigg(\frac{ 2 \big\langle \V{p}_{s,\mathrm{mva}},\V{p}^{(j)}_{\mathrm{pa}} \big\rangle}{ \big\|\V{p}_{s,\mathrm{mva}}\big\|^2 } - 1 \bigg) \ist \V{p}_{s,\mathrm{mva}} + \ist \V{p}^{(j)}_{\mathrm{pa}}
\end{align}
where $\langle \cdot, \cdot \rangle$ denotes the inner-product between two vectors and $\|\cdot\|$ denotes the Euclidean norm of a vector.

\subsection{Derivation of Transformation from \ac{va} to \ac{mva}}\label{sec:app_va_model}

To calculate \meqref{eq:nonLinearTransformationMVA}, i.e., $\V{p}_{s,\mathrm{mva}} = h_\text{mva}\big( \V{p}_{ss,\mathrm{va}}, \V{p}^{(j)}_{\mathrm{pa}} \big)$, we define the point $\V{p}_{s,\mathrm{mwp}}$ given by an intersection of reflective surface $s$ and the line between the origin $[0 \hspace{.6mm} 0]^{\T}$ and $\V{p}_{s,\mathrm{mva}}$. $\V{p}_{s,\mathrm{mwp}}$ can expressed as function of $\V{p}^{(j)}_{ss,\mathrm{va}}$, $\V{p}^{(j)}_{\mathrm{pa}}$, and $\V{p}^{(j)}_1 = [p_{1,1}^{(j)} \iist p_{1,2}^{(j)}]^{\T} = \V{p}^{(j)}_{\mathrm{pa}} - \V{p}^{(j)}_{ss,\mathrm{va}}$, i.e.,
\begin{equation}
	\V{p}_{s,\mathrm{mwp}} = \gamma_3 \iist [-p_{2,1} \iist p_{1,1}]^{\T} + \frac{\V{p}^{(j)}_{\mathrm{pa}} + \V{p}^{(j)}_{ss,\mathrm{va}}}{2} \label{eq6}
	\vspace{2mm}
\end{equation}
as well as function of $\V{p}^{(j)}_1$
\begin{equation}
	\V{p}_{s,\mathrm{mwp}} = \gamma_4 \iist \V{p}^{(j)}_1 \label{eq7}
	\vspace{2mm}
\end{equation}
where the constants $\gamma_3$ and $\gamma_4$ will be defined in what follows. Furthermore, we express the position of the \ac{va} $\V{p}_{s,\mathrm{mva}}$ as function of $\V{p}^{(j)}_1$, i.e.,
\begin{equation}
	\V{p}_{s,\mathrm{mva}} = 2 \gamma_4 \iist \V{p}^{(j)}_1. \label{eq:MVAgamma4}
\end{equation}
By combining \eqref{eq6} and \eqref{eq7}, we obtain the following expression for $\gamma_3$ and $\gamma_4$
\begin{equation}
	\gamma_3 = \frac{\gamma_4 \iist p^{(j)}_{2,1} - 1/2\ist\ist (p^{(j)}_{2,ss,\mathrm{va}} + p^{(j)}_{2,\mathrm{pa}})}{p^{(j)}_{1,1}} \label{eq:gamma3}
\end{equation}
and 
\begin{align}
	\gamma_4 = \Bigg( \frac{ \big\langle \V{p}^{(j)}_1, \V{p}^{(j)}_{\mathrm{pa}} \big\rangle \ist+\ist \big\langle \V{p}^{(j)}_1, \V{p}^{(j)}_{ss,\mathrm{va}} \big\rangle  }{\|\V{p}^{(j)}_1\|^2}\Bigg)\ist. \label{eq:gamma4}
\end{align}
By plugging \eqref{eq:gamma4} into \eqref{eq:MVAgamma4}, the nonlinear transformation from \ac{va} to \ac{mva} is given by
\begin{align}
	\V{p}_{s,\mathrm{mva}} &= h_\text{mva}\big( \V{p}_{s,\mathrm{mva}}, \V{p}^{(j)}_{\mathrm{pa}} \big) \nn \\
	&=  \Bigg( \frac{ \big\langle \V{p}^{(j)}_1, \V{p}^{(j)}_{\mathrm{pa}} \big\rangle \ist+\ist \big\langle \V{p}^{(j)}_1, \V{p}^{(j)}_{ss,\mathrm{va}} \big\rangle  }{\big\|\V{p}^{(j)}_1\big\|^2}\Bigg) \V{p}^{(j)}_1 \nn \\
	&= \Bigg( \frac{ \big\|\V{p}^{(j)}_{\mathrm{pa}} \big\|^2 \ist-\ist \big\|\V{p}^{(j)}_{ss,\mathrm{va}} \big\|^2  }{\big\|\V{p}^{(j)}_{\mathrm{pa}} - \V{p}^{(j)}_{ss,\mathrm{va}}\big\|^2}\Bigg)\ist(\V{p}^{(j)}_{\mathrm{pa}} - \V{p}^{(j)}_{ss,\mathrm{va}})\ist.
\end{align}

\section{Statistical Model} \label{sec:app_statistical_model}

In this section, we derive the expression of $f( \V{y}_{0:n}, \V{x}_{0:n}, \underline{\V{a}}_{1:n},\overline{\V{a}}_{1:n} \ist | \ist \V{z}_{1:n})$ in \meqref{eq:factorGraph} which is represented by the factor graph in \mref{Fig.}{fig:factorgraph} and provides the basis for the development of a  \ac{spa} algorithm for data fusion multipath-based \ac{slam}.

\subsection{Joint Prior PDF}

Before presenting derivations, we first define a few sets as follows: $ \mathcal{D}_{\underline{\V{a}}_n^{(j)}}\triangleq \{(s,s') \in \tilde{\mathcal{D}}^{(j)}_n : \underline{a}_{ss',n}^{(j)} \neq 0 \} $ denotes the set of existing legacy \acp{pmva}, where $\tilde{\mathcal{D}}^{(j)}_n = (0, 0) \cup \mathcal{D}^{(j)}_n $ with $\mathcal{D}^{(j)}_n \in \{ (s,s') \rmv\in\rmv \Set{S}_n \rmv\times\rmv \Set{S}_n \} = \Set{D}_{\text{S}, n}^{(j)} \cup \Set{D}_{\text{D},n}^{(j)}$, which is composed of the index sets for single-bounce propagation path $\Set{D}_{\text{S},n}^{(j)}$ and double-bounce propagation paths $\Set{D}_{\text{D},n}^{(j)}$, respectively (see \mref{Sec. }{sec:systemModel}). $\mathcal{N}_{\overline{\V{r}}_n^{(j)}}\triangleq \{m \in \{1,\dots,M_{n}^{(j)}\}: \overline{r}_{m,n}^{(j)} = 1, \overline{a}_{m,n}^{(j)} = 0 \} $ denotes the set of existing new \acp{pmva}. 
 
The joint prior \ac{pdf} of $\RV{y}_{0:n} = [\underline{\RV{y}}_{0:n}^{\T}, \overline{\RV{y}}_{1:n}^{\T}]^{\T}$, $ \underline{\RV{a}}_{1:n} $, $ \overline{\RV{a}}_{1:n} $, $\RV{x}_{1:n}$, and the number of the measurements $\RV{m}_{1:n} \triangleq \big[\rv{M}_{1} \cdots\ist \rv{M}_{n} \big]^{\T}\!$ factorizes as \cite{MeyBraWilHla:J17, MeyKroWilLauHlaBraWin:J18, LeiMeyHlaWitTufWin:J19}
\begin{align}
	f(\V{x}_{0:n},&\V{y}_{0:n}, \underline{\V{a}}_{1:n},  \overline{\V{a}}_{1:n}, \V{m}_{1:n}) 
	\nn \\ 
	&= f(\V{x}_{0:n}, \underline{\V{y}}_{0:n}, \overline{\V{y}}_{1:n}, \underline{\V{a}}_{1:n}, \overline{\V{a}}_{1:n}, \V{m}_{1:n}) \nn \\ 
	&= \Bigg(f(\V{x}_{0}) \ist \prod^{S_0}_{s = 1} \,  f(\V{y}_{s,0}) \Bigg) \ist
	 \prod_{ n' = 1 }^{n} \,  f\big(\V{x}_{n'} | \V{x}_{n'-1}\big)
	\Bigg(\prod_{s = 1}^{S_{n'-1}} f(\underline{\V{y}}_{s,n'}|\V{y}_{s,n'-1})\Bigg) %
	 \nn \\
	& \hspace{5mm} \hspace{3mm} \times   \bigg( \prod^{J}_{j' = 2} \ist\ist \prod^{S^{(j')}_{n'}}_{s' = 1}\ist f^{(j)}\big(\underline{\V{y}}^{(j')}_{s',n'} \big| \underline{\V{y}}^{(j'-1)}_{s', n'}\big) \Bigg) 
	 \Bigg( \prod^{J}_{j = 1}   f\big(\overline{\V{p}}_{\text{mva}}^{(j)} \big| \overline{\V{r}}_{n'}^{(j)}, M_{n'}^{(j)}, \V{x}_{n'}\big)\;
	 p\big(\overline{\V{r}}_{n'}^{(j)}, \underline{\V{a}}_{n'}^{(j)}, \overline{\V{a}}_{n'}^{(j)}, M_{n'}^{(j)}\big|\underline{\V{y}}_{n'}^{(j)},\V{x}_{n'}\big)
    \Bigg)\ist.
	\label{eq:jointPriorPDF_global} %
\end{align}
We determine the prior \ac{pdf} of new \acp{pmva} $ f(\overline{\V{p}}_{\text{mva}}^{(j)} | \overline{\V{r}}_{n}^{(j)}, M_{n}^{(j)}, \V{x}_n) $ and the joint conditional prior \ac{pmf} $  p(\overline{\V{r}}_{n}^{(j)}, \underline{\V{a}}_{n}^{(j)}, \overline{\V{a}}_{n}^{(j)}, M_{n}^{(j)}|\underline{\V{y}}_{n}^{(j)},\V{x}_n)$ in what follows. Before the current measurements are observed, the number of measurements $\rv{M}_{n}^{(j)}$ is random. The Poisson \ac{pmf} of the number of existing new \acp{pmva} evaluated at $ |\mathcal{N}_{\overline{\V{r}}_n^{(j)}}| $ is given by 
\begin{equation}
p\big(|\mathcal{N}_{\overline{\V{r}}_n^{(j)}}|\big) = \mu_{\mathrm{n}}^{|\mathcal{N}_{\overline{\V{r}}_n^{(j)}}|}/|\mathcal{N}_{\overline{\V{r}}_n^{(j)}}|!\iist \mathrm{e}^{\mu_{\mathrm{n}}}\ist.
\end{equation}
The prior \ac{pdf} of the new \ac{pmva} state $ \overline{\RV{x}}_{n}^{(j)} $ conditioned on $\overline{\RV{r}}_{n}^{(j)}$ and $\rv{M}_{n}^{(j)}$ is expressed as %
\begin{align}
	& f\big( \overline{\V{p}}_{m,\text{mva}}^{(j)} \big| \overline{\V{r}}_{n}^{(j)}, M_{n}^{(j)}, \V{x}_n\big) = \prod_{m \in \mathcal{N}_{\overline{\V{r}}_n^{(j)}}} f_{\mathrm{n}}\big( \overline{\V{p}}_{m,\text{mva}}^{(j)} \big| \V{x}_n\big) \hspace{-6mm} \prod_{m' \in \{ 1 , \, ... \, , M_n^{(j)}\} \setminus \mathcal{N}_{\overline{\V{r}}_n^{(j)}} } \hspace{-6mm} f_{\text{d}}\big(\overline{\V{p}}_{m',\text{mva}}^{(j)}\big)\ist.
	\label{eq:priorPDF_newPSMC}
\end{align}
The joint conditional prior \ac{pmf} of the binary existence variables of new \acp{pmva} $\overline{\RV{r}}_{n} \triangleq [\overline{\rv{r}}_{1,n}\ist\cdots$ $\overline{\rv{r}}_{\rv{M}_n,n}]$, the association vectors $ \underline{\RV{a}}_{n} $ and $ \overline{\RV{a}}_{n} $ and the number of the measurements $ \rv{M}_{n} $ conditioned on $ \RV{x}_n$ and $ \underline{\RV{y}}_{n}^{(j)} $ is obtained as \cite{MeyKroWilLauHlaBraWin:J18, MeyKroWilLauHlaBraWin:Proc2018_suppl, LeiMeyHlaWitTufWin:J19} %
\begin{align} 
	p\big(\overline{\V{r}}_{n}^{(j)}, \underline{\V{a}}_{n}^{(j)}, \overline{\V{a}}_{n}^{(j)}, M_{n}^{(j)}\big|\underline{\V{y}}_{n}^{(j)},\V{x}_n\big)
    & = \chi_{\overline{\V{r}}_{n}^{(j)}, \underline{\V{a}}_{n}^{(j)}, M_{n}^{(j)}}  \Bigg(\prod_{ m\in \mathcal{N}_{\overline{\V{r}}_n^{(j)}} } \Gamma_{\underline{\V{a}}_{n}^{(j)}}\big(\overline{r}_{m,n}^{(j)}\big) \Bigg) \Bigg(\prod_{ (s,s') \in \mathcal{D}_{\underline{\V{a}}_n} }{p}^{(j)}_{\mathrm{d} ss'} \big({\V{p}}_n,\underline{\V{y}}^{(j)}_{s,n},\underline{\V{y}}^{(j)}_{s',n}\big)  \Bigg) \nn \\
	& \hspace{2mm} \times \Psi\big(\underline{\V{a}}_n^{(j)},\overline{\V{a}}_n^{(j)}\big) \Bigg( \hspace{-0.5mm} \prod_{(s,s') \in \tilde{\mathcal{D}}_n^{(j)} \setminus \mathcal{D}_{\underline{\V{a}}_n^{(j)}} } \hspace{-4mm} \bigg( 1 - {p}^{(j)}_{\mathrm{d} ss'} \big({\V{p}}_n,\underline{\V{y}}^{(j)}_{s,n},\underline{\V{y}}^{(j)}_{s',n}\big)  \bigg)\Bigg)\ist .
	\label{eq:priorPDF_DA}
\end{align}
where binary check function $\Psi(\underline{\V{a}}_n^{(j)},\overline{\V{a}}_n^{(j)})$ that check consistency for any pair $(\underline{\V{a}}_n^{(j)}, \overline{\V{a}}_n^{(j)})$ of \ac{pmva}-oriented and measurement-oriented
association variables, read
\begin{equation}
\Psi\big(\underline{\V{a}}_n^{(j)},\overline{\V{a}}_n^{(j)}\big) \triangleq \prod_{ (s,s') \in \tilde{\mathcal{D}}^{(j)}_n }   \prod_{m = 1}^{M_n^{(j)}} \Psi\big(\underline{a}^{(j)}_{ss',n},\overline{a}^{(j)}_{m,n}\big)
\label{eq:(j)DA}
\end{equation}
and
\begin{equation}
	\Gamma_{\underline{\V{a}}_{n}^{(j)}}\big(\overline{r}_{m,n}^{(j)}\big) \triangleq 
	\begin{cases} 
		0 \ist, & \text{$\overline{r}_{m,n}^{(j)} = 1$ and $\underline{a}_{ss',n}^{(j)} = m$} \\
		%\rd{\underline{r}^{(j)}_{s,n}\, p^{(j)}_{\mathrm{d}}\big (\V{p}_n,\underline{\V{p}}^{(j)}_{s,\text{mva}}\big)}
		1
		\ist, & \text{otherwise}
	\end{cases}\ist.\label{eq:indicatorNew}
\end{equation}
The function
\begin{align}
	{p}^{(j)}_{\mathrm{d} ss'}\big(\V{p}_n,\underline{\V{y}}^{(j)}_{s,n},\underline{\V{y}}^{(j)}_{s',n}\big) 
	\triangleq \begin{cases} 
		 \underline{r}^{(j)}_{s,n}\,\underline{r}^{(j)}_{s',n}\, p^{(j)}_{\mathrm{d}}\big (\V{p}_n,\underline{\V{p}}^{(j)}_{s,\text{mva}},\underline{\V{p}}^{(j)}_{s',\text{mva}}\big) \ist, & s \neq s' \land (s,s') \neq (0,0)  \\
		 \underline{r}^{(j)}_{s,n}\, p^{(j)}_{\mathrm{d}}\big (\V{p}_n,\underline{\V{p}}^{(j)}_{s,\text{mva}}\big) \ist, & s = s' \land (s,s') \neq (0,0) \\
		 p^{(j)}_{\mathrm{d}}\big (\V{p}_n\big) \ist, & (s,s') = (0,0)
	\end{cases} \label{eq:visibilityDB}
\end{align}
provides the respective detection probability for the \ac{los}, single-bounce, and double-bounce \acp{va}.
The normalization constant $ \chi_{\overline{\V{r}}_{n}^{(j)}, \underline{\V{a}}_{n}^{(j)}, M_{n}^{(j)}} $ is given as 
\begin{align} 
	\chi_{\overline{\V{r}}_{n}^{(j)}, \underline{\V{a}}_{n}^{(j)}, M_{n}^{(j)}} = \Bigg(\frac{{\mu_{\mathrm{fp}}}^{M_{n}^{(j)}} \mathrm{e}^{-\mu_{\mathrm{n}} -{\mu_{\mathrm{fp}}}}}{M_n^{(j)}!} \Bigg) \Bigg( \bigg( \frac{\mu_{\mathrm{n}}}{\mu_{\mathrm{fp}}}\bigg)^{|\mathcal{N}_{\overline{\V{r}}_n^{(j)}}|}  {\mu_{\mathrm{fp}}}^{-|\mathcal{D}_{\underline{\V{a}}_n^{(j)}}|}  \Bigg)
	\label{eq:normConst_FAR}
\end{align}
where the left-hand-side term (in brackets) is fixed after observing the current measurements given the assumption that the mean number of newly detected \acp{pmva} $ \rv{\mu}_{{n}} $ and the mean number of false alarms $ \rv{\mu}_{\mathrm{fp}} $ are known.  The right-hand-side term can be merged with factors in the sets $ \mathcal{N}_{\overline{\V{r}}_n^{(j)}} $ and $ \mathcal{D}_{\underline{\V{a}}_n^{(j)}} $\vspace{0.7mm} respectively.
The product of the prior \ac{pdf} of new \acp{pmva} \eqref{eq:priorPDF_newPSMC} and the joint conditional prior \ac{pmf} \eqref{eq:priorPDF_DA} can be written up to the normalization constant as
\begin{align}
	 f\big( \overline{\V{p}}_{\text{mva}}^{(j)} \big| \overline{\V{r}}_{n}^{(j)}, M_{n}^{(j)}, \V{x}_n)& p(\overline{\V{r}}_{n}^{(j)}, \underline{\V{a}}_{n}^{(j)}, \overline{\V{a}}_{n}^{(j)},  M_{n}^{(j)}|\underline{\V{y}}_{n}^{(j)}, \V{x}_n\big) \nn \\
	& \propto \Bigg( \psi\big(\underline{\bm{a}}_n^{(j)}, \overline{\bm{a}}_n^{(j)}\big)  \prod_{ (s,s') \in \mathcal{D}_{\underline{\V{a}}_n}^{(j)} } \dfrac{ {p}^{(j)}_{\mathrm{d} ss'} \big({\V{p}}_n,\underline{\V{y}}^{(j)}_{s,n},\underline{\V{y}}^{(j)}_{s',n}\big)  } {{\mu_{\mathrm{fp}}}} \prod_{(s'',s''') \in \tilde{\mathcal{D}}_n^{(j)} \setminus \mathcal{D}_{\underline{\V{a}}_n^{(j)}} } \bigg(1 - {p}^{(j)}_{\mathrm{d} ss'} \big({\V{p}}_n,\underline{\V{y}}^{(j)}_{s'',n},\underline{\V{y}}^{(j)}_{s''',n}\big)\bigg)  \Bigg) \nonumber \\ 
	& \quad \times \Bigg( \prod_{ m\in \mathcal{N}_{\overline{\V{r}}_n^{(j)}} } \dfrac{\mu_{n} f_{\mathrm{n}}\big( \overline{\V{p}}_{m',\text{mva}}^{(j)}\big)}{{\mu_{\mathrm{fp}}}} \Gamma_{\underline{\V{a}}_{n}}\big(\overline{r}_{m,n}^{(j)}\big) \prod_{m' \in \{ 1 , \, ... \, , M_n^{(j)}\} \setminus \mathcal{N}_{\overline{\V{r}}_n^{(j)}} }  f_{\text{d}}\big(\overline{\V{p}}_{m',\text{mva}}^{(j)}\big) \Bigg).
	\label{eq:priorPDF_DA_newPSMC_combine}
\end{align} 
With some simple manipulations using the definitions of exclusion functions $ \Psi(\underline{\V{a}}_n^{(j)},\overline{\V{a}}_n^{(j)}) $ and $ \Gamma_{\underline{\V{a}}^{(j)}_{n}}(\overline{r}_{m,n}^{(j)}) $, Eq.~\eqref{eq:priorPDF_DA_newPSMC_combine} can be rewritten as the product of factors related to the legacy \acp{pmva} and to the new \acp{pmva} respectively, i.e.,
\begin{align}
	f( \overline{\V{p}}_{\text{mva}}^{(j)} | \overline{\V{r}}_{n}^{(j)}, M_{n}^{(j)}, \V{x}_n)&p(\overline{\V{r}}_{n}^{(j)}, \underline{\V{a}}_{n}^{(j)}, \overline{\V{a}}_{n}^{(j)}, M_{n}^{(j)}| \underline{\V{y}}_{n}^{(j)}, \V{x}_n ) \nn\\[-0mm]
	& \propto 
	\Bigg(\prod^{J}_{j = 1} \underline{q}_{\mathrm{P}1}\big( \V{p}_{n}, \underline{a}^{(j)}_{00,n} \big) \prod^{M^{(j)}_{n}}_{m' = 1} \rmv\rmv \Psi\big(\underline{a}^{(j)}_{00,n'},\overline{a}^{(j)}_{m',n'} \big) \Bigg)
	\Bigg( \ist \prod^{S^{(j)}_{n}}_{s = 1} \underline{q}_{\mathrm{S}1}\big( \underline{\V{y}}^{(j)}_{s,n}\rmv, \underline{a}^{(j)}_{ss,n},  \V{p}_{n} \big) \bigg(\prod^{M^{(j)}_{n}}_{m' = 1} \rmv\rmv \Psi\big(\underline{a}^{(j)}_{ss,n},\overline{a}^{(j)}_{m',n} \big) \rmv\rmv \bigg) %
    \nn\\[-1mm] &\hspace*{2mm}\times 
	\rmv\rmv\prod^{S^{(j)}_{n}}_{s' = 1,s' \neq s} \rmv\rmv\rmv\rmv\rmv\rmv\rmv\rmv\rmv  \underline{q}_{\mathrm{D}1}\big(\underline{\V{y}}^{(j)}_{s,n}, \underline{\V{y}}^{(j)}_{s',n},\underline{a}^{(j)}_{ss',n},  \V{p}_{n}  \big) \prod^{M^{(j)}_{n}}_{m' = 1} \rmv\rmv \Psi\big(\underline{a}^{(j)}_{ss',n},\overline{a}^{(j)}_{m',n} \big) \rmv\rmv \Bigg) 
	 \Bigg(  \prod^{M^{(j)}_{n}}_{m = 1} \overline{q}_{\mathrm{S}1}\big( \overline{\V{y}}^{(j)}_{m,n}, \overline{a}^{(j)}_{m,n}, \V{p}_{n}  \big) \rmv \Bigg)\ist.
 \label{eq:priorPDF_DA_newPSMC}
\end{align}
We note that the factor $\prod^{M^{(j)}_{n'}}_{m' = 1} \rmv\rmv \Psi(\underline{a}^{(j)}_{00,n'},\overline{a}^{(j)}_{m',n'} ) $ in \eqref{eq:jointPriorPDF_global} considers the joint data association with respect to the LOS component \cite{MeyKroWilLauHlaBraWin:J18} that is assumed to always exist (but may not always be detected).
The functions related to the \ac{pa} $\underline{q}_{\mathrm{P}1}\big( \V{p}_{n}, \underline{a}^{(j)}_{00,n}\big) $ and to the legacy \ac{pmva} states $ \underline{q}_{\mathrm{S}1}( \underline{\V{y}}^{(j)}_{s,n}\rmv, \underline{a}^{(j)}_{ss,n},  \V{p}_{n} ) \rmv\rmv = \rmv\rmv \underline{q}_{\mathrm{S}1}( \underline{{r}}^{(j)}_{s,n}, \underline{\V{p}}_{s,\text{mva}}^{(j)} \rmv, \underline{a}^{(j)}_{ss,n}, \V{p}_{n} )$ and $\underline{q}_{\mathrm{D}1}\big( \underline{\V{y}}^{(j)}_{s,n}\rmv,\underline{\V{y}}^{(j)}_{s',n}\rmv, \underline{a}^{(j)}_{ss',n},  \V{p}_{n} \big) = \underline{q}_{\mathrm{D}1}\big( \underline{\V{p}}^{(j)}_{s,\mathrm{mva}}\rmv, \underline{r}^{(j)}_{s,n}\rmv,\underline{\V{p}}^{(j)}_{s',\mathrm{mva}}\rmv,$ $\underline{r}^{(j)}_{s',n}\rmv, \underline{a}^{(j)}_{ss',n}, \V{p}_{n}  \big)\vspace*{0.4mm}$ are respectively given by
\begin{align}
	\underline{q}_{\mathrm{P}1}\big( \V{p}_{n}, \underline{a}^{(j)}_{00,n}\big)  \triangleq \begin{cases}
		\displaystyle \ist \frac{ p^{(j)}_{\mathrm{d}} (\V{p}_n) }{ \mu_{\mathrm{fp}} } \ist, 
		& \!\!\rmv \underline{a}^{(j)}_{00,n} \rmv\in\rmv \Set{M}_n^{(j)}\\[3.5mm]
		1 \!-\rmv p^{(j)}_{\mathrm{d}} (\V{p}_n) \ist, & \!\!\rmv \underline{a}^{(j)}_{00,n} \!=\rmv 0   
	\end{cases} \ist,\label{eq:factorqP} \\[-5mm]\nn
\end{align}
\begin{align}
\underline{q}_{\mathrm{S}1}(  \underline{\V{p}}_{s,\text{mva}}^{(j)} \rmv, \underline{{r}}^{(j)}_{s,n} = 1, \underline{a}^{(j)}_{ss,n},  \V{p}_{n} )\triangleq
	\begin{cases}
		\dfrac{	p^{(j)}_{\mathrm{d}}\big (\V{p}_n,\underline{\V{p}}^{(j)}_{s,\text{mva}}\big) } {{\mu_{\mathrm{fp}}}}, 		& \underline{a}^{(j)}_{ss,n} \rmv\in\rmv \Set{M}_n^{(j)}\\
		1 - p^{(j)}_{\mathrm{d}}\big (\V{p}_n,\underline{\V{p}}^{(j)}_{s,\text{mva}}\big) , &\underline{a}^{(j)}_{ss,n} \!=\rmv  0
	\end{cases} \ist ,
	\label{eq:g1} \\[-5mm]\nn
\end{align}
$\underline{q}_{\mathrm{S}1}\big( \underline{\V{p}}^{(j)}_{s,\mathrm{mva}}\rmv, \underline{r}^{(j)}_{s,n} \rmv=\rmv 0, \underline{a}^{(j)}_{ss,n},  \V{p}_{n}  \big)   \rmv\triangleq\rmv \delta_{\underline{a}^{(j)}_{ss,n}}$, 
\begin{align}
	\underline{q}_{\mathrm{D}1}\big( \underline{\V{p}}^{(j)}_{s,\mathrm{mva}}\rmv, \underline{r}^{(j)}_{s,n}\rmv=\rmv1\rmv,\underline{\V{p}}^{(j)}_{s',\mathrm{mva}}\rmv, \underline{r}^{(j)}_{s',n}\rmv=\rmv1\rmv,  \underline{a}^{(j)}_{ss',n},  \V{p}_{n}  \big)%
    \triangleq \begin{cases}
		\displaystyle \ist \frac{ p^{(j)}_{\mathrm{d}} (\V{p}_n,\underline{\V{p}}^{(j)}_{s,\text{mva}},\underline{\V{p}}^{(j)}_{s',\text{mva}})}{\mu_{\mathrm{fp}}} 
		\ist, \iist
		& \!\!\rmv a^{(j)}_{ss'} \rmv\in\rmv \Set{M}_n^{(j)} \\[3.5mm]
		1 \!-\rmv p^{(j)}_{\mathrm{d}} (\V{p}_n,\underline{\V{p}}^{(j)}_{s,\text{mva}},\underline{\V{p}}^{(j)}_{s',\text{mva}})  \ist, \iist & \!\!\rmv a^{(j)}_{ss'} \!=\rmv 0    
	\end{cases}\ist,\label{eq:factorqD}
\end{align}
and $\underline{q}_{\mathrm{D}1}\big( \underline{\V{p}}^{(j)}_{s,\mathrm{mva}}\rmv, \underline{r}^{(j)}_{s,n},\underline{\V{p}}^{(j)}_{s',\mathrm{mva}}\rmv, \underline{r}^{(j)}_{s',n},  \underline{a}^{(j)}_{ss',n},  \V{p}_{n} \big)   \rmv\triangleq\rmv \delta_{\underline{a}^{(j)}_{ss',n}}$ if any $\underline{r}^{(j)}_{s'',n} \rmv=\rmv 0$ for $s'' \in \{s,s'\}$. The function $\overline{q}_{\mathrm{S}1}\big(  \overline{\V{y}}^{(j)}_{m,\mathrm{mva}}, \overline{a}^{(j)}_{m,n}, \V{p}_{n}\big) = \overline{q}_{\mathrm{S}1}\big(  \overline{\V{p}}^{(j)}_{m,\mathrm{mva}}, \overline{r}^{(j)}_{s,n}, \overline{a}^{(j)}_{m,n}, \V{p}_{n}  \big)$ related to new \ac{pmva} states reads
\begin{align}
	\overline{q}_{\mathrm{S}1}\big(  \overline{\V{p}}^{(j)}_{m,\mathrm{mva}}, \overline{r}^{(j)}_{s,n} \rmv=\rmv1, \overline{a}^{(j)}_{m,n}, \V{p}_{n} \big) %
	\hspace{0mm} \triangleq \begin{cases}
		0    \ist, 
		& \hspace{-1mm} \overline{a}^{(j)}_{m,n} \rmv\rmv\in\rmv \tilde{\Set{D}}_{n}^{(j)} \\[1mm]
		\frac{ \mu_{\mathrm{n}}f_{\mathrm{n}}(\overline{\V{p}}^{(j)}_{m,\mathrm{mva}}\ist | \ist\V{p}_{n})}{\mu_{\mathrm{fp}}} \ist,  & \hspace{-1mm} \overline{a}^{(j)}_{m,n} \rmv=\rmv 0  
	\end{cases} \nn\\[-2mm] \label{eq:factorvNewPMVAs}\\[-7mm]\nn
\end{align}
and $\overline{q}_{\mathrm{S}1}\big(  \overline{\V{p}}^{(j)}_{m,\mathrm{mva}}, \overline{r}^{(j)}_{s,n} \rmv=\rmv 0,  \overline{a}^{(j)}_{m,n}, \V{p}_{n} \big) \rmv\triangleq\rmv  f_{\text{d}}\big( \overline{\V{p}}^{(j)}_{m,\mathrm{mva}}\big)$\vspace{0.5mm}. 

Finally, by inserting (\ref{eq:priorPDF_DA_newPSMC}) into (\ref{eq:jointPriorPDF_global}), the joint prior \ac{pdf} can be rewritten as
\begin{align}
	f(\V{x}_{0:n}, &\V{y}_{0:n}, \underline{\V{a}}_{1:n},\overline{\V{a}}_{1:n}, \V{m}_{1:n})\nn\\[-1mm]
	&\hspace{0mm} \propto \rmv \Bigg( \ist f(\V{x}_{0}) \ist \prod^{S_0}_{s = 1} \ist f(\V{y}_{s,0})  \Bigg) \rmv\rmv \prod^{n}_{n' = 1} f\big(\V{x}_{n'} | \V{x}_{n'-1}\big) 
	\Bigg(\prod^{J}_{j = 1} \underline{q}_{\mathrm{P}1}\big( \V{p}_{n'}, \underline{a}^{(j)}_{00,n'}\big) \prod^{M^{(j)}_{n'}}_{m' = 1} \rmv\rmv \Psi\big(\underline{a}^{(j)}_{00,n'},\overline{a}^{(j)}_{m',n'} \big) \Bigg) \Bigg(\prod^{S_{n'\rmv-\rmv1}}_{s' = 1} f \big(\underline{\V{y}}_{s'\rmv\rmv,n'} | \V{y}_{s'\rmv\rmv,n'-1}\big) \rmv \Bigg) \nn\\[-1mm]
	&\hspace{1mm}\times \Bigg( \prod^{J}_{j' = 2} \ist\ist \bigg(\prod^{S^{(j')}_{n'}}_{s' = 1}\ist f^{(j)}\big(\underline{\V{y}}^{(j')}_{s',n'} \big| \underline{\V{y}}^{(j'-1)}_{s', n'}\big)\bigg)\Bigg) \prod^{J}_{j = 1}\Bigg( \ist \prod^{S^{(j)}_{n'}}_{s = 1} \underline{q}_{\mathrm{S}1}\big( \underline{\V{y}}^{(j)}_{s,n'}\rmv, \underline{a}^{(j)}_{ss,n'},  \V{p}_{n'}  \big) \bigg(\prod^{M^{(j)}_{n'}}_{m' = 1} \rmv\rmv \Psi\big(\underline{a}^{(j)}_{ss,n'},\overline{a}^{(j)}_{m',n'} \big) \rmv\rmv \bigg) \nn\\[-1mm]
	& \hspace{1mm} \times \rmv\rmv\rmv \prod^{S^{(j)}_{n'}}_{s' = 1,s' \neq s}  \underline{q}_{\mathrm{D}1}\big(\underline{\V{y}}^{(j)}_{s,n'}, \underline{\V{y}}^{(j)}_{s',n'},\underline{a}^{(j)}_{ss',n'},  \V{p}_{n'}  \big) \rmv\rmv\prod^{M^{(j)}_{n'}}_{m' = 1} \rmv\rmv \Psi\big(\underline{a}^{(j)}_{ss',n'},\overline{a}^{(j)}_{m',n'} \big) \Bigg) \Bigg( \prod^{M^{(j)}_{n'}}_{m = 1} \overline{q}_{\mathrm{S}1}\big( \overline{\V{y}}^{(j)}_{m,n'}, \overline{a}^{(j)}_{m,n'}, \V{p}_{n'} \big) \rmv \Bigg)
	\label{eq:jointPriorPDF_global_factorized}\\[-5mm]\nn
\end{align}

\subsection{Joint Likelihood Function}

Assume that the measurements $\RV{z}_{n}$ are independent across $n$, the conditional \ac{pdf} of $\RV{z}_{1:n}$ given $\V{x}_{1:n}$, $\underline{\RV{y}}_{1:n}$, $\overline{\RV{y}}_{1:n}$, $ \underline{\RV{a}}_{1:n} $, $ \overline{\RV{a}}_{1:n} $, and the number of measurements $\RV{m}_{1:n}$ is given by %
\begin{align}
	&f(\V{z}_{1:n}|\V{x}_{1:n}, \underline{\V{y}}_{1:n}, \overline{\V{y}}_{1:n}, \underline{\V{a}}_{1:n}, \overline{\V{a}}_{1:n}, \V{m}_{1:n}) = \prod_{n'=1}^{n} f(\V{z}_{n'}|\V{x}_{n}, \underline{\V{y}}_{n'}, \overline{\V{y}}_{n'}, \underline{\V{a}}_{n'}, \overline{\V{a}}_{n'}, M_{n'})
	\label{eq:LHF1}\\[-5mm]\nn 
\end{align}
for $\V{z}_{n}$ and $f(\V{z}_{n}|\V{x}_{n}, \underline{\V{y}}_{n}, \overline{\V{y}}_{n}, \underline{\V{a}}_{n}, \overline{\V{a}}_{n}, M_{n}) = 0$ otherwise. Assuming that the measurements $\RV{z}_{m,n}$ are conditionally independent across $m$ given $\underline{\RV{y}}_{k,n}$, $\overline{\RV{y}}_{m,n}$, $\underline{\RV{a}}_{k,n}$, $\overline{\rv{a}}_{m,n}$, and $\rv{M}_n$ \cite{BarShalomBook:Book2001,MeyKroWilLauHlaBraWin:J18}, Eq.~\eqref{eq:LHF1} factorizes as %
\begin{align}
f(\V{z}_{1:n}|\V{x}_{1:n},&\underline{\V{y}}_{1:n}, \overline{\V{y}}_{1:n}, \underline{\V{a}}_{1:n}, \overline{\V{a}}_{1:n}, \V{m}_{1:n})\nn \\ 
& \hspace*{0mm} =
	\prod_{n' = 1}^{n} C(\V{z}_{n'})  \left( \hspace{-0.1cm}\prod_{ \hspace{0.1cm} (s,s') \in  \mathcal{D}_{\rmv\rmv\underline{\V{a}}_n^{(j)}\rmv\rmv\rmv\rmv,\underline{\V{r}}_{ss'\rmv\rmv\rmv,n}^{(j)}  }  \hspace{-0.3cm}}\hspace{-0.6cm}  \dfrac{{f}_{ss'}( \V{z}_{\underline{a}_{ss',n'},n'} ) |\ist \V{p}_n, \underline{\V{p}}^{(j)}_{s,\mathrm{mva}},  \underline{\V{p}}^{(j)}_{s',\mathrm{mva}})} {f_{\mathrm{fp}}(\V{z}_{\underline{a}_{ss',n'},n'})} \right) \left(\prod_{m \in \mathcal{N}_{\overline{\V{r}}_{n'}}} \dfrac{f( \V{z}_{m,n}^{(j)} |\ist \V{p}_n, \overline{\V{p}}^{(j)}_{m,\mathrm{mva}}) } {f_{\mathrm{fp}}(\V{z}_{m,n'})}\right)
\label{eq:LHF_global}\\[-5mm]\nn
\end{align}
where $  \mathcal{D}_{\rmv\rmv\underline{\V{a}}_n^{(j)}\rmv\rmv\rmv\rmv,\underline{\V{r}}_{ss'\rmv\rmv\rmv,n}^{(j)} }\triangleq \{(s,s') \in  \mathcal{D}_{\underline{\V{a}}_n^{(j)}} : \underline{r}_{s,n}^{(j)} \neq 0 \land \underline{r}_{s',n}^{(j)} \neq 0 \} $ considers non existent \acp{pmva} and 
\begin{align}
	{f}_{ss'}( \V{z}_{m,n}^{(j)} |\ist \V{p}_n, \underline{\V{p}}^{(j)}_{s,\mathrm{mva}},  \underline{\V{p}}^{(j)}_{s',\mathrm{mva}})
	\triangleq \begin{cases} 
	 f( \V{z}_{m,n}^{(j)} |\ist \V{p}_n, \underline{\V{p}}^{(j)}_{s,\mathrm{mva}},  \underline{\V{p}}^{(j)}_{s',\mathrm{mva}}) \ist, & s \neq s' \land (s,s') \neq (0,0)  \\
	f( \V{z}_{m,n}^{(j)} |\ist \V{p}_n, \underline{\V{p}}^{(j)}_{s,\mathrm{mva}}) \ist, & s = s' \land (s,s') \neq (0,0) \\
	f( \V{z}_{m,n}^{(j)} |\ist \V{p}_n)\ist, & (s,s') = (0,0)
	\end{cases} \label{eq:visibilityDB}
\end{align}
provides the respective likelihood function for \ac{los}, single-bounce and double-bounce \acp{va}.
Since the normalization factor $ C(\V{z}_{n}) = \prod_{m=1}^{M_n}f_{\mathrm{fa}}(\V{z}_{m,n}) $ depending on $ \V{z}_{n} $ and $ M_n $ is fixed after the current measurement $ \V{z}_{n} $ is observed and using $\RV{y}_{1:n} = [\underline{\RV{y}}_{1:n}^{\T}, \overline{\RV{y}}_{1:n}^{\T}]^{\T}$, the likelihood function in \eqref{eq:LHF_global} can be rewritten up to the normalization constant as 
\begin{align}
f(\V{z}_{1:n} |&\V{x}_{1:n}, \V{y}_{1:n}, \underline{\V{a}}_{1:n},\overline{\V{a}}_{1:n}, \V{m}_{1:n})\nn\\[-1mm]
& \propto \rmv \prod^{n}_{n' = 1}  \prod^{J}_{j = 1}  \bigg(  \underline{q}_{\mathrm{P}2}\big( \V{x}_{n'}, \underline{a}^{(j)}_{00,n'} ; \V{z}_{n'}^{(j)} \big) 
\ist \prod^{S^{(j)}_{n'}}_{s = 1} \underline{q}_{\mathrm{S}2}\big( \underline{\V{y}}^{(j)}_{s,n'}\rmv, \underline{a}^{(j)}_{ss,n'},  \V{x}_{n'} ; \V{z}_{n'}^{(j)}  \big) 
\rmv\rmv\rmv \prod^{S^{(j)}_{n'}}_{s' = 1,s' \neq s}  \underline{q}_{\mathrm{D}2}\big(\underline{\V{y}}^{(j)}_{s,n'}, \underline{\V{y}}^{(j)}_{s',n'},\underline{a}^{(j)}_{ss',n'},  \V{x}_{n'} ; \V{z}_{n'}^{(j)}   \big) \bigg) 
\nn\\ 
&\hspace{2mm} \times
 \prod^{M^{(j)}_{n'}}_{m = 1} \overline{q}_{\mathrm{S}2}\big( \overline{\V{y}}^{(j)}_{m,n'}, \overline{a}^{(j)}_{m,n'},  \V{x}_{n'} ; \V{z}_{n'}^{(j)} \big) \rmv \label{eq:LHF_global_factorized} 
\end{align} 
where the factors related to the \ac{pa} $\underline{q}_{\mathrm{P}2}\big( \V{x}_{n}, \underline{a}^{(j)}_{00,n}; \V{z}_{n}^{(j)} \big) $ is given by
\begin{align}
	\underline{q}_{\mathrm{P}2}\big( \V{x}_{n}, \underline{a}^{(j)}_{00,n}; \V{z}_{n}^{(j)} \big)  \triangleq \begin{cases}
		\displaystyle \ist \frac{ f\big( \V{z}_{m,n}^{(j)} \big|\ist \V{p}_n \big)}{  f_{\mathrm{fp}}\big( \V{z}_{m,n}^{(j)} \big)}  \ist, 
		& \!\!\rmv \underline{a}^{(j)}_{00,n} = m\rmv\in\rmv \Set{M}_n^{(j)}\\[3.5mm]
		1 \ist, & \!\!\rmv \underline{a}^{(j)}_{00,n} \!=\rmv 0   
	\end{cases}\label{eq:factorqP} \\[-5mm]\nn
\end{align}
and the factors related to legacy \ac{pmva} states $ \underline{q}_{\mathrm{S}2}( \underline{\V{y}}^{(j)}_{s,n}\rmv, \underline{a}^{(j)}_{ss,n},  \V{x}_{n} ; \V{z}_{n}^{(j)} ) \rmv\rmv = \rmv\rmv \underline{q}_{\mathrm{S}2}( \underline{{r}}^{(j)}_{s,n}, \underline{\V{p}}_{s,\text{mva}}^{(j)} \rmv, \underline{a}^{(j)}_{ss,n},  \V{x}_{n}; \V{z}_{n}^{(j)} )$ and $\underline{q}_{\mathrm{D}2}\big( \underline{\V{y}}^{(j)}_{s,n}\rmv,\underline{\V{y}}^{(j)}_{s',n}\rmv, \underline{a}^{(j)}_{ss',n},  \V{x}_{n} ; \V{z}_{n}^{(j)} \big) = \underline{q}_{\mathrm{D}2}\big( \underline{\V{p}}^{(j)}_{s,\mathrm{mva}}\rmv,\underline{r}^{(j)}_{s,n}\rmv,\underline{\V{p}}^{(j)}_{s',\mathrm{mva}}\rmv,\underline{r}^{(j)}_{s',n}\rmv, \underline{a}^{(j)}_{ss',n}, \V{x}_{n} ; \V{z}_{n}^{(j)} \big)\vspace*{0.4mm}$ are given  respectively by 
\begin{align}
	\underline{q}_{\mathrm{S}2}(  \underline{\V{p}}_{s,\text{mva}}^{(j)} \rmv, \underline{{r}}^{(j)}_{s,n} = 1, \underline{a}^{(j)}_{ss,n},  \V{x}_{n} ; \V{z}_{n}^{(j)} )\triangleq
	\begin{cases}
		\dfrac{ f\big( \V{z}_{m,n}^{(j)} \big|\ist \V{p}_n, \underline{\V{p}}^{(j)}_{s,\text{mva}} \big)}{  f_{\mathrm{fp}}\big( \V{z}_{m,n}^{(j)} \big) } \ist , 		& \underline{a}^{(j)}_{ss,n} = m \rmv\in\rmv \Set{M}_n^{(j)}\\
		1  , &\underline{a}^{(j)}_{ss,n} \!=\rmv  0
	\end{cases}
	\label{eq:g1} \\[-5mm]\nn
\end{align}
and $\underline{q}_{\mathrm{S}2}\big( \underline{\V{p}}^{(j)}_{s,\mathrm{mva}}\rmv, \underline{r}^{(j)}_{s,n} \rmv=\rmv 0, \underline{a}^{(j)}_{ss,n},  \V{x}_{n} ; \V{z}_{n}^{(j)} \big)   \rmv\triangleq\rmv 1 $ and by 
\begin{align}
	\underline{q}_{\mathrm{D}2}\big( \underline{\V{p}}^{(j)}_{s,\mathrm{mva}}\rmv, \underline{r}^{(j)}_{s,n}\rmv=\rmv1\rmv,\underline{\V{p}}^{(j)}_{s',\mathrm{mva}}\rmv, \underline{r}^{(j)}_{s',n}\rmv=\rmv1\rmv,  \underline{a}^{(j)}_{ss',n},  \V{x}_{n} ; \V{z}_{n}^{(j)} \big)%
	\triangleq \begin{cases}
		\displaystyle \ist \dfrac{ f\big( \V{z}_{m,n}^{(j)} \big|\ist \V{p}_n,\underline{\V{p}}^{(j)}_{s,\text{mva}},\underline{\V{p}}^{(j)}_{s',\text{mva}})}{ f_{\mathrm{fp}}\big( \V{z}_{m,n}^{(j)}) } 
		\ist, \iist
		& \!\!\rmv a^{(j)}_{ss'} = m \rmv\in\rmv \Set{M}_n^{(j)} \\[3.5mm]
		1  \ist, \iist & \!\!\rmv a^{(j)}_{ss'} \!=\rmv 0    
	\end{cases}\nn\\[-2mm] \label{eq:factorqD}\\[-7mm]\nn
\end{align}
and $\underline{q}_{\mathrm{D}2}\big( \underline{\V{p}}^{(j)}_{s,\mathrm{mva}}\rmv, \underline{r}^{(j)}_{s,n},\underline{\V{p}}^{(j)}_{s',\mathrm{mva}}\rmv, \underline{r}^{(j)}_{s',n},  \underline{a}^{(j)}_{ss',n},  \V{x}_{n} ; \V{z}_{n}^{(j)} \big)   \rmv\triangleq\rmv 1$  if any $\underline{r}^{(j)}_{s'',n} \rmv=\rmv 0$ for $s'' \in \{s,s'\}$. The factor related to new \ac{pmva} states $\overline{q}_{\mathrm{S}2}\big(  \overline{\V{y}}^{(j)}_{m,\mathrm{mva}}, \overline{a}^{(j)}_{m,n}, \V{x}_{n} ; \V{z}_{n}^{(j)} \big) = \overline{q}_{\mathrm{S}2}\big(  \overline{\V{p}}^{(j)}_{m,\mathrm{mva}}, \overline{r}^{(j)}_{s,n}, \overline{a}^{(j)}_{m,n}, \V{x}_{n} ; \V{z}_{n}^{(j)} \big)$ is given by
 \begin{align}
 	\overline{q}_{\mathrm{S}2}\big(  \overline{\V{p}}^{(j)}_{m,\mathrm{mva}}, \overline{r}^{(j)}_{s,n} \rmv=\rmv1, \overline{a}^{(j)}_{m,n}, \V{x}_{n} ; \V{z}_{n}^{(j)} \big) %
 	\hspace{0mm} \triangleq \begin{cases}
 		0    \ist, 
 		& \hspace{-1mm} \overline{a}^{(j)}_{m,n} \rmv\rmv\in\rmv \tilde{\Set{D}}_{n}^{(j)} \\[1mm]
 		\dfrac{ f\big( \V{z}_{m,n}^{(j)} \big|\ist \V{p}_n, \underline{\V{p}}^{(j)}_{m,\text{mva}} \big)}{  f_{\mathrm{fp}}\big( \V{z}_{m,n}^{(j)} \big) } \ist,  & \hspace{-1mm} \overline{a}^{(j)}_{m,n} \rmv=\rmv 0  
 	\end{cases} \nn\\[-2mm] \label{eq:factorvNewPMVAs}\\[-7mm]\nn
 \end{align}
 and $\overline{q}_{\mathrm{S}2}\big(  \overline{\V{p}}^{(j)}_{m,\mathrm{mva}}, \overline{r}^{(j)}_{s,n} \rmv=\rmv 0,  \overline{a}^{(j)}_{m,n}, \V{x}_{n} ; \V{z}_{n}^{(j)} \big) \rmv\triangleq\rmv 1$.

\subsection{Joint Posterior PDF}

We derive the factorization of $f(\V{y}_{0:n}, \V{x}_{0:n}, \underline{\V{a}}_{1:n}, \overline{\V{a}}_{1:n}| \V{z}_{1:n} )$ considering that the measurements $\V{z}_{1:n}$ are observed and thus fixed (consequently $M_n$ is fixed as well). By using Bayes’rule and by exploiting the fact that $\V{z}_{n}$ implies $M_n$ according to \eqref{eq:LHF_global}, we obtain \cite{MeyKroWilLauHlaBraWin:Proc2018_suppl,MeyWil:TSP2021}
\begin{align}
 f(\V{y}_{0:n}, \V{x}_{0:n}, \underline{\V{a}}_{1:n}, \overline{\V{a}}_{1:n}| \V{z}_{1:n} ) &= \sum_{M'_1 =0}^\infty \sum_{M'_2 =0}^\infty \cdots \sum_{M'_n =0}^\infty f(\V{y}_{0:n}, \V{x}_{0:n}, \underline{\V{a}}_{1:n}, \overline{\V{a}}_{1:n}, \V{m}'_{1:n} | \V{z}_{1:n} ) \nn\\
 &= \sum_{M'_1 =0}^\infty \sum_{M'_2 =0}^\infty \cdots \sum_{M'_n =0}^\infty f(\V{z}_{1:n}|\V{x}_{1:n}, \V{y}_{1:n}, \underline{\V{a}}_{1:n}, \overline{\V{a}}_{1:n}, \V{m}'_{1:n})f( \V{y}_{0:n}, \V{x}_{0:n}, \underline{\V{a}}_{1:n},\overline{\V{a}}_{1:n}, \V{m}'_{1:n})
 \label{eq:jointPosterior_factorized1}
\end{align}
Using the factorized joint prior \ac{pdf} \eqref{eq:jointPriorPDF_global_factorized} and the factorized joint likelihood function \eqref{eq:LHF_global_factorized} the joint posterior \ac{pdf} \eqref{eq:jointPosterior_factorized1} can be rearranged as
\begin{align}
	\hspace*{-1mm}
	&  f(\V{y}_{0:n}, \V{x}_{0:n}, \underline{\V{a}}_{1:n}, \overline{\V{a}}_{1:n}| \V{z}_{1:n} ) \nn \\ 
	&\hspace{0mm} \propto \rmv \bigg( \ist f(\V{x}_{0}) \ist \prod^{S_0}_{s = 1} \ist f(\V{y}_{s,0})  \bigg) \rmv\rmv \prod^{n}_{n' = 1} f\big(\V{x}_{n'} | \V{x}_{n'-1}\big) 
	\bigg(\prod^{J}_{j = 1} \underline{q}_{\mathrm{P}1}\big( \V{p}_{n'}, \underline{a}^{(j)}_{00,n'}\big) \underline{q}_{\mathrm{P}2}\big( \V{x}_{n'}, \underline{a}^{(j)}_{00,n'} ; \V{z}_{n'}^{(j)} \big)  \prod^{M^{(j)}_{n'}}_{m' = 1} \rmv\rmv \Psi\big(\underline{a}^{(j)}_{00,n'},\overline{a}^{(j)}_{m',n'} \big) \bigg)  \nn\\[-1mm]
	&\hspace{1mm}\times  \bigg(  \prod^{S_{n'\rmv-\rmv1}}_{s' = 1} f \big(\underline{\V{y}}_{s'\rmv\rmv,n'} | \V{y}_{s'\rmv\rmv,n'-1}\big) \rmv \bigg) \bigg( \prod^{J}_{j' = 2} \ist\ist \bigg(\prod^{S^{(j')}_{n'}}_{s' = 1}\ist f^{(j)}\big(\underline{\V{y}}^{(j')}_{s',n'} \big| \underline{\V{y}}^{(j'-1)}_{s', n'}\big)\bigg)\bigg) \prod^{J}_{j = 1}  \bigg( \ist \prod^{S^{(j)}_{n'}}_{s = 1} \underline{q}_{\mathrm{S}1}\big( \underline{\V{y}}^{(j)}_{s,n'}\rmv, \underline{a}^{(j)}_{ss,n'},  \V{p}_{n'}  \big) \underline{q}_{\mathrm{S}2}\big( \underline{\V{y}}^{(j)}_{s,n'}\rmv, \underline{a}^{(j)}_{ss,n'},  \V{x}_{n'} ; \V{z}_{n'}^{(j)}  \big)  \nn\\[-1mm]
	& \hspace{1mm} \times  \bigg(\prod^{M^{(j)}_{n'}}_{m' = 1} \rmv\rmv \Psi\big(\underline{a}^{(j)}_{ss,n'},\overline{a}^{(j)}_{m',n'} \big) \rmv\rmv \bigg) \rmv\rmv\rmv\rmv \rmv  \prod^{S^{(j)}_{n'}}_{s' = 1,s' \neq s} \rmv \rmv \rmv \rmv \rmv \rmv   \underline{q}_{\mathrm{D}1}\big(\underline{\V{y}}^{(j)}_{s,n'}, \underline{\V{y}}^{(j)}_{s',n'},\underline{a}^{(j)}_{ss',n'},  \V{p}_{n'}  \big)  \underline{q}_{\mathrm{D}2}\big(\underline{\V{y}}^{(j)}_{s,n'}, \underline{\V{y}}^{(j)}_{s',n'},\underline{a}^{(j)}_{ss',n'},  \V{x}_{n'} ; \V{z}_{n'}^{(j)}   \big) \bigg)  \rmv\rmv\prod^{M^{(j)}_{n'}}_{m' = 1} \rmv\rmv\rmv \rmv  \Psi\big(\underline{a}^{(j)}_{ss',n'},\overline{a}^{(j)}_{m',n'} \big) \bigg)  \nn\\[-1mm]
	& \hspace{1mm} \times  \bigg( \prod^{M^{(j)}_{n'}}_{m = 1} \overline{q}_{\mathrm{S}1}\big( \overline{\V{y}}^{(j)}_{m,n'}, \overline{a}^{(j)}_{m,n'}, \V{p}_{n'} \big) \overline{q}_{\mathrm{S}2}\big( \overline{\V{y}}^{(j)}_{m,n'}, \overline{a}^{(j)}_{m,n'},  \V{x}_{n'} ; \V{z}_{n'}^{(j)} \big)  \rmv \bigg)
	\label{eq:jointPosterior_factorized}\\[-5mm]\nn
\end{align}
The factors related to the legacy \acp{pmva} and to the new \acp{pmva} can be simplified as 
$\underline{q}_{\mathrm{P}}\big( \V{x}_{n'}, \underline{a}^{(j)}_{00,n'} ; \V{z}_{n'}^{(j)} \big)  \triangleq \underline{q}_{\mathrm{P}1}\big( \V{p}_{n'}, \underline{a}^{(j)}_{00,n'}\big) \underline{q}_{\mathrm{P}2}\big( \V{x}_{n'}, \underline{a}^{(j)}_{00,n'} ; \V{z}_{n'}^{(j)} \big) $, 
$\underline{q}_{\mathrm{S}}\big( \underline{\V{y}}^{(j)}_{s,n'}\rmv, \underline{a}^{(j)}_{ss,n'},  \V{x}_{n'} ; \V{z}_{n'}^{(j)}  \big) \triangleq \underline{q}_{\mathrm{S}1}\big( \underline{\V{y}}^{(j)}_{s,n'}\rmv, \underline{a}^{(j)}_{ss,n'},  \V{p}_{n'}  \big) \underline{q}_{\mathrm{S}2}\big( \underline{\V{y}}^{(j)}_{s,n'}\rmv, \underline{a}^{(j)}_{ss,n'},  \V{x}_{n'} ; \V{z}_{n'}^{(j)}  \big) $,
$\underline{q}_{\mathrm{D}}\big(\underline{\V{y}}^{(j)}_{s,n'}, \underline{\V{y}}^{(j)}_{s',n'},\underline{a}^{(j)}_{ss',n'},  \V{x}_{n'} ; \V{z}_{n'}^{(j)}   \big) \triangleq \underline{q}_{\mathrm{D}1}\big(\underline{\V{y}}^{(j)}_{s,n'}, \underline{\V{y}}^{(j)}_{s',n'},\underline{a}^{(j)}_{ss',n'},  \V{p}_{n'}  \big)  \underline{q}_{\mathrm{D}2}\big(\underline{\V{y}}^{(j)}_{s,n'}, \underline{\V{y}}^{(j)}_{s',n'},\underline{a}^{(j)}_{ss',n'},  \V{x}_{n'} ; \V{z}_{n'}^{(j)}   \big)$, and 
$\overline{q}_{\mathrm{S}2}\big( \overline{\V{y}}^{(j)}_{m,n'}, \overline{a}^{(j)}_{m,n'},  \V{x}_{n'} ; \V{z}_{n'}^{(j)} \big) \triangleq \overline{q}_{\mathrm{S}1}\big( \overline{\V{y}}^{(j)}_{m,n'}, \overline{a}^{(j)}_{m,n'}, \V{p}_{n'} \big) \overline{q}_{\mathrm{S}2}\big( \overline{\V{y}}^{(j)}_{m,n'}, \overline{a}^{(j)}_{m,n'},  \V{x}_{n'} ; \V{z}_{n'}^{(j)} \big) $
respectively, yielding \meqref{eq:factorGraph}. 

\section{Iterative Data Association} \label{sec:DA}

This section contains the detailed messages of Section~\mref{Sec.}{sec:DAmain}. Using the messages $\beta\big( \underline{a}_{ss',n}^{(j)} \big)$ given \mref{Sec.}{sec:betaMess} as well as in \mref{Sec.}{sec:betaPAMess} and $\xi\big(\overline{a}^{(j)}_{m,n}\big)$ given in \mref{Sec.}{sec:xiMess}, messages $\eta\big( \underline{a}_{ss',n}^{(j)} \big)$ and $\varsigma\big( \overline{a}_{m,n}^{(j)} \big)$ are obtained using loopy (iterative) BP. To keep the notation concise, we also define the sets $\Set{M}_{0,n}^{(j)} \rmv\triangleq\rmv \Set{M}_n^{(j)} \cup \{0\}$ and $ \tilde{\Set{D}}^{(j)}_{0,n} \in \tilde{\Set{D}}^{(j)}_{n} \cup \{0\}$. For each measurement, $m \!\in\! \Set{M}_n^{(j)} \rmv\triangleq\rmv \{1,$ $\dots,M^{(j)}_n\}$, messages $\nu_{m\rightarrow s}^{(p)}\big(\underline{a}_{ss',n}^{(j)}\big)$ and $\zeta_{s \rightarrow m}^{(p)}\big(\overline{a}_{m,n}^{(j)}\big)$ are calculated iteratively according to \cite{WilLau:J14,MeyKroWilLauHlaBraWin:J18}\vspace*{-0.5mm}
\begin{align}
	\nu_{m\rightarrow ss'}^{(p)}\big(\underline{a}_{ss',n}^{(j)}\big) 
	&=\! \sum_{\overline{a}_{m,n}^{(j)} \in \tilde{\Set{D}}^{(j)}_{0, n}} \!\!\xi\big( \overline{a}_{m,n}^{(j)} \big) \ist \psi\big(\underline{a}_{ss',n}^{(j)}, \overline{a}_{m,n}^{(j)}\big) \prod_{(s'',s''') \in \tilde{\Set{D}}^{(j)}_{n}\backslash\{ss'\}} \!\!\! \zeta_{s''s'''\rightarrow m}^{(p-1)}\big(\overline{a}_{m,n}^{(j)}\big)
	\label{eq:featureDARV}\\
	\zeta_{ss' \rightarrow m}^{(p)}\big(\overline{a}_{m,n}^{(j)}\big) 
	&=\! \sum_{\underline{a}_{ss',n}^{(j)} \in \Set{M}_{0,n}^{(j)} } \!\!\beta\big( \underline{a}_{ss',n}^{(j)} \big) \ist 
	\psi\big(\underline{a}_{ss',n}^{(j)}, \overline{a}_{m,n}^{(j)}\big) \prod_{m' \in \Set{M}_n^{(j)}\backslash\{m\}} \!\!\! \nu_{m'\rightarrow ss'}^{(p)}\big(\underline{a}_{ss',n}^{(j)}\big) \ist, 
	\label{eq:measurementDARV} \\[-5mm]
	\nn
\end{align}	
for $ss' \rmv\in\rmv \tilde{\mathcal{D}}_{n}^{(j)}$, $m \rmv\in\rmv \mathcal{M}_{n}^{(j)}\ist$, and iteration index $p \rmv\in\rmv \{1,\ldots, P\}$. The recursion defined by \eqref{eq:featureDARV} and \eqref{eq:measurementDARV} is initialized (for $p \!=\!  0$) by $\zeta_{ss' \rightarrow m}^{(0)}\big(\overline{a}_{m,n}^{(j)}\big) \rmv =\vspace*{-.8mm}\sum^{M_n^{(j)}}_{\underline{a}_{ss',n}^{(j)} = 0} \beta\big( \underline{a}_{ss',n}^{(j)} \big) \psi\big(\underline{a}_{ss',n}^{(j)}, \overline{a}_{m,n}^{(j)}\big)$. Then, after the last iteration $p \rmv=\rmv P\rmv$, the messages $\eta\big( \underline{a}_{ss',n}^{(j)} \big)$ and $\varsigma\big( \overline{a}_{m,n}^{(j)} \big)$ are calculated as
\begin{align}
	\eta\big( \underline{a}_{ss',n}^{(j)} \big) &=\! \prod_{m \in \Set{M}_n^{(j)}} \!\!\! \nu_{m\rightarrow ss'}^{(P)}\big(\underline{a}_{ss',n} ^{(j)}\big) \\[1mm] 
	\varsigma\big( \overline{a}_{m,n}^{(j)} \big) &=\! \prod_{ss' \in \Set{D}_{n}^{(j)}} \!\!\rmv \zeta_{ss' \rightarrow m}^{(P)}\big(\overline{a}_{m,n}^{(j)}\big) \ist.
	\label{eq:messagesDA}
\end{align}

\begin{figure}[t!]
	\centering	
	\includegraphics{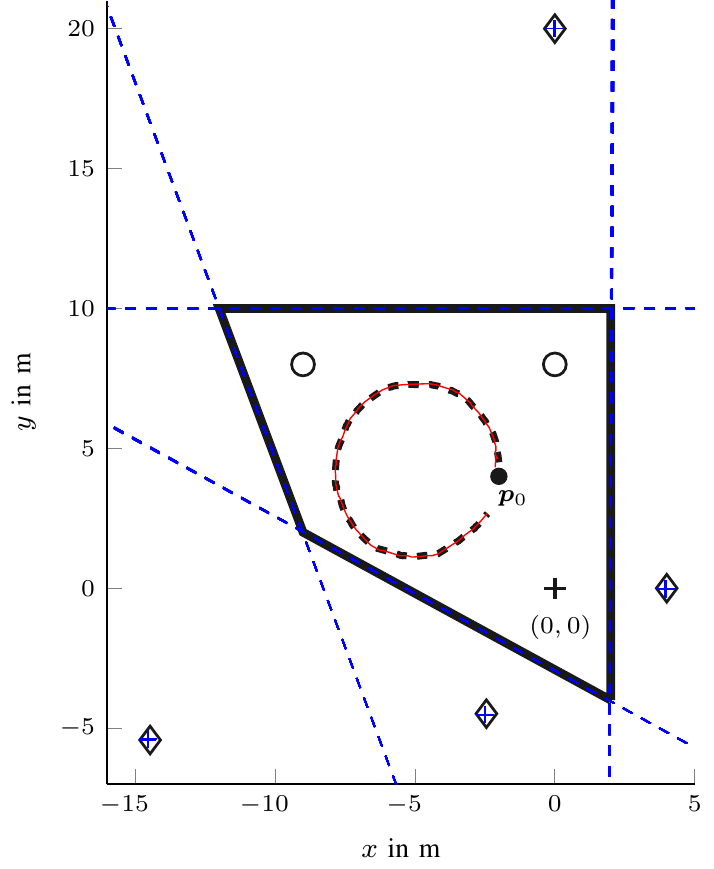}\\[1mm]
	\centering
	\includegraphics{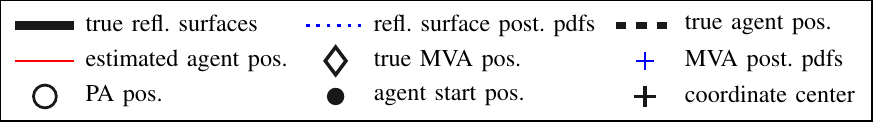}		
	\caption{Considered scenario for performance evaluation in non-rectangular room with two \acp{pa}, four reflective surfaces and corresponding \acp{mva}, as well as agent trajectory.}\label{fig:floorplan}
\end{figure}

\section{Additional Results: Performance in Non-Rectangular Room} \label{sec:app_results}

\begin{figure}[t!]
	\centering
	\subfloat[\label{fig:VAserrorA}]{\includegraphics{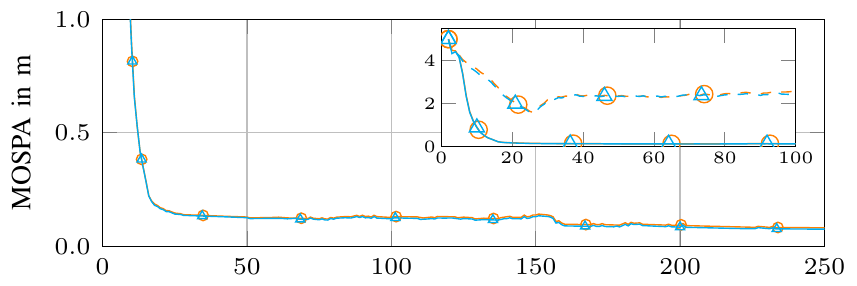}}
	\subfloat[\label{fig:MVAserrorA}]{\includegraphics{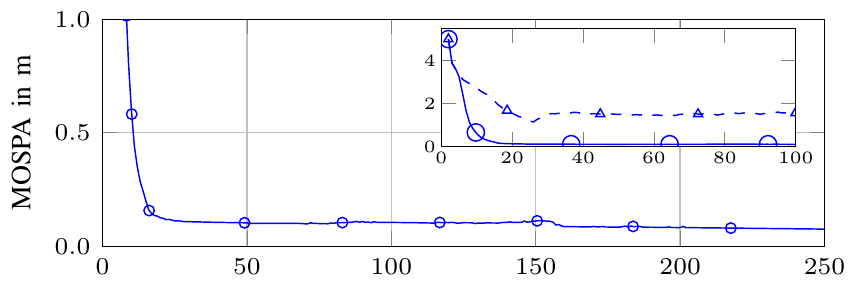}}\\[-1mm]	
	\subfloat[\label{fig:agenterrorposA}]{\includegraphics{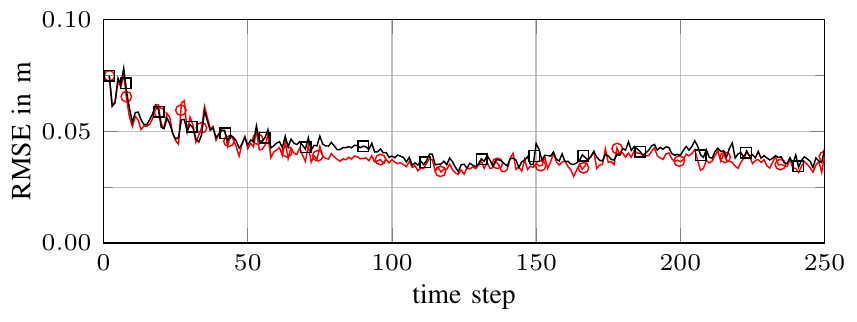}}
	\subfloat[\label{fig:agenterrororienA}]{\includegraphics{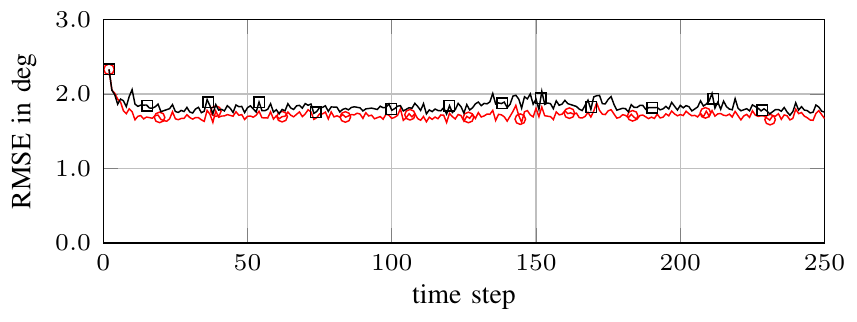}}\\[1mm]
	\centering
	\includegraphics{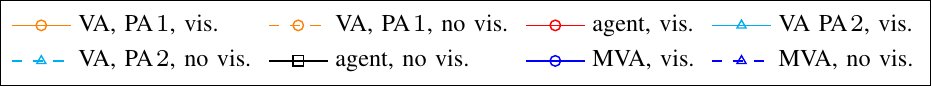}
	\vspace*{-3mm}
	\caption{Performance results: (a) MOSPA errors of the VAs of each PA, (b) MOSPA errors of the MVAs, (c) RMSEs of the mobile agent position, (d) RMSEs of the agent orientation.} \label{fig:errorsA}
\end{figure}

In this section, we provide additional simulation results using synthetic measurements. In particular, this experiment demonstrates the need for an integrated \ac{rt} to determine the available propagation paths. The setups and parameter are set according to Section~\mref{Sec.}{sec:commonSetup}.

In this experiment, we compare the complete version of the proposed \ac{mva}-based \ac{slam} algorithm (vis.), which includes the availability checks as introduced in Sec.~\mref{Fig.}{sec:Pdvisibility} and Sec.~\mref{Fig.}{sec:visibilityAlgo}, with a reduced variant, where we deactivate availability checks (no vis.), i.e., $p_{\mathrm{d}} (\V{p}_n,\V{p}^{(j)}_{s,\text{mva}}) = p_{\mathrm{d}}(\V{p}_n,\V{p}^{(j)}_{s,\text{mva}},{\V{p}}^{(j)}_{s',\text{mva}}) = p_{\mathrm{d}}$. We consider the indoor scenario shown in Figure~\ref{fig:floorplan}. The scenario consists of four reflective surfaces, i.e., $K=4$ \acp{mva}, as well as two \acp{pa} at positions $\V{p}_{\mathrm{pa}}^{(1)} = [-0.5\iist\iist 6]^{\T}$, and  $\V{p}_{\mathrm{pa}}^{(2)} = [4.2\iist\iist 1.3]^{\T}$. The acceleration noise standard deviation is $\sigma_w = 0.02\,\text{m}/\text{s}^2$.

Fig.~\ref{fig:VAserrorA} shows the \ac{mospa} errors for the two \acp{pa} all associated \acp{va}, Fig.~\ref{fig:MVAserrorA} shows the \ac{mospa} errors for all \acp{mva}, Fig.~\ref{fig:agenterrorposA} shows the \ac{rmse} of the mobile agent's position, and Fig.~\ref{fig:agenterrororienA} shows the \ac{rmse} of the mobile agent's orientation, all versus time $n$.	
As an example, Fig.~\ref{fig:floorplan} also depicts one simulation run of the complete version of the proposed method with availability checks. The posterior \acp{pdf} of the \ac{mva} positions represented by particles, the corresponding reflective surfaces, and the estimated agent tracks are also shown.
The \ac{mospa} errors in Fig.~\ref{fig:VAserrorA} and \ref{fig:MVAserrorA} of the algorithm variant with availability check converge faster and to a much smaller value than those of the variant without availability check. This is because in the scenario investigated, several \acp{va} corresponding to the left as well as the lower walls are not available over large parts of the trajectory (this is the case especially for \acp{va} of $\V{p}_{\mathrm{pa}}^{(2)}$). See also Fig.~\mref{Fig.}{fig:bounceC}, which provides a graphical explanation. Thus, the algorithm variant without availability check tends to deactivate the corresponding \acp{mva} (i.e., strongly lower the probability of existence) as some of the corresponding \acp{va}, which are expected to be detected with $p_{\mathrm{d}}$ are not observed for significant amounts of time. The \ac{rmse} of the agent position are not strongly influenced by this deactivation since still sufficient position-related information is provided by the two \acp{pa} and the remaining \acp{va}.

\section{Implementation of the Proposed \ac{mva}-based \ac{slam} Method}\label{sec:pdeudocode}

Pseudocode for one time step of the proposed \ac{mva}-based \ac{slam} method is provided in Algorithm \ref{al:mvaslam}. This pseudocode closely follows the presentation of the particle-based implementation discussed in \cite[Section~V]{Main}. Note that the existence probabilities are defined as $p^{\mathrm{e}}_{s,\text{mva}} \triangleq p(r_{s,n}=1|\V{z}_{1:n})$.

\begin{algorithm}[H]
	\SetAlgoLined
	\DontPrintSemicolon 
	\SetKwFunction{visibilityCheck}{visibilityCheck}
	\SetKwFunction{measEvalPAs}{measEvalPAs}
	\SetKwFunction{measEvalLegacyMVAs}{measEvalLegacyMVAs}
	\SetKwFunction{measEvalNewMVAs}{measEvalNewMVAs}
	\SetKwFunction{measUpdateAgent}{measUpdateAgent}
	\SetKwFunction{measUpdateLegacyMVA}{measUpdateLegacyMVA}
	\SetKwFunction{measUpdateNewMVA}{measUpdateNewMVA}
	\SetKwFunction{resamplingMVA}{resamplingMVA}
	\SetKwFunction{pruning}{pruning}
	\SetKwFunction{agentBelief}{agentBelief}
	\SetKwFunction{dataAssociation}{dataAssociation}
	\footnotesize
	\vspace{.8mm}	
	$\bigg[\big\{ \V{x}^{(i)}\big\}_{i=1}^{I},\Big\{  \big\{ \V{p}^{(i)}_{s,\text{mva}}\big\}_{i=1}^{I}, p^{\mathrm{e}}_{s,\text{mva}} \Big\}_{s=1}^{S^{(J)} +M^{(J)}} \bigg]= \text{DFSLAM}\Big[ \big\{ \V{x}^{-\ist(i)}\big\}_{i=1}^{I}, \Big\{  \big\{ \V{p}^{-\ist(i)}_{s,\text{mva}}  \big\}_{i=1}^{I}, p^{-\ist\mathrm{e}}_{s,\text{mva}} \Big\}_{k=1}^{S^{-}}\rmv, \big\{\V{z}_m \big\}_{m=1}^{M} \Big]$\\[1mm]
	
	\For{$i = 1 : I$}{
		$\tilde{\V{x}}^{(i)} \sim  f\rmv\big(\V{x}^{(i)}\big|\V{x}^{-}\big)$  \bl{\tcp*[r]{\footnotesize prediction of agent}}
	}
	\vspace{1mm}
	\For{$s = 1 : S^-$}{ \bl{\tcp*[r]{\footnotesize prediction legacy \acp{mva}}}
		\vspace{-4mm}
		Calculate $\underline{p}^{(0)\ist\mathrm{e}}_{s,\text{mva}} = p_\mathrm{s} \, p^{-\ist\mathrm{e}}_{s,\text{mva}}$\\
		\vspace{1mm}
		\For{$i = 1 : I$}{
			\vspace{0.5mm}
			Draw $\V{\omega}_{s}^{(i)}$ from a zero-mean Gaussian distribution with covariance matrix $\sigma_a^2\bold{I}_2$\\
			$\underline{\V{p}}^{(0,i)}_{s,\text{mva}}$ = $\V{p}^{(i)}_{s,\text{mva}} + \V{\omega}_{s}^{(i)}$  \bl{\tcp*[r]{\footnotesize regularize \ac{mva}}}
		} 	
	}
	\vspace{0.5mm}
	$S^{(1)} = S^-$\\[0.5mm]	
	$M^{(0)}=0$\\[0.5mm]
	\For{$j = 1 : J$}{ \vspace*{-4mm}\bl{\tcp*[r]{\footnotesize Loop \acp{pa}}}
		\For{$i = 1 : I$}{ \bl{\tcp*[r]{\footnotesize stacking \acp{mva}}}
			\vspace{-4mm}
			\For{$m = 1 : M^{(j-1)}$}{	
				$\underline{\tilde{\V{p}}}_{S^{(j-1)}+m,\text{mva}}^{(j,i)} = \overline{\V{p}}_{m,\text{mva}}^{(j-1,i)}$\\[1mm]
				$\underline{p}^{(j)\ist\mathrm{e}}_{S^{(j-1)}+m,\text{mva}} = \overline{p}^{(j)\ist\mathrm{e}}_{s,\text{mva}}$	
			}	
		}
		$S^{(j)} = S^{(j-1)} + M^{(j-1)}$\\[1mm]
		\For{$i = 1 : I$}{
			Draw $\tilde{\overline{\V{p}}}^{(j,i)}_{m,\text{mva}}$ from $f_\text{n}\big(\overline{\V{p}}^{(j)}_{m,\text{mva}}\big|\tilde{\V{x}}^{(i)}_n\big)$\\[1mm]
			Draw $\V{p}_{m,\text{va}}^{(j,i)}$ from the inverse of \meqref{eq:VADistmeas} and \meqref{eq:VAAnglemeas}\\[1mm]
			Calculate new PMVAs $\overline{\V{p}}^{(i)}_{m,\text{mva}}$ using $\V{p}_{m,\text{va}}^{(j,i)}$ and the transform in \meqref{eq:nonLinearTransformationMVA} \bl{\tcp*[r]{\footnotesize draw samples new \acp{mva}}}	
		}
		\visibilityCheck{}  \bl{\tcp*[r]{\footnotesize \acf{rt}}}
		\measEvalPAs{} \bl{\tcp*[r]{\footnotesize \acp{mva} messages to data association nodes}}
		\measEvalNewMVAs{} \bl{\tcp*[r]{\footnotesize new \acp{mva} messages to data association nodes}}
		\measEvalLegacyMVAs{} \bl{\tcp*[r]{\footnotesize legacy \acp{mva} messages to data association nodes}}
		\dataAssociation{} \bl{\tcp*[r]{\footnotesize data association}}
		\measUpdateAgent{} \bl{\tcp*[r]{\footnotesize measurement update agent}}		
		\measUpdateLegacyMVA{} \bl{\tcp*[r]{\footnotesize measurement legacy \acp{mva}}}	
		\resamplingMVA{} \bl{\tcp*[r]{\footnotesize resampling of \acp{mva}}}
		\pruning{}    \bl{\tcp*[r]{\footnotesize remove unreliable \acp{mva}}}
	}
	\agentBelief{}\bl{\tcp*[r]{\footnotesize agent belief update}}
	\For{$s = 1 : S^{(J)}$}{
		\vspace{1mm}
		$\big\{\V{p}_{s,\text{mva}}^{(i)}\big\}_{i=1}^{I} = \big\{\underline{\V{p}}_{s,\text{mva}}^{(J,i)}\big\}_{i=1}^{I}$\\[1mm]
		$p^{\ist\mathrm{e}}_{s,\text{mva}} = \underline{p}^{(J)\ist\mathrm{e}}_{s,\text{mva}}$
	}
	\For{$m = 1 : M^{(j)}$}{
		\vspace{1mm}
		$\big\{\V{p}_{S^{(J)}+m,\text{mva}}^{(i)}\big\}_{i=1}^{I} = \big\{\overline{\V{p}}_{m,\text{mva}}^{(J,i)}\big\}_{i=1}^{I}$\\[1mm]
		$p^{\ist\mathrm{e}}_{S^{(J)}+m,\text{mva}} = \overline{p}^{(J)\ist\mathrm{e}}_{m,\text{mva}}$				
	}	
	\vspace{-2mm}
	Output: $\big\{ \V{x}^{(i)}\big\}_{i=1}^{I},\Big\{  \big\{ \V{p}^{(i)}_{s,\text{mva}}\big\}_{i=1}^{I}, p^{\mathrm{e}}_{s,\text{mva}} \Big\}_{s=1}^{S^{(J)} +M^{(J)}}$
	
	\caption{Proposed Particle-Based \ac{mva} \ac{slam} Method --- Single Time Step}
	\label{al:mvaslam}
\end{algorithm}

\begin{myproc}[h]
	\SetAlgoLined
	\DontPrintSemicolon 
	\footnotesize
	\SetKwFunction{visibilityCheck}{visibilityCheck}
	\SetKwProg{proc}{Procedure}{}{}
	\proc{\visibilityCheck{}}{
		\vspace{1mm}
		\For{$i = 1 : I$}{
			\For{$s = 1 : S^{(j)}$}{	
				\vspace{1mm}
				$\V{p}_{ss,\text{va}}^{(j,i)}$ is calculated according to \meqref{eq:nonLinearTransformation} using $\tilde{\underline{\V{p}}}_{s,\text{mva}}^{(j,i)}$ and $\V{p}_{\text{pa}}^{(j)}$\\[1mm]	
				$\underline{v}_{ss}^{(i)} = f_{\text{vis}}\big(\V{x},\V{p}_{s,\text{va}}^{(j,i)} \big)$\bl{\tcp*[r]{\footnotesize recursive visibility check \cite{Bor:JASA1984,MckHam:IEEENetwork1991}}}
				\vspace{1mm}
				\For{$s' = 1 : S^{(j)}$ and $s \neq s'$}{				
					\vspace{1mm}
					$\V{p}_{ss',\text{va}}^{(j,i)}$ is calculated according to \meqref{eq:nonLinearTransformation} using $\tilde{\underline{\V{p}}}_{s,\text{mva}}^{(j,i)}$ and $\V{p}_{s,\text{va}}^{(j,i)}$\\[1mm]						
					$\underline{v}_{ss'}^{(i)} = f_{\text{vis}}\big(\V{x},\V{p}_{ss',\text{va}}^{(j,i)}\big)$  \bl{\tcp*[r]{\footnotesize recursive visibility check \cite{Bor:JASA1984,MckHam:IEEENetwork1991}}}	
				}
			}	 	
		}
	}
	\caption{Visibility Check}\label{proc:search}
\end{myproc}

\begin{myproc}[h]
	\SetAlgoLined
	\DontPrintSemicolon 
	\footnotesize
	\SetKwFunction{measEvalPAs}{measEvalPAs}
	\SetKwProg{proc}{Procedure}{}{}
	\proc{\measEvalPAs{}}{
		\vspace{1mm}
		\For{$m = 1 : M^{(j)}$}{
			\vspace{1mm}
			Calculate $\beta_{00,m}^{(j)} = \frac{1}{\mu_{\mathrm{fp}} \ist f_{\mathrm{fp}}\big(\V{z}^{(j)}_{m}\big)} \hspace{.5mm} \frac{1}{I}\sum^{I}_{i=1} p_{\mathrm{d},00}^{(j)} \ist f\big(\V{z}^{(j)}_{m}\big| \tilde{\V{x}}^{(i)} \big)$ 	
		}
	}
	\caption{measurement evaluation \acp{pa}}\label{proc:measEvalPAs}
\end{myproc}

\begin{myproc}[h]
	\SetAlgoLined
	\DontPrintSemicolon 
	\footnotesize
	\SetKwFunction{measEvalLegacyMVAs}{measEvalLegacyMVAs}
	\SetKwProg{proc}{Procedure}{}{}
	\proc{\measEvalLegacyMVAs{}}{
		\vspace{1mm}
		\For{$i = 1 : I$}{
			\vspace{1mm}
			\For{$m = 1 : M^{(j)}$}{	
				\For{$s = 1 : S^{(j)}$}{
					\vspace{1mm}
					\If{$m == 0$}{
						$\beta_{ss,m}^{(j)} = \big(1-\underline{p}^{(j)\ist\mathrm{e}}_{s}\big) + \big(1- \underline{v}_{ss}^{(i)}\iist p_{\mathrm{d},ss}^{(j)}\big)$
					}
					Calculate $\beta_{ss,m}^{(j)} = \frac{\underline{p}^{(j)\ist\mathrm{e}}_{s}}{\mu_{\mathrm{fp}} \ist f_{\mathrm{fp}}\big(\V{z}^{(j)}_{m}\big)} \hspace{.5mm} \frac{1}{I}\sum^{I}_{i=1}\iist\underline{v}_{ss}^{(i)}\iist p_{\mathrm{d},ss}^{(j)} \iist f\big(\V{z}^{(j)}_{m}\big| \tilde{\V{x}}^{(i)} ,\tilde{\underline{\V{p}}}^{(j,i)}_{s,\text{mva}}\big)$\\[1mm]
					\For{$s' = 1 : S^{(j)}$ and $s \neq s'$}{				
						\vspace{1mm}
						\If{$m == 0$}{
							Calculate $\beta_{ss',m}^{(j)} = \big(1-\underline{p}^{(j)\ist\mathrm{e}}_{s}\big)\big(1-\underline{p}^{(j)\ist\mathrm{e}}_{s'}\big)+ \big(1-\underline{v}_{ss'}^{(i)}\iist p_{\mathrm{d},ss'}^{(j)}\big)$
						}
						Calculate $\beta_{ss,m}^{(j)} = \frac{\underline{p}^{(j)\ist\mathrm{e}}_{s}\iist \underline{p}^{(j)\ist\mathrm{e}}_{s'}}{\mu_{\mathrm{fp}} \ist f_{\mathrm{fp}}\big(\V{z}^{(j)}_{m}\big)} \hspace{.5mm} \frac{1}{I}\sum^{I}_{i=1}\iist\underline{v}_{ss'}^{(i)}\iist p_{\mathrm{d},ss'}^{(j)} \iist f\big(\V{z}^{(j)}_{m}\big| \tilde{\V{x}}^{(i)} ,\tilde{\underline{\V{p}}}^{(j,i)}_{s,\text{mva}},\tilde{\underline{\V{p}}}^{(j,i)}_{s',\text{mva}}\big)$
					}
				}
			}	 	
		}
	}
	\caption{measurement evaluation legacy \acp{mva}}\label{proc:measEvalLegMVAs}
\end{myproc}

\begin{myproc}[h]
	\SetAlgoLined
	\DontPrintSemicolon 
	\footnotesize
	\SetKwFunction{measEvalNewMVAs}{measEvalNewMVAs}
	\SetKwProg{proc}{Procedure}{}{}
	\proc{\measEvalNewMVAs{}}{
		\vspace{1mm}
		\For{$i = 1 : I$}{
			\vspace{1mm}
			\For{$m = 1 : M^{(j)}$}{	
				Calculate $\zeta_{m}^{(j)} = 1 + \frac{\mu_{\mathrm{n}}}{\mu_{\mathrm{fp}} \ist f_{\mathrm{fp}}\big(\V{z}^{(j)}_{m}\big)} \hspace{.5mm} \frac{1}{I} \sum^{I}_{i=1}\iist
				f_\text{n}\big(\tilde{\overline{\V{p}}}^{(j,i)}_{m,\text{mva}}\big|\tilde{\V{x}}^{(i)}\big)\iist f\big(\V{z}^{(j)}_{m}\big| \tilde{\V{x}}^{(i)} ,\tilde{\overline{\V{p}}}^{(j,i)}_{m,\text{mva}}\big)$\\[1mm]
			}	 	
		}
	}
	\caption{measurement evaluation new \acp{mva}}\label{proc:measEvalNewMVAs}
\end{myproc}

\begin{myproc}[h]
	\SetAlgoLined
	\DontPrintSemicolon 
	\footnotesize
	\SetKwFunction{dataAssociation}{dataAssociation}
	\SetKwProg{proc}{Procedure}{}{}
	\proc{\dataAssociation{}}{
		\vspace{1mm}
		Calculate $\eta_{ss',m}^{(j)} $ and $\varsigma_{m}^{(j)}$ from $\beta_{ss',m}^{(j)}$ and $\zeta_{m,ss'}^{(j)}$ with $(s,s') \in \tilde{\mathcal{D}}^{(j)}$ and $m \in \tilde{\mathcal{M}}_0^{(j)}$ (see Section~\ref{sec:DA})
	}
	\caption{data association}\label{proc:measEvalPAs}
\end{myproc}

\begin{myproc}[h]
	\SetAlgoLined
	\DontPrintSemicolon 
	\footnotesize
	\SetKwFunction{measUpdateAgent}{measUpdateAgent}
	\SetKwProg{proc}{Procedure}{}{}
	\proc{\measUpdateAgent{}}{
		\vspace{1mm}
		Calculate $\gamma^{(j,i)}_{00} = \frac{1}{\mu_{\mathrm{fp}} \ist f_{\mathrm{fp}}\big(\V{z}^{(j)}_{m}\big)} \hspace{.5mm} \frac{1}{I}\sum_{m = 1}^{M^{(j)}} \eta_{00,m}^{(j)}\iist p_{\mathrm{d},00}^{(j)} \ist f\big(\V{z}^{(j)}_{m}\big| \tilde{\underline{\V{p}}}^{(i)}\big) + \eta_{00,0}^{(j)}\big(1- p_{\mathrm{d},00}^{(j)}\big)$\\[1mm]
		\For{$i = 1 : I$}{
			\vspace{1mm}
			\For{$s = 1 : S^{(j)}$}{
				\vspace{1mm}
				Calculate $\gamma^{(j,i)}_{ss}  = \frac{\underline{p}^{(j)\ist\mathrm{e}}_{s}}{\mu_{\mathrm{fp}} \ist f_{\mathrm{fp}}\big(\V{z}^{(j)}_{m}\big)} \hspace{.5mm} \frac{1}{I}\sum^{M^{(j)}}_{m=1}\iist\eta_{ss,m}^{(j)}\iist\underline{v}_{ss}^{(i)}\iist p_{\mathrm{d},ss}^{(j)} \iist f\big(\V{z}^{(j)}_{m}\big| \tilde{\V{x}}^{(i)} ,\tilde{\underline{\V{p}}}^{(i)}_{s,\text{mva}}\big) + \eta_{ss,0}^{(j)}\iist \underline{p}^{(j)\iist\mathrm{e}}_{s}\big(1- \underline{v}_{ss}^{(i)}\iist p_{\mathrm{d},ss}^{(j)}\big)+ \big(1-\underline{p}^{(j)\ist\mathrm{e}}_{s}\big)$\\[1mm]
				\For{$s' = 1 : S^{(j)}$ and $s \neq s'$}{				
					\vspace{1mm}
					Calculate $\gamma^{(j,i)}_{ss'} = \frac{\underline{p}^{(j)\ist\mathrm{e}}_{s}\iist \underline{p}^{(j)\ist\mathrm{e}}_{s'}}{\mu_{\mathrm{fp}} \ist f_{\mathrm{fp}}\big(\V{z}^{(j)}_{m}\big)} \hspace{.5mm} \frac{1}{I}\sum^{M^{(j)}}_{m=1}\iist\eta_{ss',m}^{(j)}\iist\underline{v}_{ss'}^{(i)}\iist p_{\mathrm{d},ss'}^{(j)} \iist f\big(\V{z}^{(j)}_{m}\big| \tilde{\V{x}}^{(i)} ,\tilde{\underline{\V{p}}}^{(i)}_{s,\text{mva}},\tilde{\underline{\V{p}}}^{(i)}_{s',\text{mva}}\big) +\eta_{ss',0}^{(j)}\iist\underline{p}^{(j)\ist\mathrm{e}}_{s}\iist\underline{p}^{(j)\ist\mathrm{e}}_{s'}\big(1- \underline{v}_{ss'}^{(i)}\iist p_{\mathrm{d},ss'}^{(j)}\big) + \big(1-\underline{p}^{(j)\ist\mathrm{e}}_{s}\big)\big(1-\underline{p}^{(j)\ist\mathrm{e}}_{s'}\big)$
				}
			} 	
		}
	}
	\caption{measurement update agent}\label{proc:measUpdateAgent}
\end{myproc}

\begin{myproc}[h]
	\SetAlgoLined
	\DontPrintSemicolon 
	\footnotesize
	\SetKwFunction{measUpdateLegacyMVA}{measUpdateLegacyMVA}
	\SetKwProg{proc}{Procedure}{}{}
	\proc{\measUpdateLegacyMVA{}}{
		\vspace{1mm}
		\For{$i = 1 : I$}{
			\vspace{1mm}
			\For{$s = 1 : S^{(j)}$}{
				\vspace{1mm}
				Calculate $\rho^{(j,i)}_{ss,1}  = \frac{\underline{p}^{(j)\ist\mathrm{e}}_{s}}{\mu_{\mathrm{fp}} \ist f_{\mathrm{fp}}\big(\V{z}^{(j)}_{m}\big)} \hspace{.5mm} \frac{1}{I}\sum^{M^{(j)}}_{m=1}\iist\eta_{ss,m}^{(j)}\iist\underline{v}_{ss}^{(i)}\iist p_{\mathrm{d},ss}^{(j)} \iist f\big(\V{z}^{(j)}_{m}\big| \tilde{\V{x}}^{(i)} ,\V{p}^{(i)}_{s,\text{mva}}\big) + \eta_{ss,0}^{(j)}\iist\underline{p}^{(j)\ist\mathrm{e}}_{s}\big(1- \underline{v}_{ss}^{(i)}\iist p_{\mathrm{d},ss}^{(j)}\big)$\\[1mm]
				Calculate $\rho^{(j)}_{ss,0}  =\eta_{ss,0}^{(j)}$\\[1mm]
				\For{$s' = 1 : S^{(j)}$ and $s \neq s'$}{				
					\vspace{1mm}
					Calculate $\rho^{(j,i)}_{ss',1}  = \frac{\underline{p}^{(j)\ist\mathrm{e}}_{s}\underline{p}^{(j)\ist\mathrm{e}}_{s'}}{\mu_{\mathrm{fp}} \ist f_{\mathrm{fp}}\big(\V{z}^{(j)}_{m}\big)} \hspace{.5mm} \frac{1}{I}\sum^{M^{(j)}}_{m=0}\iist\eta_{ss',m}^{(j)}\iist\underline{v}_{ss'}^{(i)}\iist p_{\mathrm{d},ss'}^{(j)} \iist f\big(\V{z}^{(j)}_{m}\big| \tilde{\V{x}}^{(i)} ,\V{p}^{(i)}_{s,\text{mva}},\V{p}^{(i)}_{s',\text{mva}}\big) + \eta_{ss',0}^{(j)}\iist \underline{p}^{(j)\mathrm{e}}_{s}\iist\underline{p}^{(j)\ist\mathrm{e}}_{s'}$ $\times\big(1- \underline{v}_{ss'}^{(i)} \iist p_{\mathrm{d},ss'}^{(j)}\big)$\\[1mm]
					Calculate $\rho^{(j)}_{ss',0}  =\eta_{ss',0}^{(j)}$\\[1mm]
				}
			}
			Calculate $\gamma^{(j,i)}_{s,1} = \rho^{(j,i)}_{ss,1} + \prod^{S_n^{(j)}}_{s' = 1, s\neq s'} \rho^{(j,i)}_{ss',1}$ \\[1mm]
			Calculate $\gamma^{(j)}_{s,0} = \rho^{(j)}_{ss,0} + \prod^{S_n^{(j)}}_{s' = 1, s\neq s'} \rho^{(j)}_{ss',0}$
		}
		\For{$s = 1 : S^{(j)}$}{
			$\underline{p}^{(j)\ist\mathrm{e}}_{s} = \frac{\sum_{i=1}^{(I)}\gamma^{(j,i)}_{s,1}}{\gamma^{(j)}_{s,0} + \sum_{i=1}^{(I)} \gamma^{(j,i)}_{s,1}}$\\[1mm]
			$\gamma^{(j,i)}_{s} = \frac{\gamma^{(j,i)}_{s,1}}{\sum_{i=1}^{(I)} \gamma^{(j,i)}_{s,1}}$
		}
	}
	\caption{measurement update legacy \acp{mva}}\label{proc:measUpdateLegPMVA}
\end{myproc}

\begin{myproc}[h]
	\SetAlgoLined
	\footnotesize
	\DontPrintSemicolon 
	\SetKwFunction{measUpdateNewMVA}{measUpdateNewMVA}
	\SetKwProg{proc}{Procedure}{}{}
	\proc{\measUpdateNewMVA{}}{
		\vspace{1mm}
		\For{$i = 1 : I$}{
			\vspace{1mm}
			\For{$m = 1 : M^{(j)}$}{
				\vspace{1mm}
				Calculate $\phi^{(j,i)}_{m,1}  = \frac{1}{\mu_{\mathrm{fp}} \ist f_{\mathrm{fp}} \big(\V{z}^{(j)}_{m}\big)}\iist\frac{1}{I}\iist\zeta_{m,0}^{(j)}\iist f\big(\V{z}^{(j)}_{m}\big| \tilde{\V{x}}^{(i)} ,\overline{\V{p}}^{(j,i)}_{m,\text{mva}}\big) $\\[1mm]
				Calculate $\phi^{(j)}_{m,0}  = \sum_{\tilde{\mathcal{D}}^{(j)}_{0}} \varsigma_{m,ss'}^{(j)}$\\[1mm]	
				$\overline{p}^{(j)\ist\mathrm{e}}_{m} = \frac{\sum_{i=1}^{I}\phi^{(j,i)}_{m,1}}{\phi^{(j)}_{m,0} + \sum_{i=1}^{I} \phi^{(j,i)}_{m,1}}$\\[1mm]
				$\phi^{(j,i)}_{m} = \frac{\phi^{(j,i)}_{m,1}}{\sum_{i=1}^{I} \phi^{(j,i)}_{m,1}}$
			}
		}
	}
	\caption{measurement update new \acp{mva}}\label{proc:measUpdateNewPMVA}
\end{myproc}

\begin{myproc}[h]
	\SetAlgoLined
	\DontPrintSemicolon 
	\footnotesize
	\SetKwFunction{resamplingMVA}{resamplingMVA}
	\SetKwProg{proc}{Procedure}{}{}
	\proc{\resamplingMVA{}}{
		\vspace{1mm}
		\For{$s = 1 : S^{(j)}$}{
			\vspace{1mm}
			$\Big[\Big\{\frac{1}{I}\rmv,\underline{\V{p}}_{s,\text{mva}}^{(j,i)}\Big\}_{i=1}^I\Big]=\text{resampling}\Big(\Big\{\gamma^{(j,k)}_{s}\rmv,\tilde{\underline{\V{p}}}_{s,\text{mva}}^{(j,k)}\Big\}_{k=1}^I\Big)$ 
			\hspace{27mm} \bl{\tcp*[r]{\scriptsize systematic resampling \cite{AruMasGorCla:TSP2002}}}
		}
		\For{$m = 1 : M^{(j)}$}{
			\vspace{1mm}
			$\Big[\Big\{\frac{1}{I}\rmv,\overline{\V{p}}_{m,\text{mva}}^{(j,i)}\Big\}_{i=1}^I\Big]=\text{resampling}\Big(\Big\{\phi^{(j,i)}_{m}\rmv,\tilde{\overline{\V{p}}}_{m,\text{mva}}^{(j,k)}\Big\}_{k=1}^I\Big)$
			\hspace{25mm} \bl{\tcp*[r]{\scriptsize systematic resampling \cite{AruMasGorCla:TSP2002}}}				
		}
	}
	\caption{resampling of \acp{mva}}\label{proc:measEvalPAs}
\end{myproc}

\begin{myproc}[h]
	\SetAlgoLined
	\DontPrintSemicolon 
	\footnotesize
	\SetKwFunction{pruning}{pruning}
	\SetKwProg{proc}{Procedure}{}{}
	\proc{\pruning{}}{
		\vspace{1mm}
		$S_\text{pr} =0$\\[1mm]
		\For{$s = 1 : S^{(j)}$}{
			\vspace{1mm}
			\If{$\underline{p}^{(j)\ist\mathrm{e}}_{s} < p_{\text{pr}}$ }{
				$\underline{p}^{(j)\ist\mathrm{e}}_{s} = [\iist]$\\[1mm]
				$S_\text{pr} = S_\text{pr} + 1$\\[1mm]
				\For{$i = 1 : I$}{
					$\underline{\V{p}}_{s,\text{mva}}^{(j,i)} = [\iist]$
				}
			}
		}
		$S^{(j)} = S^{(j)} - S_\text{pr}$\\[1mm]
		$M_\text{pr} = 0$\\[1mm]
		\For{$m = 1 : M^{(j)}$}{
			\vspace{1mm}
			\If{$\overline{p}^{(j)\ist\mathrm{e}}_{m} < p_{\text{pr}}$ }{
				$\underline{p}^{(j)\ist\mathrm{e}}_{s} = [\iist]$\\[1mm]
				$M_\text{pr} = M_\text{pr} + 1$\\[1mm]
				\For{$i = 1 : I$}{
					$\overline{\V{p}}_{m,\text{mva}}^{(j,i)} = [\iist]$
				}
			}				
		}
		$M^{(j)} = M^{(j)} - M_\text{pr}$
	}
	\caption{pruning of \acp{mva}}\label{proc:measEvalPAs}
\end{myproc}

\FloatBarrier
\begin{myproc}[h]
	\SetAlgoLined
	\DontPrintSemicolon 
	\scriptsize
	\SetKwFunction{agentBelief}{agentBelief}
	\SetKwProg{proc}{Procedure}{}{}
	\proc{\agentBelief{}}{
		\vspace{1mm}
		\For{$i = 1 : I$}{
			\vspace{1mm}
			$\gamma^{(i)} = \prod_{ss' \in \tilde{\mathcal{D}}^{(J)}}\gamma^{(J,i)}_{ss'} $\\[1mm]
			$\gamma^{(i)} = \frac{\gamma^{(i)}}{\sum_{i=1}^{(I)} \gamma^{(i)}}$\\[1mm]
			$\Big[\Big\{\frac{1}{I}\rmv,\V{x}^{(i)}\Big\}_{i=1}^I\Big]=\text{resampling}\Big(\Big\{\gamma^{(i)}\rmv,\tilde{\V{x}}^{(i)}\Big\}_{k=1}^I\Big)$
			\hspace{25mm} \bl{\tcp*[r]{\scriptsize systematic resampling \cite{AruMasGorCla:TSP2002}}}
		}
		
	}
	\caption{agent belief calculations \acp{mva}}\label{proc:measUpdateLegPMVA}
\end{myproc}

\bibliographystyle{IEEEtran}
\bibliography{IEEEabrv,StringDefinitions,Books,References,XRefs}